\documentclass[longauth]{aa} % for a printer version

\usepackage{graphicx}
\usepackage{txfonts}
\usepackage{hyperref}
\usepackage{color}
\usepackage{epstopdf}

\providecommand{\sorthelp}[1]{}

\newcommand{\eq}[1]{\begin{equation}#1\end{equation}}
\usepackage{amssymb}
\usepackage{xspace}

\newcommand{\bs}{\ensuremath{\boldsymbol{s}}\xspace}

\newcommand{\bx}{\ensuremath{\boldsymbol{x}}\xspace}

\newcommand{\bI}{\ensuremath{\boldsymbol{I}}\xspace}

\newcommand{\bSigma}{\ensuremath{\boldsymbol{\Sigma}}\xspace}

\newcommand{\btheta}{\ensuremath{\boldsymbol{\theta}}\xspace}

\newcommand{\FWHM}{\ensuremath{\mathrm{FWHM}\xspace}}

\newcommand{\tr}{\ensuremath{\mbox{tr}}\xspace}

\newcommand{\var}{\mathrm{var}}

\newcommand{\CRB}{{\cal{B}}}
\newcommand{\CRBm}{\boldsymbol{\cal{B}}}
\newcommand{\calA}{{\cal{A}}}

\newcommand{\calL}{{\cal{L}}}
\newcommand{\calN}{{\cal{N}}}
\newcommand{\calS}{{\cal{S}}}
\newcommand{\emm}[1]{\ensuremath{#1}}
\newcommand{\emr}[1]{\emm{\mathrm{#1}}}
\newcommand{\chem}[1]{\emr{#1}} 
\newcommand{\unit}[1]{\emr{\,#1}}
\newcommand{\hd}{\ensuremath{\mathrm{H}_2}\xspace}
\newcommand{\ohd}{\ensuremath{\mathrm{o}-\mathrm{H}_2}\xspace}
\newcommand{\phd}{\ensuremath{\mathrm{p}-\mathrm{H}_2}\xspace}
\newcommand{\opr}{\ensuremath{\mathrm{OPR}}\xspace}
\newcommand{\Eup}{\ensuremath{E_{\mathrm{up}}\xspace}}
\newcommand{\nhd}{\ensuremath{n_{\mathrm{H}_2}}}

\newcommand{\ncriteff}{\ensuremath{n_\mathrm{crit}^\mathrm{eff}}}
\newcommand{\ncritthin}{\ensuremath{n_\mathrm{crit}^\mathrm{thin}}}

\newcommand{\Pth}{\ensuremath{P_\mathrm{th}}}

\newcommand{\Tkin}{\ensuremath{T_\mathrm{kin}}}
\newcommand{\Tpeak}{\ensuremath{T_\mathrm{peak}}}
\newcommand{\Tex}{\ensuremath{T_\mathrm{ex}}}
\newcommand{\Texl}{\ensuremath{T_{\mathrm{ex},l}}}
\newcommand{\TCMB}{\ensuremath{T_\mathrm{CMB}}}

\newcommand{\pc}{\unit{pc}} 
\newcommand{\mpc}{\unit{mpc}}

\newcommand{\Kpccm}{\unit{K\,cm^{-3}}} 
\newcommand{\pccm}{\unit{cm^{-3}}} 
\newcommand{\pscm}{\unit{cm^{-2}}} 
\newcommand{\kms}{\unit{km\,s^{-1}}}

\newcommand{\K}{\unit{K}}
\newcommand{\GHz}{\unit{GHz}} 
\newcommand{\Kkms}{\unit{K\,km\,s^{-1}}}

\newcommand{\el}{\chem{e^{-}}}
\newcommand{\HH}{\chem{H_2}}
\newcommand{\oHH}{\chem{o-H_2}}
\newcommand{\pHH}{\chem{p-H_2}}

\newcommand{\CO}{\chem{CO}}
\newcommand{\thCO}{\chem{^{13}CO}}
\newcommand{\twCO}{\chem{^{12}CO}} 
\newcommand{\CeiO}{\chem{C^{18}O}}
\newcommand{\HCOp}{\chem{HCO^{+}}}
\newcommand{\HthCOp}{\chem{H^{13}CO^{+}}}

\newcommand{\HCN}{\chem{HCN}}

\newcommand{\Jone}{\chem{(1-0)}}
\newcommand{\Jtwo}{\chem{(2-1)}}
\newcommand{\Jthree}{\chem{(3-2)}}
\newcommand{\Jfour}{\chem{(4-3)}}

\newcommand{\nHH}{\emm{n_{\HH}}}

\newcommand{\cbrace}[1] {\emm{\left\{ #1 \right\}}}

\newcommand{\set}[1]{\emm{\calS\cbrace{#1}}}

\newcommand{\abs}[1]{\emm{\left| #1 \right|}} % Absolute value

\begin{document}

\title{Bias versus variance when fitting multi-species molecular lines with
  a non-LTE radiative transfer model} %
\subtitle{Application to the estimation of the gas temperature and volume
  density}

\author{%
  % Paper team
  Antoine Roueff\inst{\ref{IM2NP}} %
  \and Jérôme Pety\inst{\ref{IRAM},\ref{LERMA/PARIS}} %
  \and Maryvonne Gerin\inst{\ref{LERMA/PARIS}} %
  \and Léontine E. Ségal\inst{\ref{IRAM},\ref{IM2NP}} %
  \and Javier R. Goicoechea\inst{\ref{CSIC}} %
  \and Harvey S. Liszt\inst{\ref{NRAO}} %
  \and Pierre Gratier \inst{\ref{LAB}} %
  % Non-permanent researchers by alphabetical order
  \and Ivana Beslic\inst{\ref{LERMA/PARIS}} %
  \and Lucas Einig\inst{\ref{IRAM},\ref{GIPSA-Lab}} %
  \and Mathilde Gaudel\inst{\ref{LERMA/PARIS}} %
  \and Jan H. Orkisz\inst{\ref{Chalmers}} %
  \and Pierre Palud\inst{\ref{CRISTAL},\ref{LERMA/MEUDON}} %
  \and Miriam G. Santa-Maria\inst{\ref{CSIC}} %
  \and Victor de Souza Magalhaes\inst{\ref{IRAM}} %
  \and Antoine Zakardjian\inst{\ref{IRAP}}
  % Reminder of consortium by alphabetical order
  \and S\'ebastien Bardeau\inst{\ref{IRAM}} %
  \and Emeric Bron\inst{\ref{LERMA/MEUDON}} %
  \and Pierre Chainais\inst{\ref{CRISTAL}} %
  \and Simon Coudé\inst{\ref{WORC},\ref{CfA}} %
  \and Karine Demyk\inst{\ref{IRAP}} %
  \and Viviana V. Guzman\inst{\ref{Catholica}} %
  \and Annie Hughes\inst{\ref{IRAP}} %
  \and David Languignon\inst{\ref{LERMA/MEUDON}} %
  \and François Levrier\inst{\ref{LPENS}} %
  \and Dariusz C. Lis\inst{\ref{JPL}} %
  \and Jacques Le Bourlot\inst{\ref{LERMA/MEUDON}} %
  \and Franck Le Petit\inst{\ref{LERMA/MEUDON}} %
  \and Nicolas Peretto\inst{\ref{UC}} %
  \and Evelyne Roueff\inst{\ref{LERMA/MEUDON}} %
  \and Albrecht Sievers\inst{\ref{IRAM}} %
  \and Pierre-Antoine Thouvenin\inst{\ref{CRISTAL}}%
}

\institute{%
  Université de Toulon, Aix Marseille Univ, CNRS, IM2NP, Toulon, France,
  \email{antoine.roueff@univ-tln.fr}. \label{IM2NP} %
  \and IRAM, 300 rue de la Piscine, 38406 Saint Martin d'H\`eres,
  France. \label{IRAM} %
  \and LERMA, Observatoire de Paris, PSL Research University, CNRS,
  Sorbonne Universit\'es, 75014 Paris, France. \label{LERMA/PARIS} %
  \and Instituto de Física Fundamental (CSIC). Calle Serrano 121, 28006,
  Madrid, Spain. \label{CSIC} %
  \and National Radio Astronomy Observatory, 520 Edgemont Road,
  Charlottesville, VA, 22903, USA. \label{NRAO} %
  \and Laboratoire d'Astrophysique de Bordeaux, Univ. Bordeaux, CNRS, B18N,
  Allee Geoffroy Saint-Hilaire,33615 Pessac, France. \label{LAB} %
  \and Univ. Grenoble Alpes, Inria, CNRS, Grenoble INP, GIPSA-Lab,
  Grenoble, 38000, France. \label{GIPSA-Lab} %
  \and Chalmers University of Technology, Department of Space, Earth and
  Environment, 412 93 Gothenburg, Sweden. \label{Chalmers} %
  \and Univ. Lille, CNRS, Centrale Lille, UMR 9189 - CRIStAL, 59651
  Villeneuve d’Ascq, France. \label{CRISTAL} %
  \and LERMA, Observatoire de Paris, PSL Research University, CNRS,
  Sorbonne Universit\'es, 92190 Meudon, France. \label{LERMA/MEUDON} %
  \and Institut de Recherche en Astrophysique et Planétologie (IRAP),
  Université Paul Sabatier, Toulouse cedex 4, France. \label{IRAP} %
  \and Instituto de Astrofísica, Pontificia Universidad Católica de Chile,
  Av. Vicuña Mackenna 4860, 7820436 Macul, Santiago,
  Chile. \label{Catholica} %
  \and Laboratoire de Physique de l’Ecole normale supérieure, ENS,
  Université PSL, CNRS, Sorbonne Université, Université de Paris, Sorbonne
  Paris Cité, Paris, France. \label{LPENS} %
  \and Jet Propulsion Laboratory, California Institute of Technology, 4800
  Oak Grove Drive, Pasadena, CA 91109, USA. \label{JPL} \and Department of
  Earth, Environment, and Physics, Worcester State University, Worcester,
  MA 01602, USA \label{WORC} \and Harvard-Smithsonian Center for
  Astrophysics, 60 Garden Street, Cambridge, MA, 02138, USA. \label{CfA}
  \and School of Physics and Astronomy, Cardiff University, Queen's
  buildings, Cardiff CF24 3AA, UK. \label{UC} %
} %

%%%%%%%%%
% ABSTRACT
%%%%%%%%%

\abstract
% context heading (optional)
{ Robust radiative transfer techniques are requisite for efficiently
  extracting the physical and chemical information from molecular
  rotational lines.}
% aims heading (mandatory)
{ We study several hypotheses that enable robust estimations of the column
  densities and physical conditions when fitting one or two transitions per
  molecular species. We study the extent to which simplifying assumptions
  aimed at reducing the complexity of the problem introduce estimation
  biases and how to detect them.}
% methods heading (mandatory)
{ We focus on the CO and \HCOp{} isotopologues and analyze maps of a 50
  square arcminutes field. We used the RADEX escape probability model to
  solve the statistical equilibrium equations and compute the emerging line
  profiles, assuming that all species coexist. Depending on the considered
  set of species, we also fixed the abundance ratio between some species
  and explored different values.  We proposed a maximum likelihood
  estimator to infer the physical conditions and considered the effect of
  both the thermal noise and calibration uncertainty. We analyzed any
  potential biases induced by model misspecifications by comparing the
  results on the actual data for several sets of species and confirmed with
  Monte Carlo simulations. The variance of the estimations and the
  efficiency of the estimator were studied based on the Cramér-Rao lower
  bound.}
% results heading (mandatory)
{ Column densities can be estimated with 30\% accuracy, while the best
  estimations of the volume density are found to be within a factor of two.
  Under the chosen model framework, the peak \twCO{} \Jone{} is useful for
  constraining the kinetic temperature.  The thermal pressure is better and
  more robustly estimated than the volume density and kinetic temperature
  separately.  Analyzing CO and \HCOp{} isotopologues and fitting the full
  line profile are recommended practices with respect to detecting possible
  biases.}
% conclusion
{Combining a non-local thermodynamic equilibrium model with a rigorous
  analysis of the accuracy allows us to obtain an efficient estimator and
  identify where the model is misspecified.  We note that other
  combinations of molecular lines could be studied in the future. }


\date{}
\maketitle{}

%%%%%%%
% INTRO
%%%%%%%

\section{Introduction}

Recent progress in developing receiver and backend technologies has now
made it possible for a significant proportion of a giant molecular cloud to
be imaged with enough sensitivity to simultaneously map dozens of molecular
lines. \citet{Nishimura:2017} and \citet{Watanabe:2017} observed the W3(OH)
and W51 molecular clouds with the Nobeyama 45m telescope over a significant
fraction of the $87-112$~GHz spectral range, while \citet{Barnes:2020}
studied the W49N massive star-forming region with the IRAM 30m telescope in
two spectral ranges: $86.1 - 99.9$~GHz and $101.8 - 115.6$~GHz.  We studied
the Orion B star-forming cloud with the IRAM 30\,m telescope within the
framework of the project named Outstanding Radio-Imaging of OrioN-B
(ORION-B, PI: Jérôme Pety \& Maryvonne Gerin). The resulting data cube
covers $13\times18\pc$ on the sky and a bandwidth of 40\,GHz at a typical
resolution of 50\mpc{} and 0.6\kms{} and at a typical sensitivity of 0.1\,K
in the main beam temperature scale, as described by \citet{pety17} in their
discussion of results from a subset of $5.6\times7.5\pc$.  These scales are
derived assuming a distance to the Orion B cloud of 400\pc, as discussed
in \citet{pety17} and confirmed by more recent analyses based on Gaia
distances who found a distance between 397 and 410\pc{}
\citep{Zucker2021,Cao2023}. Extracting physical information from such large
spectroscopic datasets requires the use of robust and fast radiative
transfer methods that can be applied with limited human supervision. Large
imaging datasets with thousands of pixels usually include bright spectral
lines in a specific spectral range, for instance, the 3\,mm spectral
window. The number of mapped lines for a given species is then usually
limited to one.  Extracting quantitative information therefore represents a
different challenge as compared to studies of specific objects where
several transitions of a few molecular species can be obtained on a
restricted field of view.  For this reason, and because the method is
simple and fast, the local thermodynamic equilibrium (LTE) framework is
often used to provide estimations of molecular column densities and
excitation temperature over the covered large scale maps.  As shown by
\citet{roueff21}, this simple hypothesis may lead to bias in the derived
parameters as the excitation temperatures of the three main CO
isotopologues are different, the excitation temperature of \twCO{} being
typically twice as large as that of \thCO{} and \CeiO{}. In general, the
excitation temperature of the three CO isotopologues are different with
$\Tex(\twCO) \geq \Tex(\thCO) \geq \Tex(\CeiO)$. If it is not taken
properly into account, this difference may induce significant errors in the
derivation of molecular column densities and relative abundances.

Furthermore, the hypothesis that the level populations of the considered
species are at thermal equilibrium is often proven to be invalid because
the volume densities are too low for collisions to be fully effective in
balancing the populations of the energy levels \citep[see,
e.g.,][]{shirley15}. Non-LTE excitation and radiative transfer models have
thus been developed for decades to analyze the molecular line emission and
have been widely used in different contexts, most often by adding a priori
information on some properties of the considered source (e.g., its
structure) to limit the parameter space that is meant to be explored.  When
studying a specific object, an analytical model of the geometry, density
structure, and velocity field of the object is often specified. The
information in the line intensities is then used to fit the velocity field,
volume density, temperature structure, and the molecular column
densities. Additional hypotheses on the temperature can also be made,
either by using additional information (e.g., the kinetic temperature equal
to the dust temperature) or by fixing the kinetic temperature to some given
value. For instance, \citet{Tafalla:2021} used an isothermal model with
simple parametric laws for the variation of the density, velocity
dispersion, and molecular abundances as a function of the position to
analyze their survey of the Perseus molecular cloud. Outputs of
thermo-chemical models such as the Meudon PDR code~\citep{lepetit2006} can
also be used as inputs for radiative transfer calculations. Using the
volume density, kinetic temperature, and relative abundances of CS and CO
isotopologues from a plane-parallel PDR model of the Horsehead nebula,
\citet{Goico:2006} computed the expected molecular emission in an edge-on
cloud with a nonlocal and non-LTE Monte Carlo radiative transfer model, and
set the derived constraints on the sulfur abundance.  These examples show
that this approach works well for known sources with many observations
across a broad range of wavelengths accessible because it requires
substantial a priori information.
 
For sources with less available information, especially when a simple
geometry cannot be specified, fitting molecular line emission may lead to
ambiguities and uncertainties in the derived parameters. In their analysis
of the $(1-0)$ lines of \twCO{}, \thCO, and \CeiO{} in the Orion A
molecular cloud, \citet{Castets:1990} showed that the correlation between
the estimated kinetic temperature and the volume density may lead to
ambiguities and large uncertainties in these parameters. For this
particular cloud, using the peak \twCO{} intensity as a constraint for the
kinetic temperature offers relatively poor results because different values
of the kinetic temperature ought to be used for the different CO
isotopologues. \citet{Castets:1990} assumed that these different values of
the kinetic temperature are spatially uniform across the cloud and deduced
them from observations of the ammonia inversion lines towards small
sub-regions of the map.  Hence, these examples show that even with a
non-LTE model, the inversion of the radiative transfer equation is not
solely based on the information provided by the considered molecular lines;
rather, it makes use of additional a priori information, which may be
derived from the analysis of additional datasets or from a geometric or
theoretical model. The uncertainties on the derived parameters are then
more difficult to evaluate as they depend both on the quality of the data
and the quality of the model, including the additional a priori
assumptions.

The question of accuracy of the retrieval of physical conditions from a set
of observed noisy lines was discussed by \citet{Tunnard:2016} for the
observations of CO or HCN lines in distant galaxies.  In this study, the
authors used relatively optimistic observation conditions with a noise
level set either to 0 or to 10\% of the line intensity.  They showed that
the accuracy of the estimated parameters (density, temperature, and
pressure) is not better than half a dex (i.e., a factor of three) and that
using isotopologue lines helps to increase the accuracy of the
retrieval. For the considered species and lines, they show that better
results are obtained when the isotopologue abundance ratio is not fixed
because introducing a fixed value may bias the results and lead to
satisfactory fits even though the fitted parameters have significant
offsets from their true values. \citet{Tunnard:2016} analyzed CO and HCN
separately and did not attempt to combine the information from these two
species.

To study the bias in the gas column density determination based on CO \Jone
\, rotational line emission, \citet{Teng:2023} discussed how the \Jone,
\Jtwo, and \Jthree{} CO isotopologue emission, and the CO column density in
three nearby galaxies depend on the physical conditions.  Using a large
grid of RADEX models, they derived (for each pixel) the volume density,
kinetic temperature, CO column density per unit line width, and the
isotopic ratio between \twCO{} and \thCO{} and between \thCO{} and
\CeiO{}. An additional constraint on the path length along the line of
sight was added to remove models with low volume density and very large CO
column densities. In these models, the CO abundance relative to \HH{} is
fixed. The fitted results are used to estimate the \HH{} column density
from the CO column density and the conversion factor between the \twCO{}
\Jone{} intensity and the \HH{} column density for each pixel. They showed
that the variations of this conversion factor are mainly related to
variations of the CO optical depth and secondly to variations of the
kinetic temperature. The necessity to constrain the length along the line
of sight illustrates once again the presence of ambiguities in the
derivation of the physical conditions and the possibility of non physical
results. An anti-correlation between the volume density and kinetic
temperature is clearly seen in their likelihood distributions.

As in the case of LTE models, non-LTE models may also suffer from bias
induced by a priori information, but such effects are not sufficiently
documented in studies of non-LTE radiative transfer modeling. The present
paper focuses on the quantitative uncertainties obtained for the volume
density and kinetic temperature derived for the Horsehead nebula, when
analyzing the 3\,mm spectrum of the different isotopologues of CO and
\HCOp.  As shown by \citet{bron18} the combination of the \Jone{} lines of
the three main CO isotopologues and of \HCOp{} is sufficient to identify
the different regimes of volume density and FUV illumination across the
mapped area. It is now necessary to transform this qualitative study into
quantitative maps of physical parameters, using non-LTE radiative transfer
models.

Non-LTE models can be developed for different source geometries. The
simplest formulation, which is implemented in local escape probability
codes such as RADEX \citep{vandertak07}, uses a uniform 1D source
description. It solves the combined radiative transfer and molecular
excitation in a local way, where local means that molecules interact with
the local radiation field and are fully decoupled from other radiation
emission along any other position of the cloud.  The emergent radiation
results from the combination of all locally produced photons.  More
sophisticated radiative transfer codes can treat the non-local interaction
between the excitation of molecular levels at a given position of the cloud
and the radiation field from any other cloud position.  for instance the
MOLPOP-CEP code \citep{Asensio:2018} or codes based on Monte Carlo methods
\citep{Bernes:1979,Goico:2006,Brinch:2010}.  These more sophisticated
models naturally take into account the presence of density, kinetic
temperature, or relative abundance gradients along the line of sight.  Such
non-local methods are however more complicated to implement, they are
numerically heavier, and they require more {a priori} information about the
structure of the considered cloud. They provide more accurate results than
non-LTE local models, especially in the case when radiative effects become
dominant; for instance, in the case of the radiative transport of optically
thick lines in a medium where the volume density is much lower than the
critical density.

The article is organized as follows. Section~\ref{sec:methods} presents the
tools that will be used in this paper: the RADEX radiative transfer code,
the Cramér-Rao bound (CRB) used to infer a reference precision, and a
maximum likelihood estimator used to fit simultaneously several lines for
each line of sight.  After a succinct presentation of the data,
Sect.~\ref{sec:data} offers a comparison of the estimated parameters when
fitting the data under different assumptions.  We discuss the tradeoff
between variance and bias on two experiments in
Sect.~\ref{sec:MC}. Section~\ref{sec:CRB} presents a parametric study of
the achievable precision on the column densities, kinetic temperature,
volume density, and thermal pressure. Section~\ref{sec:discussion}
discusses the best strategy to maximize the precision without biasing the
results. Appendices~\ref{sec_Fisher_calc} and~\ref{sec:mle} present details
about the computation of the Cramér-Rao bound and the maximum likelihood
estimator. Appendices~\ref{app:data:supplement}
and~\ref{app:crb:supplement} show supplemental figures related to the
discussion in Sects.~\ref{sec:data} and~\ref{sec:CRB}.

%%%%%%%%%
% METHOD
%%%%%%%%%

\newcommand{\TabCollisionFile}{%
  \begin{table}
    \caption{Origin of the collision coefficients used in this article.}
    \centering %
    \begin{tabular}{ccll}
      \hline
      \hline
      Species & Collider & Database & Reference \\
      \hline
      $\twCO$   & \HH & LAMDA\tablefootmark{1} & \citet{yang2010}         \\
      $\thCO$   & \HH & LAMDA                  & \citet{yang2010}         \\
      $\CeiO$   & \HH & LAMDA                  & \citet{yang2010}         \\
      $\HCOp$   & \HH & EMAA\tablefootmark{2}  & \citet{denisalpizar2020} \\
      $\HCOp$   & \el & EMAA                   & \citet{Fuente:2008}      \\
      $\HthCOp$ & \HH & EMAA                   & \citet{denisalpizar2020} \\
      $\HthCOp$ & \el & EMAA                   & \citet{Fuente:2008}      \\
      \hline
    \end{tabular}
    \tablefoot{%
      \tablefoottext{1}{Leiden Atomic and Molecular Database (LAMDA) at
        \url{https://home.strw.leidenuniv.nl/~moldata/}}. %
      \tablefoottext{2}{Excitation of Molecules and Atoms for Astrophysics
        (EMAA) at \url{https://emaa.osug.fr}.}}
    \label{tab:collision:file}
  \end{table}
}

\section{Methods}
\label{sec:methods}

This section introduces the astrophysical framework (the non-LTE escape
probability radiative transfer code RADEX and the associated concept of
critical density) and the statistical method (the Cramér-Rao bound and our
maximum likelihood fitter) used throughout this paper.

\subsection{RADEX:\ A non-LTE radiative transfer model to estimate the gas
  volume density and kinetic temperature}
\label{sec:methods:radex}

\TabCollisionFile{}

The non-LTE radiative transfer code RADEX~\citep{vandertak07} uses an
escape probability formalism to compute the emerging radiation from an
homogeneous medium of simple geometry. In this article, we use the uniform
sphere geometry. Below, we clarify the associated notations and equations for
a single species.

As recalled recently in~\citet{roueff21}, the brightness temperature, $s,$ at
observed frequency $\nu$ around the rest frequency of a line $(\nu_{l})$
can be written as
\begin{equation}
  s(\nu)=
  \left\{ J(\Texl,\nu_l)-J(\TCMB,\nu)\right\}
  \left[1-\exp(-\Psi_l(\nu))\right].
  \label{eq_s}
\end{equation}
In this equation
\begin{itemize}
\item $J$ is defined as
  \begin{equation}
    J(T,\nu) = \frac{c^2}{2k\nu^2} B(T,\nu) = \frac{h \nu}{k}
    \frac{1}{\exp{\frac{h \nu}{k T}} - 1},
    \label{eq_J}
  \end{equation}
  where $B(T,\nu)$ is the spectral distribution of the radiation of a
  blackbody at temperature, $T$, and $h$, $k$ are the Planck and Boltzmann
  constants, respectively.
\item As we assume that the only background source of emission is the
  cosmic microwave background (CMB), we compute $J$ at its temperature
  ($\TCMB=2.73\K$).
\item $\Texl$ is the excitation temperature related to the ratio of the
  population in the upper ($n_\emr{up}$) and lower ($n_\emr{low}$) levels
  of the line, $l,$
  \begin{equation}
    \frac{n_\emr{up}}{n_\emr{low}} = \frac{g_\emr{up}}{g_\emr{low}}
    \exp\left[ - \frac{h\nu_l}{k\Texl} \right].
    \label{eq_Tex}
  \end{equation}
  The population of the different energy levels is locally computed by
  solving the statistical balance between radiative and collisional
  excitation  and de-excitation of the important energy
  levels. Table~\ref{tab:collision:file} lists the origin of the
  collisional coefficients used in this article.
\item The term $\left[1-\exp(-\Psi(\nu))\right]$ in Eq.~\ref{eq_s}
  represents the emission (or absorption) by the emitting (or absorbing)
  medium along the line of sight (LoS).  The function $\Psi$ is the opacity
  profile that corresponds to the integrated opacity through the whole
  slab.  The RADEX code computes the radiative transfer assuming that the
  opacity profile has a rectangular shape with a specified full width at
  half maximum (FWHM). To be able to fit the line profiles, we assume that
  each line $l$ is described by a Gaussian profile
  \begin{equation}
    \Psi_l(\nu) = \Psi_l(V) = 
    \tau_l\,\exp\left(-\frac{(V-C_V)^2}{2\sigma_V^2}\right),
    \label{eq_Psi}
  \end{equation}
  where $\tau_l$ is the line opacity computed by RADEX, $C_V$ and
  $\sigma_V$ are the systemic velocity and the velocity dispersion of the
  source along the studied LoS. In this equation, the velocity
  and frequency are related by the Doppler formula expressed in the radio
  convention as $\nu_l=\nu\,\left(1-\frac{V}{c}\right)$.  The FWHM used by
  RADEX is related to $\sigma_V$ through
  \begin{equation}
    \sigma_V=\frac{\FWHM}{\sqrt{8\ln 2}}.
    \label{eq_FWHM}
  \end{equation}
  The difference between the integrated opacities of the rectangular and
  Gaussian profiles is small, using
  \begin{equation}
    \int \Psi_l(V)dV %
    = \sqrt{\frac{2\pi}{8\ln 2}} \, \tau_l \, \FWHM %
    \sim 1.06 \, \tau_l \, \FWHM.
  \end{equation}
\end{itemize}
Furthermore, assuming that $J(T_\emr{CMB},\nu)= J(T_\emr{CMB},\nu_l)$ for
$\nu$ close to $\nu^\emr{red}_{l}$, Eq.~\ref{eq_s} can be simplified as
\begin{equation}
  s(V)=
  \left\{ J(\Texl,\nu_l)-J(\TCMB,\nu_l)\right\}
  \left[1-\exp(-\Psi_l(V))\right]
  \label{eq2_s}
.\end{equation}

The physical conditions are thus characterized by five unknown parameters:
the kinetic temperature, $\Tkin$, the volume density, $\nhd$, the column
density, $N,$ of the considered species, the velocity dispersion, $\sigma_V$,
and the mean velocity, $C_V$, along the considered LoS. The
emerging spectrum is computed through Eq.~\ref{eq_Psi} and
Eq.~\ref{eq2_s}, where the excitation temperature, $\Texl$, and opacity,
$\tau_l$, are computed with the RADEX code. RADEX uses
$\{\Tkin,\nhd,N,\FWHM\}$ as input parameters plus two additional parameters
that we fixed as follows.
\begin{itemize}
\item The ortho-to-para ratio ($\opr$) of the molecular hydrogen is defined
  as
  \begin{equation}
    n(\hd)=n(\phd)+n(\ohd)
    \quad\text{and}\quad
    \opr=\frac{n(\ohd)}{n(\phd)}.
  \end{equation}
  Introducing the numerical values of the level energies of molecular
  hydrogen, we obtain the thermal equilibrium value as
  \begin{equation}
    \opr=\frac
    {9 e^{-\frac{170.502}{\Tkin}}
      +21 e^{-\frac{1015.153}{\Tkin}}
      +33 e^{-\frac{2503.870}{\Tkin}}}
    {1+5e^{-\frac{509.850}{\Tkin}}
      +9 e^{-\frac{1681.678}{\Tkin}}}.
  \end{equation}
  This equation assumes that the \HH{} level population is at thermal
  equilibrium and that $\Tkin<500\K$.
\item The electron density $n(\emr{e^-})$ is derived from the gas volume
  density with
  \begin{equation}
    x(\emr{e^-})=\frac{n(\emr{e^-})}{n(\hd)}=
    \max\left(2.10^{-4}\sqrt{\frac{100}{n(\hd)}}, 10^{-8}\right).
    \label{eq:electron}
  \end{equation}
  This formula takes into account the decrease of the ionization as a
  function of density due to recombination. It includes a lower limit
  consistent with chemical model predictions in dense and dark
  regions~\citep{bron21}.
\end{itemize}
Inelastic collisions with electrons contribute to the excitation of
molecular rotational levels as long as the electron fraction remains higher
than the ratio of the downward collision rates with \HH{} and with
electrons. For \HCOp{}, collisions with electrons will be effective for an
electron fraction larger than about $3 \times 10^{-5}$ at a kinetic
temperature of 30\K. With the adopted scaling law (Eq.~\ref{eq:electron})
this corresponds to densities lower than $\sim 4\,000\pccm$. For \CO{},
only collisions with neutral species are included, because collisions with
electrons are only efficient compared to collisions with \HH{} or helium
for high dipole molecules like \HCOp{} or \HCN{} at the typical electron
fractions found in GMCs~\citep{liszt2012,goldsmith2017,santa-maria2023}.
  
\subsection{The notion of critical density to help with the interpretation
  of the results}

The critical density is the volume density at which (de-)excitations by
collisions equal radiative (de-)excitations. It is thus a crucial
characteristic of each line, which allows us to assess the typical density
above which the LTE regime is reached. As both collisional and radiative
excitation processes contribute, the critical density is usually defined in
the context of an optically thin line, $\ncritthin$, and then modified to
take into account radiative excitation~\citep{shirley15}.  The effective
critical density $\ncriteff$, including line trapping effects, can be
calculated for each line with
\begin{equation}
  \ncriteff=\ncritthin \, \beta
  \label{eq_ncriteff}
  \quad
  \text{where}
  \quad
  \beta=\frac{1.5}{\tau}\left[1-\frac{2}{\tau^2}
    +\left(\frac{2}{\tau}+\frac{2}{\tau^2}\right)e^{-\tau}
  \right].
\end{equation}
The value \ncritthin{} is computed as the sum of the Einstein coefficients,
$A$, divided by the sum of collisional de-excitation rates.  The parameter
$\beta$ is the average escape probability in the case of a uniform
sphere~\citep[the default case in RADEX, see][]{vandertak07}.

\subsection{The Cramér-Rao bound to provide a reference precision}

Once the physical model and its associated assumptions are clearly stated,
the Cramér-Rao bound delivers a lower bound of the variance of the
estimation of each fitted parameter. This means that all the information
will have been extracted from the data when we find an unbiased estimator
of any given physical parameter whose variance is equal to the Cramér-Rao
bound.

Based on a physical model and the associated assumptions, one can calculate
the Fisher matrix (defined in Eq.~\ref{eq_Fisher} of
Appendix~\ref{sec_Fisher_calc}) and then numerically compute its inverse to
get the Cramér-Rao bound matrix
\eq{ \CRBm(\btheta)=\bI_F^{-1}(\btheta).}
Noting $\theta_1,\, \theta_2, ...$ the unknown parameters of the problem,
the $i^{\mathrm{th}}$ diagonal term of the CRB matrix provides a lower
bound on any unbiased estimator $\widehat\theta_i$ of
$\theta_i$~\citep[see, e.g.,][for an example]{roueff21}
\eq{%
  \var \left(\widehat\theta_i\right) \geq \left[\CRBm(\btheta)\right]_{ii}.
  \label{eq_var_leq_CRB}
}
Thus, $\CRB^{1/2}(\theta_i)=\sqrt{\left[\CRBm(\btheta)\right]_{ii}}$ is
equivalent to a standard deviation, which can be interpreted as a precision
of reference for each physical environments (e.g., diffuse gas, dense core,
etc.) which takes into account the statistics of the noise that disturbs the
observation. The CRB is a local property. For example, for a Gaussian
noise, the CRB is the curvature of the negative log-likelihood (NLL) at its
minimum.  Thus, the CRB is useful as a first guess of the precision that
can be expected, but it is not sensitive to the presence of other local
minima, which may induce ``abnormal'' estimations in low signal-to-noise ratio
(S/N) cases~\citep{garthwaite95}.

Appendix~\ref{sec_Fisher_calc} details the computation of the Fisher matrix
for our use of the RADEX model, when the unknown parameters are
$\btheta=[\log \Tkin,\log n_{H_2},\log N,\sigma_V,C_V]$.
Section~\ref{sec:CRB} analyzes the evolution of $\CRB^{1/2}(\log N)$,
$\CRB^{1/2}(\log\nHH)$ and $\CRB^{1/2}(\Tkin)$ as a function of the volume
density and the column density, while the other parameters remain fixed. As
the thermal pressure may be better constrained than the volume density or
kinetic temperature, it is useful to also compute the Cramér-Rao bound for
the thermal pressure. Formally, since $\log \Pth=\log \Tkin + \log(\nHH)$,
one simply has
\begin{equation}
  \CRB(\log \Pth)=\CRB(\log \Tkin)+\CRB(\log \nHH)+2 \CRB(\log\Tkin,\log\nHH),
\end{equation}
where $\CRB(\log\Tkin,\log\nHH)$ is the off-diagonal term of the CRB
matrix. This term can be interpreted as the covariance between the
$\log \Tkin$ and $\log \nHH$ estimations.

\subsection{A maximum likelihood estimator to fit the data}
\label{sec:methods:mle}

In addition to the RADEX and noise models, we need an algorithm to fit the
data and deliver the associated estimation $\widehat\btheta$ of the
physical parameters $\btheta$. Each pixel is fitted independently from all
the other ones.  We will classically fit the observed data $\bx$ by looking
for the estimated $\widehat\btheta$ that maximizes the log-likelihood
function
\begin{equation}
  \widehat\btheta=\arg \max_{\btheta} \left(
    \sum_{l=1}^L \calL(\btheta;\bx_l) \right),
  \label{eq_MLE}
\end{equation}
where $l$ is the line index of the $L$ observed lines.
For a given pixel, our algorithm can be decomposed in four steps, as
described below.
\begin{description}
\item[\textbf{Signal detection and estimation of the centroid velocity}.]
  We first compute the noise level of the spectrum in channels devoid of
  signal. Then, we detect the presence of a peak in the intensity for each
  analyzed line. The initial centroid velocity for all the lines is
  estimated as the peak location on the summation of $\thCO$ and $\CeiO$
  line spectra. This is a tradeoff between fully optically thin lines, and
  one with high S/N.  We then decide whether the pixel of the image is
  further fitted. Since all species are assumed to share the same centroid
  velocity, we only require that more than half of the lines are
  detected. For example, for the set $\set{\CeiO,\HthCOp}$, we only require
  that both $\CeiO$ lines are detected.
\item[\textbf{Initialization of the fit estimation}.] We initialize the
  estimation of the other physical parameters on the point that minimizes
  the negative log-likelihood (NLL) on a grid that regularly samples the
  fitted parameters. Here it is crucial to take into account the a priori
  constraints that are assumed, for instance, the same kinetic temperature
  and volume density (see Sect.~\ref{sec:mle:initialization} for
  details). This step yields an initial estimation of all parameters
  $\btheta$. It may also induce some implicit {a priori} information
  depending on the chosen grid boundaries.
\item[\textbf{Maximization of the likelihood}.] We then refine our initial
  estimation with a standard iterative gradient descent. The iterator is
  stopped when one of the following conditions is achieved: 1) The
  difference of NLL between two consecutive iterations is smaller than
  $10^{-6}$; or 2) the number of iterations reaches $30$; or 3) the NLL
  minimum does not decrease for five consecutive iterations. This may
  happen because of numerical instability.
\item[\textbf{Estimation of the confidence interval}.] Once the physical
  parameters have been obtained, we compute the Cramér-Rao bound for each
  parameter $\CRBm(\widehat\theta_i)$ to provide a reference interval of
  confidence. Although we replaced $\theta_i$ by $\widehat\theta_i$, this
  interval of confidence remains mostly valid as long as the model is
  valid, and as soon as the estimator is efficient (i.e., it is unbiased
  and the variance is equal to the Cramér-Rao bound).
\end{description}
Additional implementation details about these implementations are provided
in Appendix~\ref{sec:mle}.

%%%%%%%%%%%%%%%%%%%%%%%%%%%%%
% DATA
%%%%%%%%%%%%%%%%%%%%%%%%%%%%%

% Presentation of the data

\newcommand{\FigPeakAreaSpectra}{%
  \begin{figure*}
    \centering %
    \includegraphics[width=\linewidth]{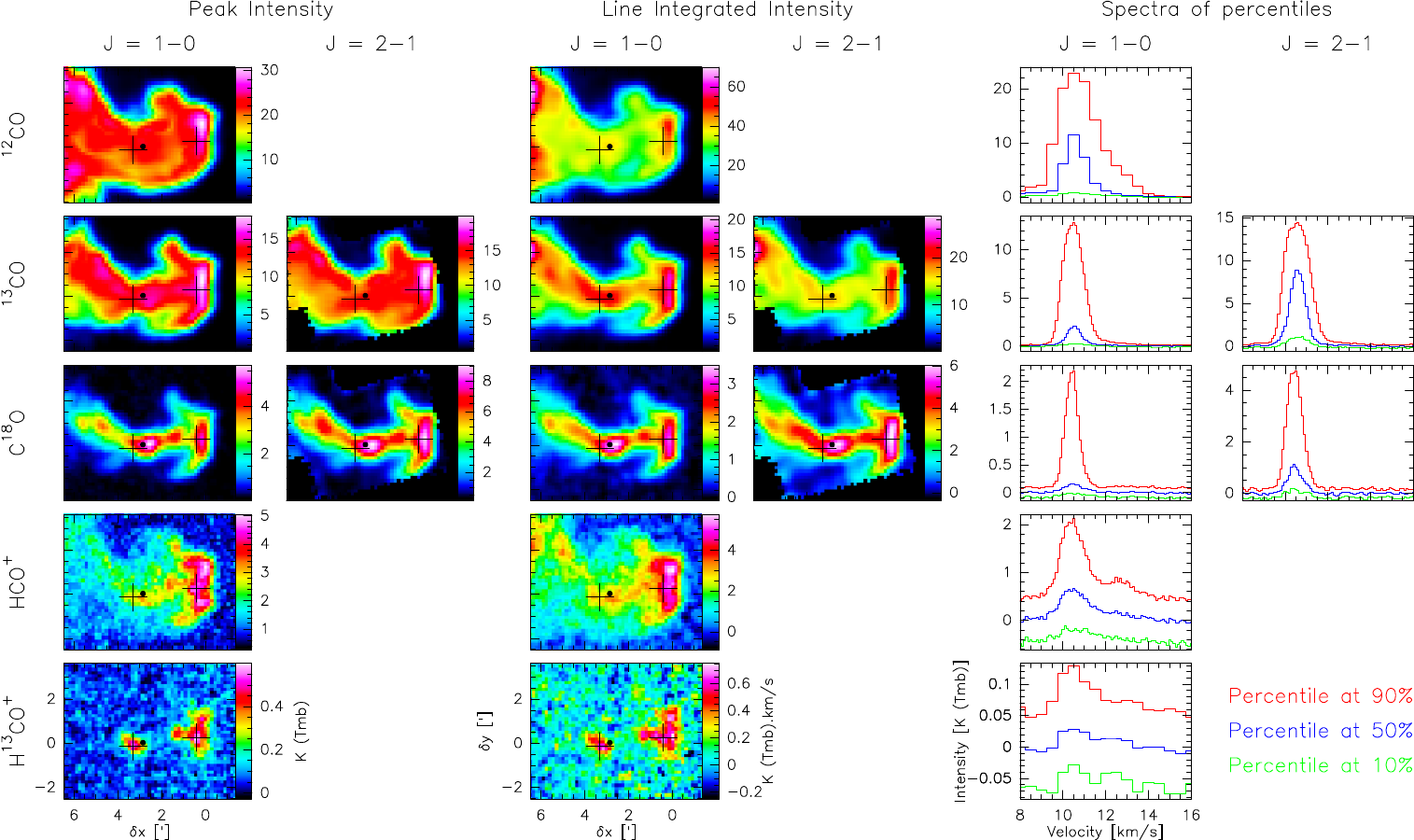}
    \caption{Presentation of the data towards the Horsehead
      pillar. \textbf{Left and middle panels:} Spatial distribution of the
      peak intensity and of the integrated intensity. The used projection
      is the Azimuthal one, centered on
      RA$=05^\emr{h}40^\emr{m}54.27^\emr{s}$, Dec$=-02^\circ{}28'00.00''$
      in the ICRS frame, and with a position angle of $14^\circ$.  The two
      crosses show the positions of the two dense cores whose coordinates
      are in Table~\ref{tab:coord}. The black dot shows the position of the
      line of sight studied in more detailed in
      Sect.~\ref{sec:MC}. \textbf{Right panel:} The green, blue, and red
      spectra show the spectra of percentiles at 10, 50, and 90\%,
      respectively, computed over the associated field of view.}
    \label{fig:data:peak:area:spectra}
  \end{figure*}
}

% Column densities

\newcommand{\FigDataHistoNthCO}{%
  \begin{figure*}
    \centering %
    \includegraphics[width=0.9\linewidth]{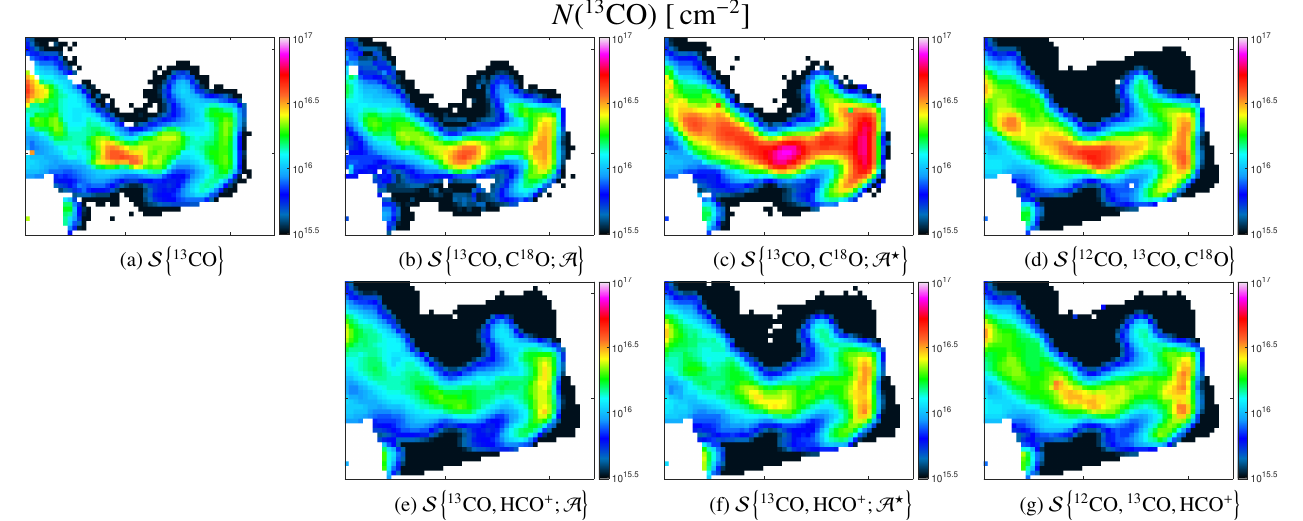}
    \\[\bigskipamount]
    \includegraphics[width=0.9\linewidth]{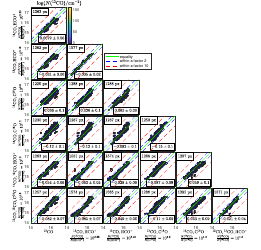}
    \caption{Comparison of the estimated column density of \thCO{} on the
      Horsehead pillar data for different sets of species and/or a
      priori. In all cases $\Tkin$, $\nHH$, $C_V$ and $\sigma_V$ are
      assumed to be the same for all lines. \textbf{Top:} Maps of estimated
      $N(\thCO)$. Table~\ref{tab_set_of_species} lists the details of each
      studied case. \textbf{Bottom:} Joint histograms of $\log N(\thCO)$
      estimations as a function of different sets of species and there
      associated {a priori} hypotheses, shown as the column and row labels.
      In each panel, the green, blue, and red lines show the identity
      function, and ratios of values within a factor 2 or 10,
      respectively. The top left and bottom right legends give the number
      of pixels used to compute the histogram, and the mean and standard
      deviation of the distance to the green line.}
    \label{fig:data:N13CO}
  \end{figure*}
}

\newcommand{\FigDataRMSETab}{%
  \begin{figure*}
    \centering %
    \includegraphics[width=\linewidth]{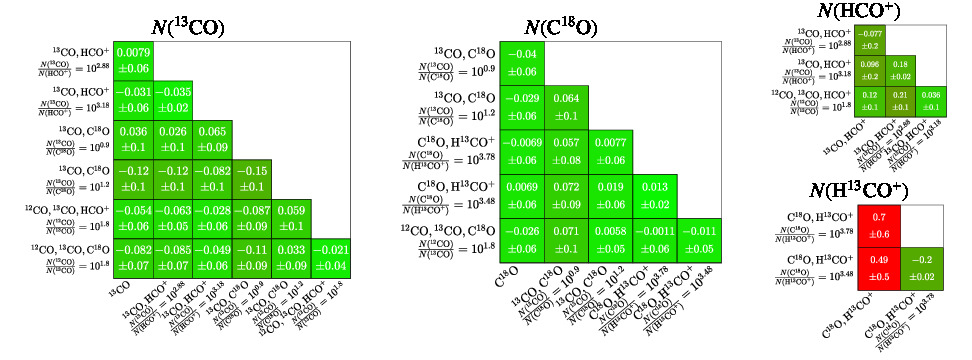}
    \includegraphics[width=\linewidth]{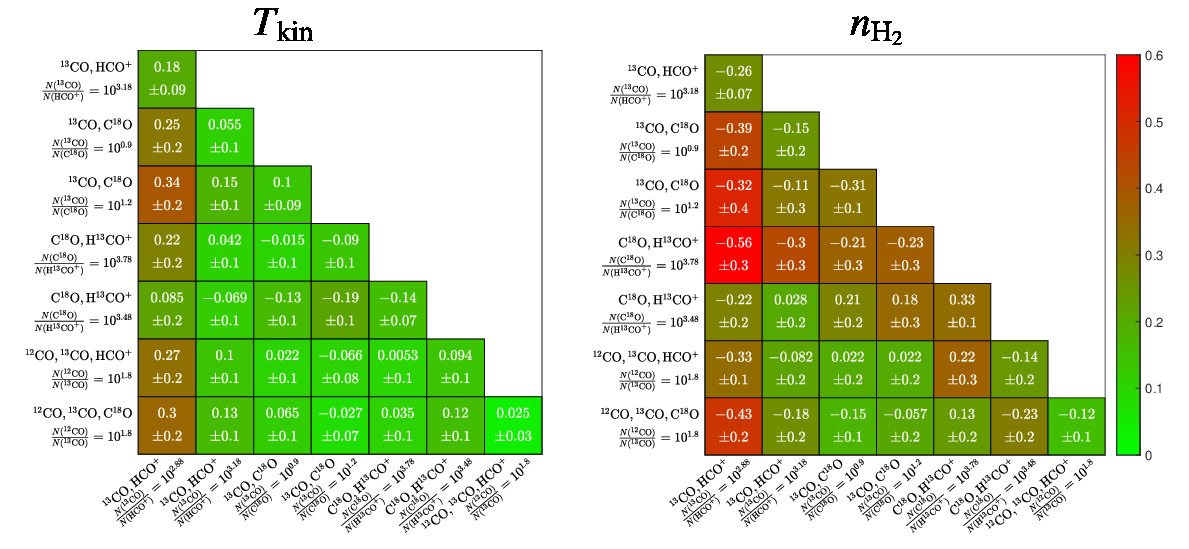}
    \caption{Mean and standard deviation of the distance in log-space
      between the estimations (i.e.,
      $\left|\widehat\theta_x-\widehat\theta_y\right|/\sqrt{2}$ where
      $\theta=\log N$, or $\log\Tkin$ or $\log\nHH$) as a function of
      different sets of species and their associated a priori
      hypotheses, shown as the column and row labels. In all cases $\Tkin$,
      $\nHH$, $C_V$ and $\sigma_V$ are assumed to be the same for all
      lines. Table~\ref{tab_set_of_species} lists the details. The cells
      are color-coded with the values of the root mean square error
      $= \sqrt{\emr{mean}^2+\emr{sdev}^2}$.  From left to right, the top
      graphs show the results for $N(\thCO),$ $N(\CeiO),$ $N(\HCOp),$ and
      $N(\HthCOp),$ and the bottom graphs the results for the kinetic
      temperature and volume density. This figure summarizes the joint
      histograms shown in Fig.~\ref{fig:data:N13CO}, \ref{fig:data:NC18O},
      \ref{fig:data:NHCOp:N13HCOp}, \ref{fig:data:Tkin}, and
      \ref{fig:data:n}.}
    \label{fig:data:rmse:tab}
  \end{figure*}
}

\newcommand{\FigDataHistoAbundances}{%
  \begin{figure*}
    \centering %
    \includegraphics[width=0.78\linewidth]{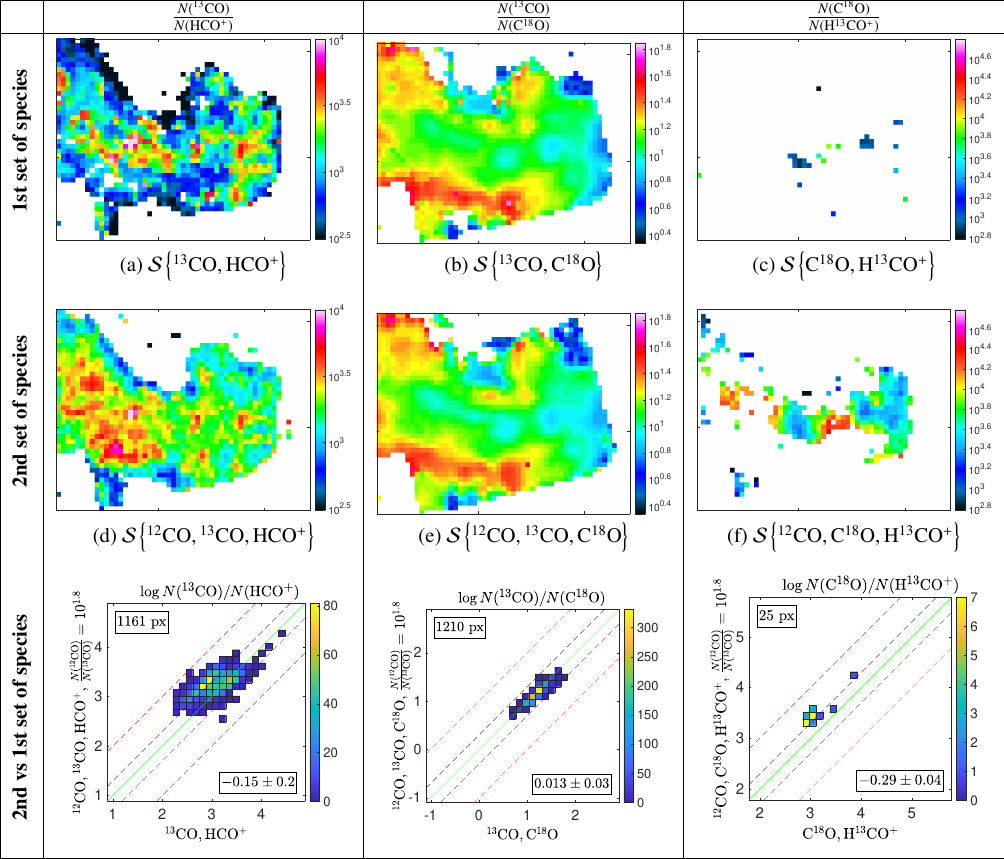}
    \caption{Maps and joint histograms of relative abundances for
      $\log\cbrace{N(\thCO)/N(\HCOp)}$ \textbf{(left)},
      $\log\cbrace{N(\thCO)/N(\CeiO)}$ \textbf{(middle)}, and
      $\log\cbrace{N(\CeiO)/N(\HthCOp)}$ \textbf{(right)}. The layout of
      the figure is similar to Fig.~\ref{fig:data:N13CO}.}
    \label{fig:data:abundances}
  \end{figure*}
}

\newcommand{\TabCoord}{%
  \begin{table}
    \centering %
    \caption{Coordinates of lines of sight marked on
      Fig.~\ref{fig:data:peak:area:spectra}.}
    \begin{tabular}{ccc}
      \hline
      \hline
      Line of sight    & R.A.           & Dec. \\
      \hline
      Dense core \#1   & $05^\emr{h}40^\emr{m}55.60^\emr{s}$ & $-02^\circ{}27'38.01''$ \\
      Dense core \#2   & $05^\emr{h}41^\emr{m}07.21^\emr{s}$ & $-02^\circ{}27'19.37''$ \\
      Studied position & $05^\emr{h}41^\emr{m}05.31^\emr{s}$ & $-02^\circ{}27'16.88''$ \\
      \hline
    \end{tabular}
    \label{tab:coord}
  \end{table}
}

\newcommand{\TabLines}{%
  \begin{table}
    \centering %
    \caption{Properties of the observed lines.}
    \resizebox{\linewidth}{!}{%
      \begin{tabular}{lccccr}
        \hline
        \hline
        Line\tablefootmark{1}  & $\nu$ & $\Eup$   & $A$           & $dV$\tablefootmark{2} & Noise\tablefootmark{3} \\
                               & GHz   & \unit{K} & \unit{s^{-1}} & \unit{km/s}           & mK    \\
        \hline
        \twCO{}   \Jone{} & 115.3 &      5.5 &     7.203e-08 & 0.5          &  91 \\
        \thCO{}   \Jone{} & 110.2 &      5.3 &     6.294e-08 & 0.1          &  74 \\
        \thCO{}   \Jtwo{} & 220.4 &     15.9 &     6.038e-07 & 0.1          & 155 \\
        \CeiO{}   \Jone{} & 109.8 &      5.3 &     6.266e-08 & 0.1          &  72 \\
        \CeiO{}   \Jtwo{} & 219.6 &     15.8 &     6.011e-07 & 0.1          & 121 \\
        \HCOp{}   \Jone{} &  89.2 &      4.3 &     4.187e-05 & 0.1          & 318 \\
        \HthCOp{} \Jone{} &  86.8 &      4.2 &     3.853e-05 & 0.5          &  45 \\
        \hline
      \end{tabular}}
    \tablefoot{%
      \tablefoottext{1}{Line frequencies, upper energy levels and Einstein A
        coefficients are taken from the Cologne Data Base for Molecular
        Spectroscopy (CDMS)~\citep{muller01}} %
      \tablefoottext{2}{Channel spacing after resampling.} %
      \tablefoottext{3}{Median noise $\sigma_b$ in the Horsehead after
        resampling and smoothing. Angular resolution after smoothing is
        $31\unit{''}$.
      }%
    }
    \label{tab_lines}
  \end{table}
}

\newcommand{\TabSetOfSpecies}{
  \begin{table}
    \centering %
    \caption{Sets of species, the assumed chemistry, the number of
      available lines, and the number of unknown parameters.}
    \resizebox{\linewidth}{!}{%
      \begin{tabular}{lccc}
        \hline  \hline
        Set of species & Relative abundance & $\#$ lines & $\#$ unknowns \\
        \hline
        % One species
        $\set{\thCO}$ or $\set{\CeiO}$   & -                                       & 2 & 5 \\
        \hline
        % Two species
        $\set{\thCO,\HCOp}$              & free                                    & 3 & 6 \\
        $\set{\thCO,\HCOp;\calA}$        & $\frac{N(\thCO)}{N(\HCOp)} = 10^{2.88}$ & 3 & 5 \\
        $\set{\thCO,\HCOp;\calA^\star}$  & $\frac{N(\thCO)}{N(\HCOp)} = 10^{3.18}$ & 3 & 5 \\
        $\set{\thCO,\CeiO}$              & free                                    & 4 & 6 \\
        $\set{\thCO,\CeiO;\calA}$        & $\frac{N(\thCO)}{N(\CeiO)} = 10^{0.90}$ & 4 & 5 \\
        $\set{\thCO,\CeiO;\calA^\star}$  & $\frac{N(\thCO)}{N(\CeiO)} = 10^{1.20}$ & 4 & 5 \\
        $\set{\CeiO,\HthCOp}$            & free                                    & 3 & 6 \\
        $\set{\CeiO,\HthCOp;\calA}$      & $\frac{N(\CeiO)}{N(\HthCOp)} = 10^{3.78}$ & 3 & 5 \\
        $\set{\CeiO,\HthCOp;\calA^\star}$& $\frac{N(\CeiO)}{N(\HthCOp)} = 10^{3.48}$ & 3 & 5 \\
        \hline
        % Three species
        $\set{\twCO,\thCO,\HCOp}$        & $\frac{N(\twCO)}{N(\thCO)} = 10^{1.80}$ & 4 & 6 \\
        $\set{\twCO,\thCO,\CeiO}$        & $\frac{N(\twCO)}{N(\thCO)} = 10^{1.80}$ & 5 & 6 \\
        \hline
      \end{tabular}
    } %
    \tablefoot{%
      When more than one species is fitted, we assume that they share the
      values of $(\nhd, \Tkin, C_V,$ and $\sigma_V)$.}%
    \label{tab_set_of_species}
  \end{table}
}

\section{Quantifying the variability of the estimation of physical
  parameters on actual data}
\label{sec:data}

\TabLines{} %
\FigPeakAreaSpectra{} %
\TabCoord{} %

In this prototype study, we used the RADEX model to fit the spectra of a
combination of the \Jone{} and \Jtwo{} transitions for some of the
following species: \twCO, \thCO, \CeiO, \HCOp, and
\HthCOp{}. Table~\ref{tab_lines} lists the properties of each line
(frequency, upper energy level and Einstein coefficient) as well as the
properties of the observed spectra (spectral spacing and thermal noise).
The studied field of view covers about 50 square arcminutes towards the
Horsehead Pillar at the South West of Orion B, which is illuminated by the
$\sigma$-Ori star.

\subsection{Data}

\subsubsection{Acquisition and data reduction}

The data were obtained at the IRAM 30m telescope. The \Jone{} lines were
acquired in 2013-2020 in the context of the ORION-B large
program~\citep{pety2022,gerin2023}, using a combination of the EMIR
receivers and FTS spectrometers. The \Jtwo{} lines towards the Horsehead
pillar were acquired in 2006 (PI: N.~Peretto) using the ABCD generation of
receivers and the VESPA auto-correlator. They have been first presented and
used in~\citet{roueff21}.

The data reduction is described in~\citet{pety17} and~\citet{roueff21}.  It
uses the standard methods provided by the \texttt{GILDAS}\footnote{See
  \url{http://www.iram.fr/IRAMFR/GILDAS} for more information about the
  GILDAS software~\citep{pety05}.}\texttt{/CLASS} software.  The data cubes
were gridded on the same astrometric grid. They were smoothed to the lowest
achieved angular resolution (i.e., $31''$) in order to ease the multi-line
fit towards the same LoS without having to take care of different angular
resolutions. The velocity spacings are listed in Table~\ref{tab_lines}. We
used two velocity spacings ($0.1$ and $0.5\kms$) depending on the native
spectral resolution of the spectrometer used for each line.

\subsubsection{Presentation of the data}

Figure~\ref{fig:data:peak:area:spectra} summarizes the data with a
combination of images of the peak and integrated intensities as well as
percentiles of the line intensities as a function of the velocity among the
considered pixels. The spatial shape of the Horsehead pillar varies with
the species. The \twCO{} \Jone{} line is detected towards most of the field
of view, while the rarer CO isotopologues better show the spine of the
Horsehead. The \HthCOp{} \Jone{} line is mainly detected towards the two
known dense cores of the
Horsehead~\citep[e.g.][]{wardthompson06}. Table~\ref{tab:coord} lists the
specified locations on the image. The two crosses show the positions of
these two dense cores. The position of the western dense core is the
Horsehead dense core position defined in \citet{pety2007} and extensively
studied in the framework of the Horsehead WHISPER project~\citep[see,
e.g.,][]{guzman2014}. The cross for the eastern dense core points towards
the local maximum of the \HthCOp{} \Jone{} line emission. The \HCOp{}
\Jone{} line exhibits a mixed behavior with some filamentary structure
embedded in more diffuse emission that has a similar shape as the \twCO{}
\Jone{} line.

The line profiles mostly show a main velocity component centered near
$V_\emr{LSR} \sim 10.5\kms$. The \twCO, \HCOp{}, and \HthCOp{} \Jone{}
lines exhibit a second velocity component or wing around
$V_\emr{LSR} \sim 13.0\kms$. As the presence of a single velocity component
is required to be consistent with the hypothesis that the lines are emitted
by the same layer of molecular gas along the LoS, we only concentrate on
the $10.5\kms$ velocity component. Moreover this $10.5\kms$-velocity
component is clearly detected for all the lines.

The \thCO{} \Jtwo{} and \CeiO{} \Jtwo{} lines have similar properties as
their \Jone{} transitions regarding their spatial and spectral shapes, but
with minor differences due to excitation effects. Finally, most of the
observed lines have high S/N with the noteworthy exception of \HthCOp{}
\Jone{}.

\subsection{Fitting assumptions}

\TabSetOfSpecies{} %

\subsubsection{Simplifying physical and chemical assumptions}

As detailed in Section~\ref{sec:methods:radex}, the considered non-LTE
model is characterized by five unknown parameters ($\nhd$, $\Tkin$, $N$,
$\sigma_V$ and $C_V$) per modeled species. The purpose of this article is
to analyze the possibility of estimating those parameters as a function of
the LoS from observations of few lines (1 or 2), from several species. We
consider here different sets of species containing from one to three
species among: \twCO, \thCO, \CeiO, \HCOp, and \HthCOp{}.  Since the number
of unknown parameters is too large compared to the available data,
assumptions are required in order to reduce the number of unknowns with
respect to the number of fitted lines.  Table~\ref{tab_set_of_species}
lists the cases considered in this section. For all species sets, we assume
that the gas emitting the different lines share the same kinetic
temperature, volume density, centroid velocity and velocity dispersion. In
other words, we assume that their emissions come from the same gas cell in
the cloud. We also often assume that at least one of the column density
ratios is known and thus fixed. We thus have between 5 and 6 unknowns.  The
first five ones are the kinetic temperature, volume density, centroid
velocity and velocity dispersion, and the column density of either \thCO{}
or \CeiO{}. For some sets, the sixth unknown is either the
$\frac{N(\HCOp{})}{N(\thCO{})}$ or the $\frac{N(\HthCOp{})}{N(\CeiO{})}$
relative abundance.

Considering a single species with two lines, for instance, the \thCO{} or \CeiO{}
isotopologues, allows us to get an idea of their column densities without
the assumption of co-location. For sets containing two species, we
associate species that share a good probability of being emitted from similar
regions: \set{\thCO,\HCOp} for diffuse regions, \set{\thCO,\CeiO} for
denser filaments, and \set{\CeiO, \HthCOp} for dense cores. We add the
\twCO{} peak value to check its influence.

The \Jone{} lines of the \HCOp{} isotopologues have higher critical
densities than the CO isotopologues' ground-state transitions because the
dipole moments of these molecular ions is much larger than the ones of
neutral species~\citep{shirley15}. These lines may thus bring additional
information towards the two dense cores of the Horsehead pillar. We use the
two main isotopologues to assess the impact of their different optical
depths and S/Ns.

Since this paper focuses on a region with $\nhd \ga 10^3\pccm$, we only use
the spectrum of $\twCO{}$ at its peak value. Indeed, the line profile
delivered by the non-LTE model with a simple Gaussian opacity profile does
not suitably describe the profile of extremely optically thick lines such
as \twCO{} \Jone{}~\citep{leung76}.

We add a $\calA$ symbol in the set name when we fix the relative
abundance. In this case, the relative abundance is defined by using the
typical isotopic ratios for Orion $^{12}$C/$^{13}$C = 63, $^{16}$O/
$^{18}$O = 510, {N(\thCO{})}/N(H$_2$)=$2.3\times 10^{-6}$ and
{N(\HCOp{})}/N(H$_2$)=$3\times 10^{-9}$ \citep[and references
therein]{gerin2019,roueff21}.
\begin{equation}
  \begin{array}{ll}
    N(\twCO)/N(\thCO) = 10^{1.8}  \sim 60,
    \\
    N(\thCO)/N(\CeiO) = 10^{0.9}  \sim 8,
    \\
    N(\thCO)/N(\HCOp) = 10^{2.88} \sim 800,
    \\
    N(\CeiO) / N(\HthCOp) = 10^{3.78} \sim 6000.
  \end{array}
  \label{eq_abundance}
\end{equation}
We also varied the last three abundance ratios by $\pm 0.3\,$dex in order
to assess the impact of variations of the assumed chemistry. These cases
are marked with the $\calA^\star$ symbol.

\subsubsection{Noise model}

The data are affected by thermal noise and calibration uncertainty.  We
thus use the following model of data
\begin{equation}
  x(\nu)=c\,s(\nu)+b(\nu),
  \label{eq_x}
\end{equation}
where $x$ is the observed data, $s$ the source intensity, $b$ is the
thermal noise, and $c$ represents the calibration uncertainty.  We assume
that $b$ and $c$ are independent random variables with normal distribution:
$b\sim \calN(0,\sigma_b^2)$ and $c\sim \calN(1,\sigma_c^2)$~\citep[see][for
more details]{einig23}.

The standard deviations associated to the calibration uncertainty are fixed
as
\begin{equation}
  \sigma_c =
  \left\{%
    \begin{array}{ll}
      5\% & \text{for 3\,mm lines}, \\
      10\% & \text{for 1\,mm lines}.
    \end{array}
  \right.
  \label{sec_sigma_c}
\end{equation}
The median values of the thermal standard deviation measured on the data
are listed in Table~\ref{tab_lines}.

\subsubsection{Choice of likelihood and S/N saturation}
\label{sec:data:likelihood}

When quantifying the theoretical precision of estimations (see
Sect.~\ref{sec:CRB}), we take into account both the thermal noise and
calibration uncertainty. However, we neglect the calibration uncertainty
and we instead set the S/N to a maximum value of 10, when fitting the data
with the maximum likelihood estimator.

On one hand, including the multiplicative factor in the estimation leads to
fits where the potential shortcomings of the physical assumptions are
incorrectly attributed to calibration errors, because the fit then cares
too much about the shape of the spectra, and too little about its
amplitude. This is counter-productive for real data analysis.  We thus
choose the standard negative likelihood when fitting data $(\bx)$, i.e.,
\begin{equation}
  \calL(\btheta;\bx)= - \sum_{n} \frac{\left(x_n-s_n\right)^2}{2\sigma_{b}^2},
\end{equation}
where $s_n$ is a sampled version of Eq.~\ref{eq2_s} and thus is a
function of the unknown parameters.  On the other hand, we artificially set
$\sigma_b$ such that the S/N, defined as
\begin{equation}
  S/N=\frac{\max_\nu s(\nu)}{\sigma_b},
  \label{eq_S/N}
\end{equation}
has a maximum value of 10. This artificial saturation is required to ensure
that lines with a S/N between 3 and 10 (e.g., \HthCOp{} \Jone) have a non
negligible weight compared to lines with S/N of 100 (e.g., the \thCO{}
lines). The value of 10 also ensures that calibration errors that are lower
than 10\% are considered as noise, and contribute to the error budget
together with the additive noise.

\subsection{Results}

\FigDataHistoNthCO{}%
\FigDataRMSETab{} %
\FigDataHistoAbundances{}%

We systematically check here the stability of the fitted parameters as a
function of the set of species fitted and the associated a priori
hypotheses.  In the following figures, we select one parameter at a
time. We look at chemical (column densities and relative abundances, i.e.,
ratios of column densities) and physical (kinetic temperature, volume
density, and thermal pressure) parameters. We focus on reliable estimations
defined as
\begin{equation}
  \CRB^{1/2}(\widehat\theta)\leq 0.3
  \quad \mbox{for} \quad
  \theta=\log N_i, \, \log\Tkin, \, \mbox{and} \, \log\nHH.
  \label{eq:bad:CRB}
\end{equation}

For each parameter, we present two figures: 1) the maps of the fitted
parameter on the Horsehead pillar data for different sets of species and/or
a priori (e.g., the top of Fig.~\ref{fig:data:N13CO}); 2) joint
histograms of this parameter estimations.  The bottom of
Fig.~\ref{fig:data:N13CO} compares the estimations of $N(\thCO)$ obtained
for 7 different sets of species and/or a priori defined in
Table~\ref{tab_set_of_species}: $\set{\thCO}$, $\set{\thCO,\HCOp;\calA}$,
$\set{\thCO,\HCOp;\calA^*}$, $\set{\thCO,\CeiO;\calA}$,
$\set{\thCO,\CeiO;\calA^*}$, $\set{\twCO,\thCO,\HCOp}$ and
$\set{\twCO,\thCO,\CeiO}$.  There are 21 pairs for 7 different fits. The
figure thus appears as a lower triangle of size $6\times 6$. For
conciseness sake, we only show in the main paper these two kinds of figures
for the case of the column density of \thCO{}. For the other parameters, we
put these figures in Appendix~\ref{app:data:supplement}, and we synthesize
the joint histograms by their distance in log-space to the line of slope
1. More precisely, we compute the mean and the standard deviation of
$\abs{ \widehat\theta_x - \widehat\theta_y}/\sqrt{2}$, where $\theta_x$ and
$\theta_y$ are the estimated parameters plotted along the $x$ and $y$ axes
of the histograms.  We present these values in a tabular way where we put
in the same lower triangle the values of the mean and standard deviation,
color-coded by the root mean square error, RMSE
$= \sqrt{\emr{mean}^2+\emr{sdev}^2}$. Figure~\ref{fig:data:rmse:tab} shows
the associated ``tables'' for $N(\thCO),$ $N(\CeiO),$ $N(\HCOp),$ and
$N(\HthCOp),$ \Tkin{}, and \nHH{}. A non zero mean indicates a systematic
shift compared to the line of slope 1, while a large standard deviation
indicates a large dispersion of the distances to the line whose slope is
the mean, and a large RMSE a large dispersion of the distances to the line
of slope 1. Quantitatively the values can also be interpreted with the help
of Table~\ref{tab:conversion}. Values of $\pm0.1$, $\pm0.3$, and $\pm0.6$
correspond to multiplicative uncertainty of the order of 20\%, a factor 2,
and 4 respectively.

\subsubsection{One detailed case: The column density of \thCO{}}

The top of Fig.~\ref{fig:data:N13CO} shows the maps of the column densities
of \thCO{} fitted on the Horsehead pillar under different hypotheses. In
this case, almost all pixels deliver a column density with a good CRB
reference precision whatever the set of species and of a priori used during
the fit. However, the shape of the maps of the column density varies
depending on the chosen set of lines. If we take the case where we only fit
the \thCO{} lines as reference (panel a), we see different
behaviors. First, adding the \CeiO{} lines increases $N(\thCO)$ towards the
two dense cores or the surrounding filaments depending on the assumed
chemistry. Second, adding the \HCOp{} lines decreases $N(\thCO)$ towards
the dense core in the Horsehead throat but increases it towards the dense
core on the top of the head. The effect is more or less pronounced
depending on the assumed chemistry. Third, adding the peak temperature of
the \twCO{} \Jone{} line regularizes the overall shape of the column
density map.

These trends are quantitatively confirmed by the joint histograms shown in
the bottom of Fig.~\ref{fig:data:N13CO}. The most consistent values of the
column densities are obtained in the cases \set{\thCO,\HCOp;\calA} and
\set{\thCO,\HCOp;\calA^\star} where only the assumed chemistry is
changed. This is followed by the pairs (\set{\twCO,\thCO,\CeiO} and
\set{\twCO,\thCO,\HCOp}), and (\set{\thCO} and \set{\thCO,\HCOp,\calA}).
The less consistent values of the column densities are obtained for the
pairs (\set{\thCO,\CeiO;\calA} and \set{\thCO,\CeiO;\calA^\star}).

Overall, the column densities of \thCO{} are defined within a typical
multiplicative accuracy of 30\%. Adding \CeiO{} (resp. \HCOp) gives more
weight to the dense cores (resp. more diffuse envelope) when fitting the
column density of \thCO.

\subsubsection{Column densities and relative abundances}

The top row of Fig.~\ref{fig:data:rmse:tab} and the associated figures in
Appendix~\ref{app:data:supplement} synthesize the results on column
densities. The column densities of \CeiO{} are mostly consistent within
15\%. That is a factor $\sim 2$ better than \thCO{}. Accordingly, the maps
of $N(\CeiO)$ show spatial variations that are similar in all tested cases
(see Fig.~\ref{fig:data:NC18O}). The noteworthy exception are the couples
(\set{\thCO,\CeiO;\calA} and \set{\thCO,\CeiO;\calA^\star}), probably
because the column density of \CeiO{} is sensitive to the choice of the
$N(\thCO)/N(\CeiO)$ abundance ratio.

The column densities of \HCOp{} and \HthCOp{} are more scattered and biased
depending on the fitted case. The worst case is $N(\HthCOp)$ because the
associated line has the lowest S/N of all. But $N(\HCOp)$ also shows a
large scatter. We relate this to a possible evolution of the \HCOp{}
chemistry between dense cores and more diffuse gas, implying an evolution
of its relative abundance.

Figure~\ref{fig:data:abundances} illustrates the stability of the relative
abundances depending on the case. The relative abundance of \CeiO{} and
\thCO{} is pretty insensitive to the presence or absence of the peak
intensity of the \twCO{} \Jone{} line during the fit.  In contrast, the
relative abundances of (\thCO{} and \HCOp{}) or (\CeiO{} and \HthCOp) are
sensitive (both in bias and scatter) to the presence or absence of the peak
intensity of the \twCO{} \Jone{} line. This may be due to the fact that the
peak intensity mostly constrains the kinetic temperature. As the CO
isotopologue line intensities are mostly sensitive to the thermal pressure
in the considered parameter space, the constraint on the temperature has an
effect on the volume density determination. With their high dipole moment
(thus higher critical densities) the \HCOp isotopologue excitation is
sensitive to the volume density.

\subsubsection{Kinetic temperature}

The determination of the kinetic temperature presents large variations
depending on the chosen set of lines. As shown in Fig.~\ref{fig:data:Tkin},
using both the \Jone{} and \Jtwo{} lines of \thCO{} and \CeiO{} enables a
correct determination of the kinetic temperature in the high column density
filament. In addition, the choice of the fixed relative abundance between
these species has little impact on the \Tkin{} estimation. In contrast, the
same set of lines leads to unrealistic high values of \Tkin{} outside the
filament.  The combination of either \thCO{} and \HCOp{} or \CeiO{} and
\HthCOp{} performs poorly. The determination of the kinetic temperature is
sensitive to the assumption made on the relative abundances. The \Tkin{}
estimations seem to be biased as the correlation with the determinations
from \thCO{} and \CeiO{} shows some scatter and presents a significant
offset.

Adding \twCO{} to the set of lines improves the \Tkin{} estimation because
this sets an upper limit to the value of \Tkin{} of about $30\K$.  The
resulting temperature maps are more extended and present smooth
variations. The two sets perform about equally well and lead to similar
estimations across the map with no systematic difference. However, using
the peak of the \twCO{} \Jone{} emission introduces a bias in the
estimation as this line is produced in the external regions of the
considered sight-line, where the kinetic temperature may not be
representative of the conditions along the full LoS.

\subsubsection{Volume density}

The volume density is the most difficult parameter to estimate. The scatter
can reach a factor of ten and different sets of lines can lead to
determinations with a systematic offset of up to 0.5\,dex, a factor of
three.  In the high column density regions, the sets combining \CeiO{} and
\HthCOp{} perform well, but the absolute value of the volume density scales
with the fixed relative abundance. Indeed, the determination are well
correlated but offset from each other by 0.34\,dex, which corresponds to
the change in the fixed relative abundances.  A similar effect is seen for
the sets combining \thCO{} and \HCOp{}, which leads to well correlated but
offsets density estimations. For pixels where both sets can be used and the
hypotheses on abundances are consistent, it is encouraging to see that
\thCO{} and \HCOp{}, and \CeiO{} and \HthCOp{} lead to similar estimations
of the volume density with no significant offset, even if a significant
scatter of about 0.2\,dex remains.
 
The estimations of the volume density made from the sets combining CO
isotopologues only, seem to reach a maximum value near 10$^{4.3}\pccm$ (see
Fig.~\ref{fig:data:n}). This may be related to the limited range of kinetic
temperature and the lower critical density of the considered CO
isotopologue transitions as compared with those of the high dipole moment
species \HCOp{} and \HthCOp{}.

\subsubsection{Thermal pressure}

The estimation of the thermal pressure is more accurate and less biased
than that of $\Tkin$ and $\nhd$ separately (see Fig.~\ref{fig:data:Pth}).
In particular the hypothesis on relative abundance has little impact on the
estimation of the thermal pressure, while we have seen that it has a strong
influence on the volume density. Estimations made with or without \twCO{}
are consistent in most cases. The thermal pressure for the pixels
associated with dense cores where \CeiO{} and \HthCOp{} are detected with
high S/N is somewhat higher when using \CeiO{} and \HthCOp{} only as
compared with sets including \twCO{} but the effect is small.  The better
accuracy of the thermal pressure estimations can be understood by looking
at the shape of the NLL function in the $(\nHH,\Tkin)$ parameter
space. This is discussed in Sect.~\ref{sec:MC}.

\subsection{Intermediate summary}

In this section, we have systematically compared the results of line fits
with the RADEX non-LTE radiative transfer model, on the same dataset with
the same fitting algorithm. The main difference between the different
experiments has been the number and the set of fitted lines. The fitted
parameters have been the column densities of the \thCO{} and \HCOp{}
isotopologues, and the associated kinetic temperature and volume
density. We have also studied the derived relative abundances and thermal
pressure.

This systematic comparison emphasizes the potential variability of the
estimated parameters depending on the chosen set of fitted species and
lines. The joint histograms of the same fitted parameters as a function of
the fitting assumptions show not only important scatter around the line of
slope 1 but also biases with respect to this line. The maps of fitted
parameters show regular patterns, suggesting that the noise level is not
the cause of the found variations. The volume density is the least accurate
estimated parameter, followed by the kinetic temperature. The thermal
pressure estimates are more accurate than the volume density and kinetic
temperature. The column densities and, above all, the ratio of column
densities are the most accurate fitted parameters.

We do not address here whether the quality of the fit allows us to define
which set of species and associated assumptions best fits the data for two
reasons. First, we here explore the impact of the typical analysis of line
data where the best solution is selected by minimizing the negative
log-likelihood of the fit for a fixed set of species. Second, fitting a
subset of lines necessarily leads to smaller $\chi^2$, but this does not
imply that estimations are better. Third, comparing the quality of fit
performed with different numbers of unknown parameters requires more
advanced statistical tools than just the $\chi^2$ value to select the
``best'' estimations. For instance, when the abundance are estimated
instead of being fixed, the $\chi^2$ is necessarily smaller, but this does
not imply that the estimation is better, notably because of the induced
estimation variance. This topic will be studied in a future paper.

%%%%
% MC
%%%%

\newcommand{\FigMCspectra}{%
  \begin{figure}
    \centering{} %
    \includegraphics[width=\linewidth]{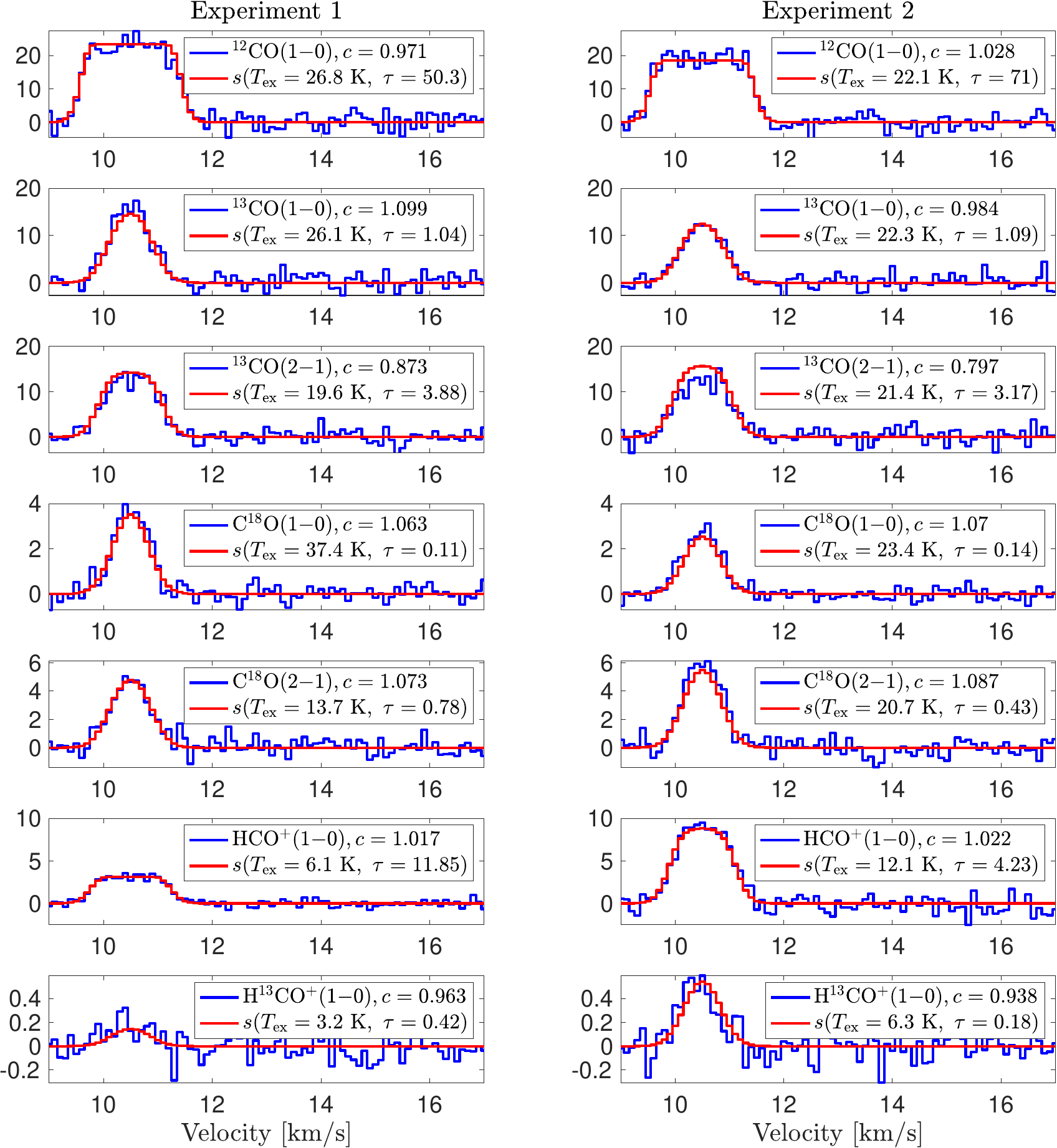}
    \caption{Comparison of the ideal (in red) line profiles with noisy (in
      blue) ones for experiment \#1 and \#2 (see
      Table~\ref{tab:exp:def}). In the legends, $c$ is the estimated
      calibration factor (see Eq.~\ref{eq_x}).}
    \label{fig:mc:spectra}
  \end{figure}
}

\newcommand{\FigMCCOiso}{%
  \begin{figure*}
    \centering{} %
    \includegraphics[width=\linewidth]{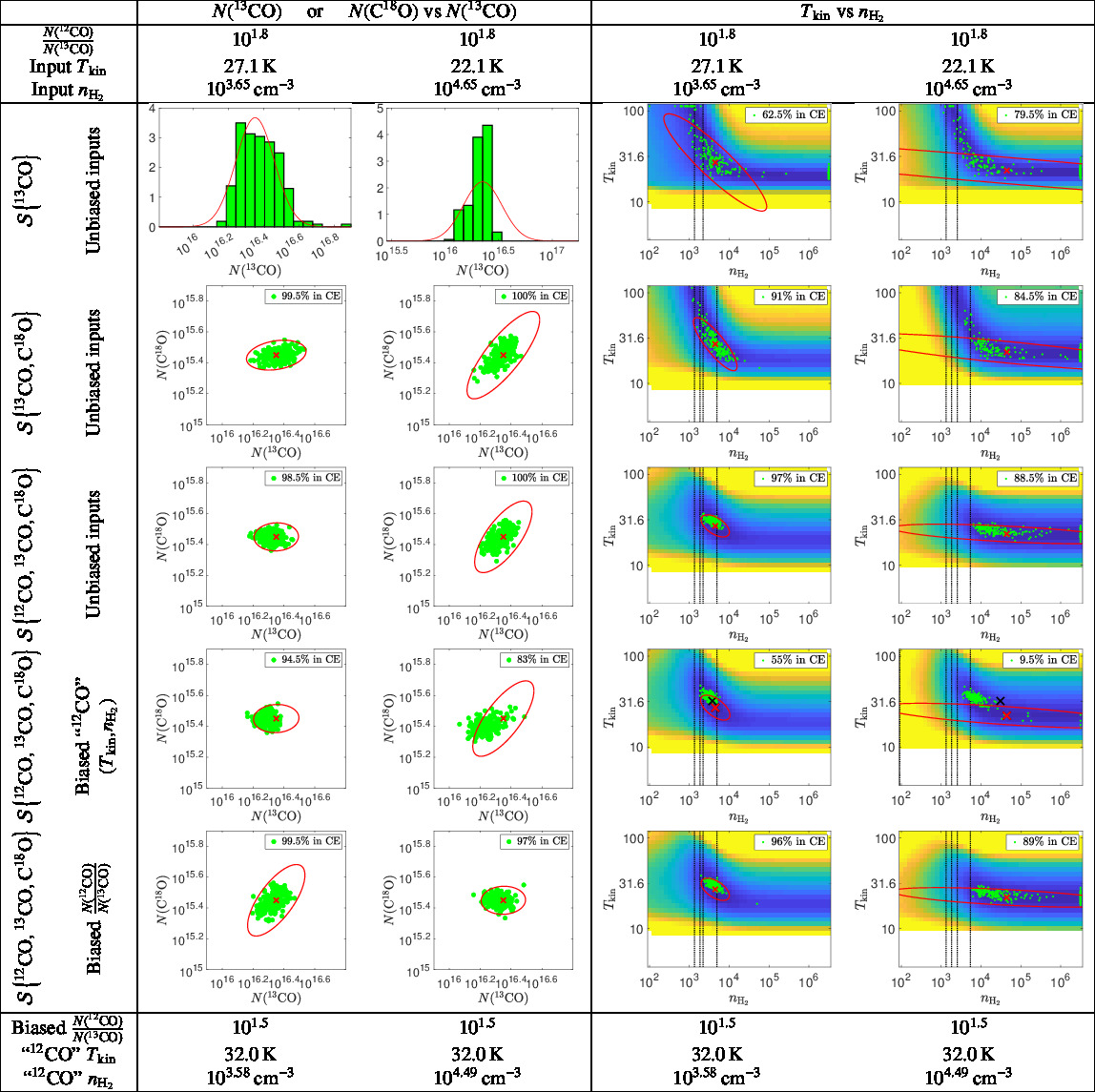}
    \caption{Comparison of the performances of our maximum likelihood
      estimator for the estimation of the column densities (columns \#2 and
      \#3) and the couple $(\Tkin,\nHH)$ (columns \#4 and \#5) and for two
      examples of kinetic temperature and volume density.  The first
      example (columns \#2 and \#4) at medium density and warm
      temperature. The second one (columns \#3 and \#5) at high density and
      colder temperature. In the first and second rows we estimate the
      parameters with the \thCO{} lines, and the (\thCO, \CeiO) lines,
      respectively. In the last three rows we estimate the parameters with
      the lines of the three main CO isotopologues. In the last two rows we
      biased one of the parameters for the \twCO{} RADEX model to test the
      effect of incorrect physical assumptions in the estimator.  The
      kinetic temperature and volume density of the \twCO{} RADEX model is
      biased in the fourth row, while the ratio of abundances of \twCO{}
      and \thCO{} is biased to a smaller values than assumed in the
      estimator on the fifth row. In all panels, the modeled and biased
      values of the estimated parameters are shown with the red and black
      crosses, respectively. The red ellipses, computed with the Cramér-Rao
      bound, are confidence ellipses (CE) of efficient estimations with a
      probability of $99\%$. The green points shows the estimated values
      for 200 simulated spectra. The color images on columns \#3 and \#4
      show the values of the negative log-likelihood for the couple
      $(\Tkin,\nHH)$ that our estimator assumes to be common to all
      analyzed species. The vertical dotted line are the effective critical
      density of the different transitions.}
    \label{fig:mc:co:iso}
  \end{figure*}
}

\newcommand{\FigMCthCOHCOp}{%
  \begin{figure*}
    \centering{} %
    \includegraphics[width=\linewidth]{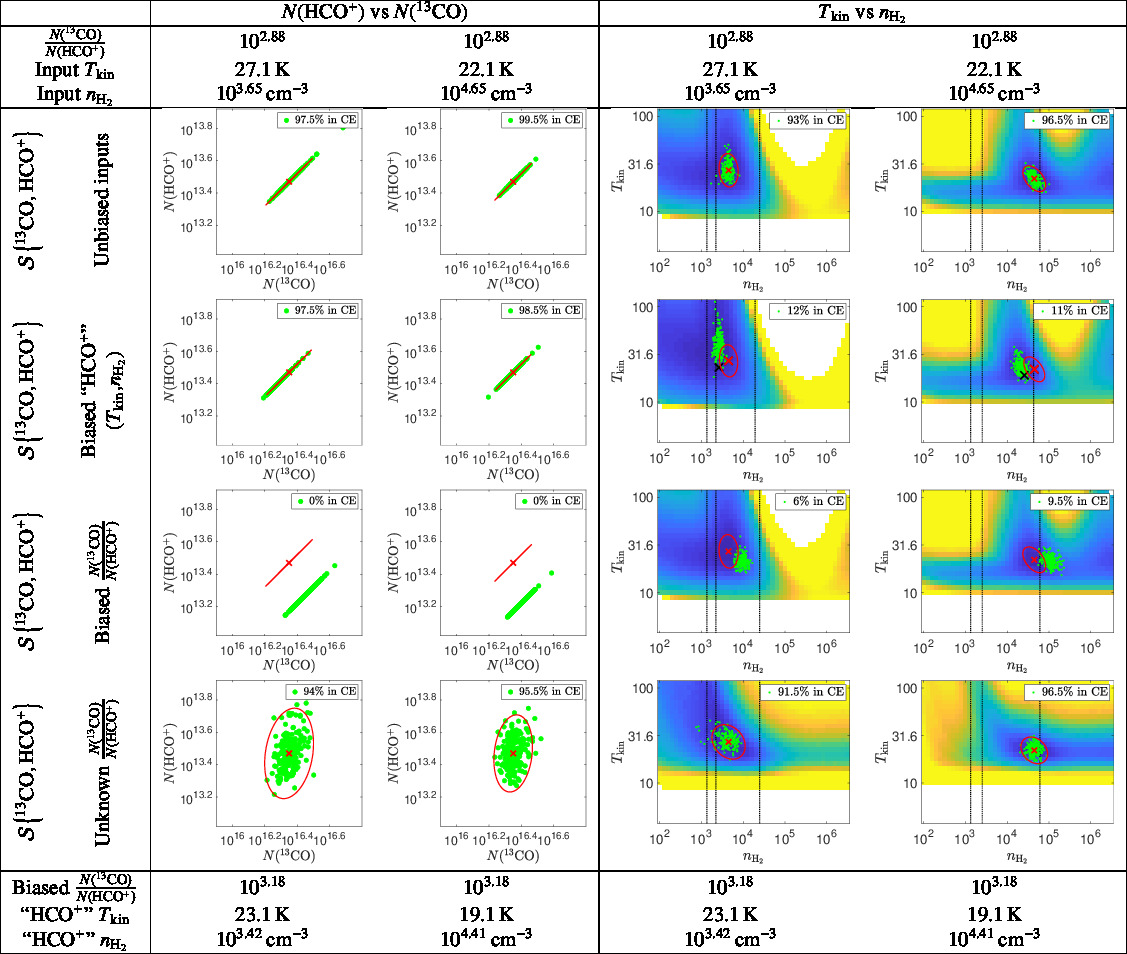}
    \caption{Same as Fig.~\ref{fig:mc:co:iso} except that 1) the couple of
      species used to estimate the parameters are \thCO{} and \HCOp{}, and
      2) the biases on the RADEX model are imposed on the \HCOp{} species.}
    \label{fig:mc:13co:hcop}
  \end{figure*}
}

\newcommand{\FigMCCeiOHthCOp}{%
  \begin{figure*}
    \centering{} %
    \includegraphics[width=\linewidth]{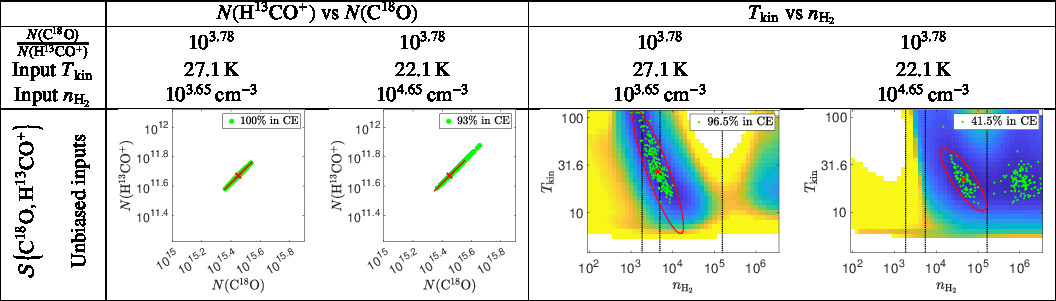}
    \caption{Same as Fig.~\ref{fig:mc:co:iso} except that 1) the couple of
      species used to estimate the parameters are \CeiO{} and \HthCOp{},
      and 2) the RADEX model are not biased.}
    \label{fig:mc:c18o:h13cop}
  \end{figure*}
}

\newcommand{\TabExpDef}{%
  \begin{table*}
    \centering{}
    \caption{Input physical parameters for the two experiments}
    \begin{tabular}{lccccccccc}
      \hline
      \hline
      Exp. & $N(\twCO)$   & $N(\thCO)$   & $N(\CeiO)$   & $N(\HCOp)$   & $N(\HthCOp)$ & \Tkin{} & $n(\HH)$    & $\sigma_V$ & $C_V$ \\
           & \pscm{}      & \pscm{}      & \pscm{}      & \pscm{}      & \pscm{}      & \K{}    & \pccm{}     & \kms{}     & \kms{}     \\
      \hline                           
      \# 1 & $10^{18.15}$ & $10^{16.35}$ & $10^{15.45}$ & $10^{13.47}$ & $10^{11.67}$ & 27.1    & $10^{3.65}$ & 0.32       & 10.5       \\
      \# 2 & idem         & idem         & idem         & idem         & idem         & 22.1    & $10^{4.65}$ & idem       & idem       \\
      \hline
    \end{tabular}
    \label{tab:exp:def}
  \end{table*}
}

\newcommand{\TabExU}{%
  \begin{table*}
    \centering{}
    \caption{Line characteristics for experiment \#1.}
    \begin{tabular}{lcccccccccc}
      \hline
      \hline
      Species   & Line & $\sigma_b$ & $\Tpeak$ & Peak S/N & $\tau$ & $\Tex$   & \multicolumn{2}{c}{\ncritthin} & \multicolumn{2}{c}{\ncriteff} \\ 
                &      & \unit{K}   & \unit{K} & --       & --     & \unit{K} & \multicolumn{2}{c}{\pccm}      & \multicolumn{2}{c}{\pccm} \\
      \hline
                &      &            &          &          &        &          & \oHH{} & \pHH{}                & \oHH{} & \pHH{} \\
      \hline
      \twCO{}   & \Jone{} & 2.3  & 23    & 10   & 50    & 27   & $10^{3.34}$ & $10^{3.35}$ & $10^{1.81}$ & $10^{1.82}$ \\
      \thCO{}   & \Jone{} & 1.5  & 15    & 10   &  1.0  & 26   & $10^{3.28}$ & $10^{3.29}$ & $10^{3.12}$ & $10^{3.13}$ \\
      \thCO{}   & \Jtwo{} & 1.4  & 14    & 10   &  3.9  & 19   & $10^{3.77}$ & $10^{3.81}$ & $10^{3.3}$  & $10^{3.35}$ \\
      \CeiO{}   & \Jone{} & 0.36 &  3.6  & 10   &  0.11 & 38   & $10^{3.28}$ & $10^{3.29}$ & $10^{3.26}$ & $10^{3.27}$ \\
      \CeiO{}   & \Jtwo{} & 0.48 &  4.8  & 10   &  0.78 & 14   & $10^{3.77}$ & $10^{3.81}$ & $10^{3.65}$ & $10^{3.69}$ \\ 
      \HCOp{}   & \Jone{} & 0.31 &  3.1  & 10   & 12    &  6.1 & $10^{5.26}$ & $10^{5.3}$  & $10^{4.36}$ & $10^{4.39}$ \\ 
      \HthCOp{} & \Jone{} & 0.10 &  0.15 &  1.5 &  0.41 &  3.2 & $10^{5.23}$ & $10^{5.26}$ & $10^{5.16}$ & $10^{5.19}$ \\
      \hline
    \end{tabular}
    \tablefoot{The peak S/N is defined in Eq.~\ref{eq_S/N}.}
    \label{tab:exp:1}
  \end{table*}
}

\newcommand{\TabExD}{%
  \begin{table*}
    \centering{}
    \caption{Line characteristics for experiment \#2.}
    \begin{tabular}{lcccccccccc}
      \hline
      \hline
      Species   & Line & $\sigma_b$ & $\Tpeak$ & Peak S/N & $\tau$ & $\Tex$   & \multicolumn{2}{c}{\ncritthin} & \multicolumn{2}{c}{\ncriteff} \\ 
                &      & \unit{K}   & \unit{K} & --       & --     & \unit{K} & \multicolumn{2}{c}{\pccm}      & \multicolumn{2}{c}{\pccm} \\
      \hline
                &      &            &          &          &        &          & \oHH{} & \pHH{}                & \oHH{} & \pHH{} \\
      \hline
      \twCO{}   & \Jone{} & 1.9  & 19    & 10   & 71    & 22   & $10^{3.33}$ & $10^{3.35}$ & $10^{1.65}$ & $10^{1.67}$ \\
      \thCO{}   & \Jone{} & 1.2  & 12    & 10   &  1.1  & 22   & $10^{3.27}$ & $10^{3.29}$ & $10^{3.11}$ & $10^{3.13}$ \\
      \thCO{}   & \Jtwo{} & 1.6  & 16    & 10   &  3.2  & 21   & $10^{3.77}$ & $10^{3.81}$ & $10^{3.36}$ & $10^{3.4}$ \\
      \CeiO{}   & \Jone{} & 0.26 &  2.5  & 10   &  0.14 & 23   & $10^{3.27}$ & $10^{3.29}$ & $10^{3.25}$ & $10^{3.26}$ \\
      \CeiO{}   & \Jtwo{} & 0.55 &  5.5  & 10   &  0.43 & 21   & $10^{3.76}$ & $10^{3.8}$  & $10^{3.7}$ & $10^{3.74}$ \\
      \HCOp{}   & \Jone{} & 0.88 &  8.8  & 10   &  4.2  & 12   & $10^{5.25}$ & $10^{5.28}$ & $10^{4.75}$ & $10^{4.69}$ \\ 
      \HthCOp{} & \Jone{} & 0.10 &  0.54 &  5.4 &  0.18 &  6.3 & $10^{5.21}$ & $10^{5.25}$ & $10^{5.18}$ & $10^{5.22}$ \\
      \hline
    \end{tabular}
    \tablefoot{The peak S/N is defined in Eq.~\ref{eq_S/N}.}
    \label{tab:exp:2}
  \end{table*}
}

\section{On the importance of prior assumptions on the bias-variance
  tradeoff}
\label{sec:MC}

\TabExpDef{} %
\TabExU{} %
\TabExD{} %

In the previous section, we tried to estimate the gas volume density and
kinetic temperature in addition to the molecular column densities, when
fitting different combinations of the CO and \HCOp{} main isotopologue
lines together. This requires some additional a priori knowledge linking
the parameters so that the number of unknowns becomes smaller than the
number of constraints, and the associated confidence interval become
``reasonably'' small. In particular, we make the hypothesis that the same
region along the LoS dominates the emission of all lines, with a unique
volume density and kinetic temperature. We also make chemical assumptions
in the form of fixed relative abundances. However, a priori knowledge may
turn into estimation biases if the assumptions happen to be incorrect. This
problem is the well-known variance vs bias tradeoff. In this section, we
will study this tradeoff on simulated data in order to identify potential
sources of bias.

\subsection{Description of the experiments and associated line
  characteristics}
\label{sec:mc:experiments}

\FigMCspectra{} %

To do this, we choose input physical conditions based on the estimations
computed in Sect.~\ref{sec:data} on actual data towards the Horsehead
nebula. We choose a pixel at the edge of the central dense core (black
point in Fig.~\ref{fig:data:peak:area:spectra}), for which the fit on
actual data shows different values of kinetic temperature and volume
density, depending on the set of species studied. In particular, analyzing
just the \thCO{} and \CeiO{} lines provides nonphysical solutions (large
kinetic temperature and small volume density compared to standard dense
core values). Meaningful values for kinetic temperature and volume density
are only obtained when adding different species.

Table~\ref{tab:exp:def} lists the parameters for two experiments sampling
typical conditions for this pixel. In experiment \#1, the gas is warm
($\sim 27\K$) and moderately dense ($\sim 4.5\times10^{3}\pccm$). In
experiment \#2, the gas is colder ($\sim 22\K$) and denser
($\sim 4.5\times10^{4}\pccm$). Tables~\ref{tab:exp:1} and~\ref{tab:exp:2}
list the characteristics of the lines for experiment \#1 and \#2,
respectively. In this section, calibration errors and thermal noise are
treated as follows. On one hand, when generating the spectra, we use the
model described in Sect.~\ref{sec:data} to simulate data with calibration
errors and thermal noise. We directly use a thermal noise level that will
give a maximum S/N value of 10, except for the low S/N line $\HthCOp$.  On
the other hand, when fitting the spectra, we use the likelihood that only
takes into account the additive noise as in the fit of the actual data in
Sect.~\ref{sec:data}. As explained there, this allows the low S/N lines to
contribute to the fit.

Figure~\ref{fig:mc:spectra} compares the associated typical profiles of the
spectra. For each experiment, we have computed 200 Monte-Carlo realizations
with white Gaussian drawings for the thermal noise and calibration
error. We fitted the obtained spectra with the maximum likelihood estimator
described in Sect.~\ref{sec:methods:mle}. We thus sample the probability
distribution function of the fitted parameters.

In experiment \#1, the peak S/N of \HthCOp{} \Jone{} is low $(\sim 1.5)$
and this line has a probability of detection of $63\%$, whereas all the
lines are detected in experiment \#2, i.e., the peak S/N $\ga 5$.

The \twCO{} \Jone{} line has a large opacity $(\tau >50)$ in both
experiments and its excitation temperature is equal to the kinetic
temperature. This can be understood as a consequence of the significantly
large volume density of the gas with respect to the line's effective
critical density. Very large opacities deliver flat-top line profiles that
are not found in actual observations of \twCO{} \Jone{} lines in
interstellar clouds. The absence of flat topped \twCO{} profiles shows that
even \twCO{} probes into the cloud instead of just its surface because of
the large velocity gradients along the LoS. This example shows that just
fitting the integrated line emission is not sufficient. Estimating the
goodness of fit requires a comparison between the modeled and measured line
profiles. For this reason, we only used the peak emission of the \twCO{}
\Jone{} line, but we did fit the whole profile for the other species,
including \thCO{} and \CeiO.

The \thCO{} and \CeiO{} lines have different excitation regimes depending
on the experiment. In experiment \#2, where the volume density is one order
of magnitude larger than the (thin and effective) critical densities, both
the \Jone{} and \Jtwo{} lines are close to the local thermodynamic
equilibrium (LTE), i.e., their excitation temperature is close to the
kinetic temperature. However, in experiment \#1, where the volume density
is of the same order of magnitude as the (thin and effective) critical
densities, only the \thCO{} \Jone{} line is close to LTE.  The \thCO{} and
\CeiO{} \Jtwo{} lines are sub-thermally excited, and the \CeiO{} \Jone{}
line is supra-thermally excited. Such a regime of suprathermal excitation
of the low-$J$ CO lines has already been pointed out
by~\citet{leung76}. The \HCOp{} and \HthCOp{} \Jone{} lines are
sub-thermally excited in both experiments.

The \CeiO{} \Jone{} and \Jtwo{} lines, as well as the \HthCOp{} \Jone{}
line, are optically thin in both experiments. The \thCO{} lines have
similar opacities (1 and $3-4$ for the \Jone{} and \Jtwo{} lines) in both
experiments. The opacity of the \HCOp{} \Jone{} line is three times larger
in experiment \#1 than in experiment \#2 where the lowest energy levels are
much less populated because the density is higher.

\subsection{Structure of the figures}
\label{sec:mc:figures}

Figures~\ref{fig:mc:co:iso} to~\ref{fig:mc:c18o:h13cop} show the comparison
between the Cramér-Rao bound reference precision and the parameters fitted
on Monte-Carlo drawings of noisy RADEX models for experiments \#1 and
\#2. In the three figures, the top row describes the RADEX input
parameters. The next rows show the results obtained when fitting different
sets of lines. We first show the results when the assumptions are
consistent between the model used to generate the data and the fitted
spectra. We then show the results when these fitting assumptions are
``biased'' (e.g., simplified) with respect to the (true) generative
models. In these cases, the bottom row lists the values of the parameters
of the generative models, which are modified compared to the top row, i.e.,
these changes are ignored when fitting the spectra.

The first column describes the set of lines used and the kind of
assumptions made (biased or unbiased) during the fits. The next two columns
then compare the estimated column densities for both experiments. The
estimated column densities are shown as scatter plots, except for the
second row of Fig.~\ref{fig:mc:co:iso} where only one species is fitted. In
this specific case, we compare the histogram from the Monte-Carlo
experiment (in green) with the normal density of an unbiased efficient
estimator (i.e., with a mean equal to the true input value, and a variance
equal to the CRB of $\log N$) in red. The last two columns show the scatter
plots of the estimated kinetic temperature and volume density.

In these figures, the green dots correspond to the estimated parameters
obtained on the 200 independent realizations of the Monte-Carlo
experiments. The red ellipse is a confidence region centered on the true
parameters. Its size and orientation is computed from the Cramér-Rao bound
matrix.  The percentage of realizations that lie in this confidence ellipse
is mentioned in the legend of the graph. Any unbiased efficient estimator
will deliver $99\%\pm 1\%$ of its estimations inside the confidence
ellipse.

When the temperature and volume density vary among the species, the red
cross indicates the true values for \thCO{} and/or \CeiO{}, and the black
cross indicates the true values for the other species (\twCO{} or
\HCOp{}). This allows us to visualize the effect of the biased a priori
knowledge on the fit.  The axes and associated uncertainties are computed
on the logarithm of the estimated parameters. In
Fig.~\ref{fig:mc:13co:hcop} and~\ref{fig:mc:c18o:h13cop}, where the
abundance ratios of the column densities is assumed fixed during the fit,
the estimated values lie on a straight lines, and the confidence ellipses
become straight lines as well. In other words, the estimation of these
column densities are perfectly correlated, as expected.

The \Tkin{} vs \nHH{} scatter plots are overlaid on the variations of the
negative log-likelihood as a function of \Tkin{} and \nHH{}. In all
generality, the negative log-likelihood variations lie in a space of
dimension larger than two (the true value of estimated
parameters). However, as explained in Sect.~\ref{sec:mle:initialization},
the variations of the negative log-likelihood can be projected on the 2D
sub-space for the couple of parameters $(\Tkin{},\nHH{})$ because the
maximum likelihood estimator assumes that all species have the same
$\Tkin{}$ and $\nHH{}$.  The images thus show the projection of the 4D
matrix of negative log-likelihood onto the plane $(\nHH,\Tkin)$ by choosing
the column density ($N$) and velocity dispersion ($\sigma_V$) that maximize
it for each value of $(\nHH,\Tkin)$.

This projection of the negative log-likelihood is shown for the ideal
realization of the RADEX model without Gaussian noise. While the shape of
the negative log-likelihood varies slightly with the noise realization, its
minimum location that yields the values of the estimated parameters may
change drastically depending on the specific noise realization. This
explains why the estimated parameters shown as green dots explore the blue
valleys in these images. Finally, the effective critical densities of the
different lines are overlaid as dotted vertical lines.

\subsection{Results for the column densities}
\label{sec:mc:results:cd}

Figures~\ref{fig:mc:co:iso} to~\ref{fig:mc:c18o:h13cop} show that the
column densities are derived with a low variance (high certainty). Biased
hypotheses on kinetic temperature and volume densities do not influence the
column density values much. In contrast, biased hypotheses on abundance
ratios bias the derived column densities.

For most of the cases, more than $95\%$ of the estimated column densities
(\thCO{}, \CeiO{}, \HCOp, or \HthCOp) are in the confidence ellipse,
illustrating the good performance of the Maximum Likelihood Estimator
(MLE). The variance on the column density reduces as the number of
assumptions implemented in the estimator increases. For some scenario
(e.g., first row of Figure~\ref{fig:mc:co:iso}, experiment \#2), enforcing
the volume density to be smaller than $10^{6.5}\pccm$ provides column
density estimations with smaller variance than the one predicted by the
CRB.

The effects of biased assumptions depend on the case.  On one hand,
assuming a too low kinetic temperature or a too high volume density for
\twCO{} does slightly bias the estimated column densities of \thCO{} and
\CeiO{} (fifth row of Fig.~\ref{fig:mc:co:iso}), but they remain inside the
CRB ellipses. Similarly, assuming an incorrect abundance ratio for
$N(\twCO)/N(\thCO)$ does not impact the values of the column densities of
\thCO{} or \CeiO{} much. This is probably due to the fact that only the
peak intensity of the \twCO{} \Jone{} line is used to constrain the kinetic
temperature.  On the other hand, assuming an abundance ratio for
$N(\thCO)/N(\HCOp)$ twice as low as the true values artificially increases
the column density of \thCO{} and decreases the column density of
\HCOp{}. The bias introduced on $N(\thCO)$ is smaller than the one on
$N(\HCOp)$.

\subsection{Results for the kinetic temperature and the volume density}
\label{sec:mc:results:tkin:nHH}

The situation for the physical conditions (the kinetic temperature and
volume density) is different. In this section, we first compare the shape
of the negative log-likelihood as a function of the kinetic temperature and
volume density in the different cases. We then analyze the variances and
biases of these parameters for the two experiments separately.

\subsubsection{Shape of the negative log-likelihood image}

The shape of the NLL significantly depends on the combination of
lines/species used, in particular on their effective critical
densities. When fitting only the CO isotopologue lines
(Fig.~\ref{fig:mc:co:iso}), the minimum of the NLL has the shape of a
valley at approximately constant thermal pressure $(\Tkin \propto 1/\nHH)$
below a density of $\sim 10^4\pccm$ and constant temperature above this
density. When fitting \set{\thCO,\HCOp} (Fig.~\ref{fig:mc:13co:hcop}), the
shape of the NLL minimum (blue) region is more complex. For experiment \#1,
where the modeled gas has $\Tkin = 27.1\K$ and $\nHH = 10^{3.65}\pccm$
(i.e., a thermal pressure of $\sim 1.2 \times 10^5\K\pccm$), it is bounded
at an approximately low constant temperature of $\sim 15\K$ and at an
approximately constant pressure of $\sim 2\times10^5\K\pccm$.  However,
the chance of getting a second local NLL minimum at a $\nHH \ga 10^6\pccm$
exists and it increases in experiment \#2, where the modeled gas has
$\Tkin = 22.1\K$ and $\nHH = 10^{4.65}\pccm$ (i.e., a thermal pressure of
$\sim 10^6\K\pccm$). This implies that some ambiguity between a low and a
high volume density solution can occur. The shape is qualitatively similar
for \set{\CeiO,\HthCOp}. However, many estimations are localized on the
second local maxima in experiment \#2 because the two local maxima gets
closer to each other compared to experiment \#1 (see columns 3 and 4 of
Fig.~\ref{fig:mc:c18o:h13cop}).

\subsubsection{Experiment \#1}

The maximum likelihood estimator (MLE) only becomes efficient (most of the
dots gather inside the CRB ellipse) and the variation of the estimation
decreases quickly when relevant information is delivered by adding lines.

On Fig.~\ref{fig:mc:co:iso}, the confidence ellipse for \nHH{} covers 2
orders of magnitude when studying \set{\thCO}, one order of magnitude when
studying \set{\thCO,\CeiO}, and a factor $1.9$ when studying
\set{\twCO,\thCO,\CeiO}. The confidence ellipse for \Tkin{} covers about 1
order of magnitude when studying either \set{\thCO} or \set{\thCO,\CeiO},
and a factor $1.4$ when studying \set{\twCO,\thCO,\CeiO}. Adding the
information about the peak temperature of a very optically thick line at
LTE as \twCO{} \Jone{} allows us to constrain the temperature.

A combination of optically thick and thin lines improves the determination
of the density by an order of magnitude (case \set{\thCO,\CeiO}). But the
volume density only becomes well constrained at the same time as the
temperature when adding the fit of the \twCO{} \Jone{} peak temperature.
Figure~\ref{fig:mc:13co:hcop} indicates that the fitting of
\set{\thCO,\HCOp} delivers small uncertainty intervals compared to, e.g.,
the fitting of \set{\thCO,\CeiO}. Increased precision here is not a
question of the S/N, as the S/N of $\CeiO$ is higher than that of $\HCOp$,
but of difference in the critical density. Indeed, the effective critical
density of the \HCOp{} \Jone{} line is more than one order of magnitude
greater than those for the two \thCO{} lines (see vertical lines).

Having lines of different critical densities is, however, not always
sufficient. In Fig.~\ref{fig:mc:c18o:h13cop}, the uncertainties of the
kinetic temperature and volume density are about one order of magnitude
each when fitting \set{\CeiO,\HthCOp} whose critical densities differ by
more than one order of magnitude. In this case, all lines are optically
thin, implying a degeneracy between temperature and density, the pressure
being relatively well constrained. We verified that the $(\Tkin,\nHH)$
degeneracy remains when increasing the S/N of the \HthCOp{} line by
decreasing the noise level $\sigma_b$ by a factor 10.

Assuming that the spatial portion of the LoS that dominates the emission of
the lines is identical for the \twCO{} line and its rare isotopologues
slightly biases the estimation to high kinetic temperatures and low volume
densities (row 4 of Fig.~\ref{fig:mc:co:iso}). This confirms that the
kinetic temperature is above all constrained by the peak temperature of the
\twCO{} \Jone{} line. In the test where the generative model assumes that
the \HCOp{} line emission comes from gas at a slightly different
temperature and volume density than the gas producing the \thCO{} emission
(2nd row of Fig.~\ref{fig:mc:13co:hcop}), the kinetic temperature becomes
highly biased. As we know that dense cores and filaments are surrounded by
a more diffuse, lower density and warmer envelope, a multi-layer model
could be more suited in this case. This will be the subject of another
article.

Assuming an $N(\thCO)/N(\HCOp)$ abundance ratio lower than reality by a
factor 2 positively biases the estimated volume density by a factor 2 as
shown in the fourth row of Fig.~\ref{fig:mc:13co:hcop}. The main difference
with the impact of biases on the column densities is that a majority of the
estimations lies outside the CRB ellipses.

\FigMCCOiso{} %
\FigMCthCOHCOp{} %
\FigMCCeiOHthCOp{} %
\clearpage %

\subsubsection{Experiment \#2}
\label{sec:mc:experiment:2}

Several significant differences exist between Experiment \#1 and Experiment
\#2.  While the accuracy of the temperature increases significantly when
adding the constraint from the \twCO{} \Jone{} peak temperature, no
information on the volume density can be recovered in experiment \#2, when
using only the CO isotopologue lines. The CRB interval of confidence for
$\nHH$ covers the full range of explored densities. This is due to the fact
that all lines get close to the LTE state (see the excitation temperature
in Table~\ref{tab:exp:2}) where only the kinetic temperature determines the
excitation of the species.

The fifth column of Fig.~\ref{fig:mc:13co:hcop} shows that improving the
$\nHH$ estimation requires us to analyze a line or species with a higher
effective critical density than the considered CO isotopologue lines. But
here again, this is only true when some of the fitted lines are optically
thin and others are optically thick in order to lift the degeneracy between
kinetic temperature (kinetic energy of collisions) and volume density
(number of collisions) during the excitation of the species. Moreover,
Fig.~\ref{fig:mc:c18o:h13cop} shows that there exist ambiguities between
the two local minima of the NLL at the true volume density of the
generative model or at a much higher density value.

Just like experiment \#1, an underestimation of the a priori
knowledge on the abundance ratio between $\thCO$ and $\twCO$ leads to an
overestimation of $\nHH$, but the induced bias, which can be observed in
the fourth row of Fig.~\ref{fig:mc:co:iso}, is much more important here:
almost a factor 10.  Not only is the volume density biased, but even worse,
the observed variance is small. This means that the observer obtains a
region with low spatial variance for the estimated parameters when the
maximum likelihood estimator is applied to several pixels, thus giving an
incorrect impression of the robustness of the results. Such an example
shows that a physical or chemical misspecification in the hypotheses of the
model used to fit the data can lead to incorrect estimates of the fitted
parameters, most notably the kinetic temperature and/or the volume density.

One possible explanation of this bias is that since
\mbox{$\nHH\leq \ncriteff$,} the non-LTE excitation model feeds the line
upper level with more collisions to compensate for the underestimation of
$N(\HCOp)$ in order to produce a sufficiently bright $\HCOp$ \Jone{} line.

\subsection{Intermediate summary}

We have studied the estimation of the column densities, kinetic
temperatures, and volume densities when fitting different combinations of
the CO and \HCOp{} isotopologue lines with the RADEX escape probability
model, on two example lines of sight. While the Cramér-Rao bound is a lower
limit for the variance of any unbiased estimator, we here confirm that this
lower limit can be reached by a ``standard'' maximum likelihood estimator,
provided that the assumptions used during the fitting process are
consistent (i.e., unbiased) with respect to reality (generative model).

Estimations of the column density are efficient on the considered examples,
even when the uncertainty (CRB) on $\nHH$ or $\Tkin$ are large. This means
that the escape probability models provide robust estimations of column
densities allowing for the production of maps in a wide variety of situations,
which can be quantitatively characterized a priori with
$\CRB^{1/2}(\log N) < 0.1$. Assuming an incorrect abundance ratio for, e.g.,
$N(\thCO)/N(\HCOp)$ biases the estimation of $\log(N(\HCOp))$ more than the
estimation of $\log(N(\thCO))$.

Likewise, we confirm that CRB precisions for the volume density and kinetic
temperature can be achieved by a ``standard'' maximum likelihood estimator
when $\CRB^{1/2}(\log \Tkin) < 0.1$, and $\CRB^{1/2}(\log \nHH)<0.1$. A
good estimation of the kinetic temperature and volume density requires a
combination of high-enough S/N lines of different opacities (thick and
thin) and effective critical densities.  On one hand, this is delivered by
the fit of \set{\twCO,\thCO,\CeiO} in experiment \#1, where
$\nHH = 10^{3.65}\pccm$ and $\Tkin = 27.1\K$. In this case, a perfect
knowledge of $N(\twCO)/N(\thCO)$ is not required, but an incorrect
assumption about the gas temperature that emits the \twCO{} line biases
both the kinetic temperature and the volume density.  This is consistent
with the fact that the $\twCO$ peak is mostly useful to regularize the
estimation of $\Tkin$.  On the other hand, fitting \set{\twCO,\thCO,\CeiO}
in experiment \#2, where $\nHH = 10^{4.65}\pccm$ and $\Tkin = 22.1\K$,
fails to constrain the volume density, in agreement with the fact that the
low-$J$ lines of the CO isotopologues have almost reached LTE.

Fitting a set of lines that have different critical densities enlarges the
range of $(\Tkin,\nHH)$ values for which the kinetic temperature and volume
density can be estimated as long as some of the lines are optically thin
and others optically thick. An incorrect assumption on the abundance ratio
between \thCO{} and \HCOp{} biases the estimation of the temperature and
density because the corresponding \Jone{} lines have different effective
critical densities.

%%%%%%%
% CRB
%%%%%%%

\newcommand{\FigCRBoneSpecies}{%
  \begin{figure}
    \centering %
    \includegraphics[width=\linewidth]{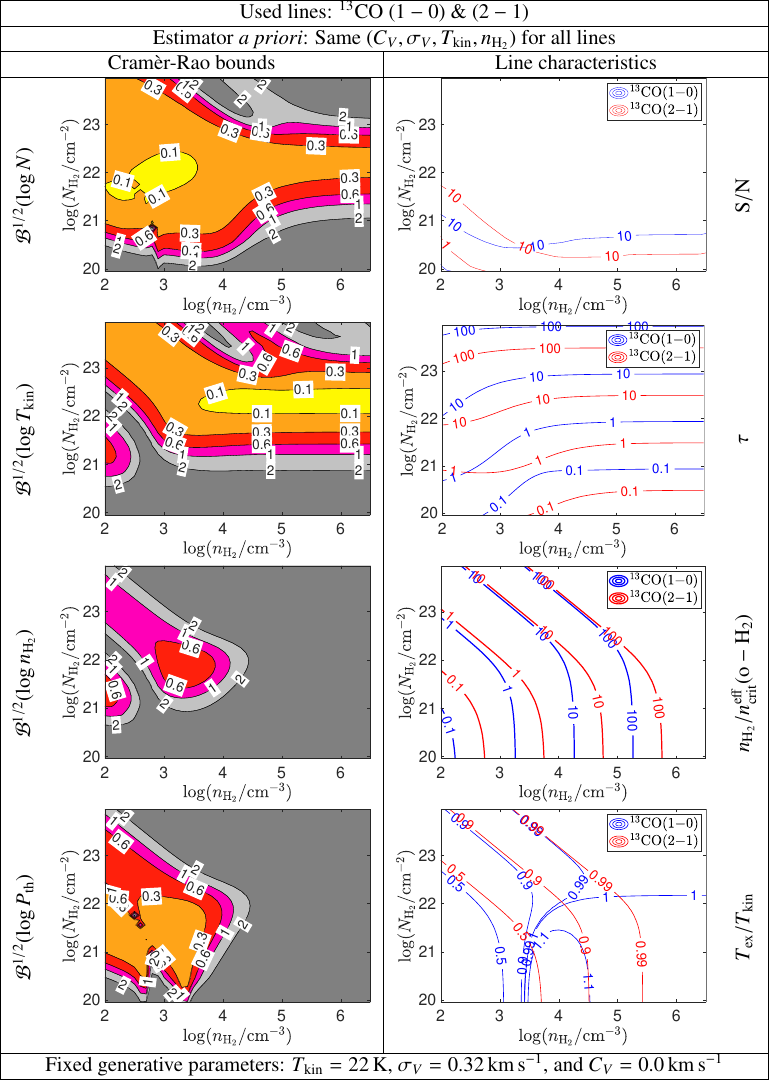}
    \caption{Comparison of the CRB reference precisions \textbf{(left
        column)} on the column density, kinetic temperature, volume
      density, and thermal pressure, when fitting the \thCO{} \Jone{} and
      \thCO{} \Jtwo{} lines, with these lines main characteristics
      \textbf{(right column)}: S/Ns, opacities, ratios of the volume
      density to the effective critical densities, and ratios of the
      excitation temperature over the kinetic temperature. The blue and red
      curves show the results for the \Jone{} and \Jtwo{} lines,
      respectively. The values of the parameters that are fixed when
      generating the RADEX spectra are listed in the bottom row. The
      a priori knowledge that are used when computing the CRB
      precisions are listed in the second row.}
    \label{fig:crb:one:species}
  \end{figure}
}

\newcommand{\FigCRBspectra}{%
  \begin{figure}
    \centering %
    \includegraphics[width=\linewidth]{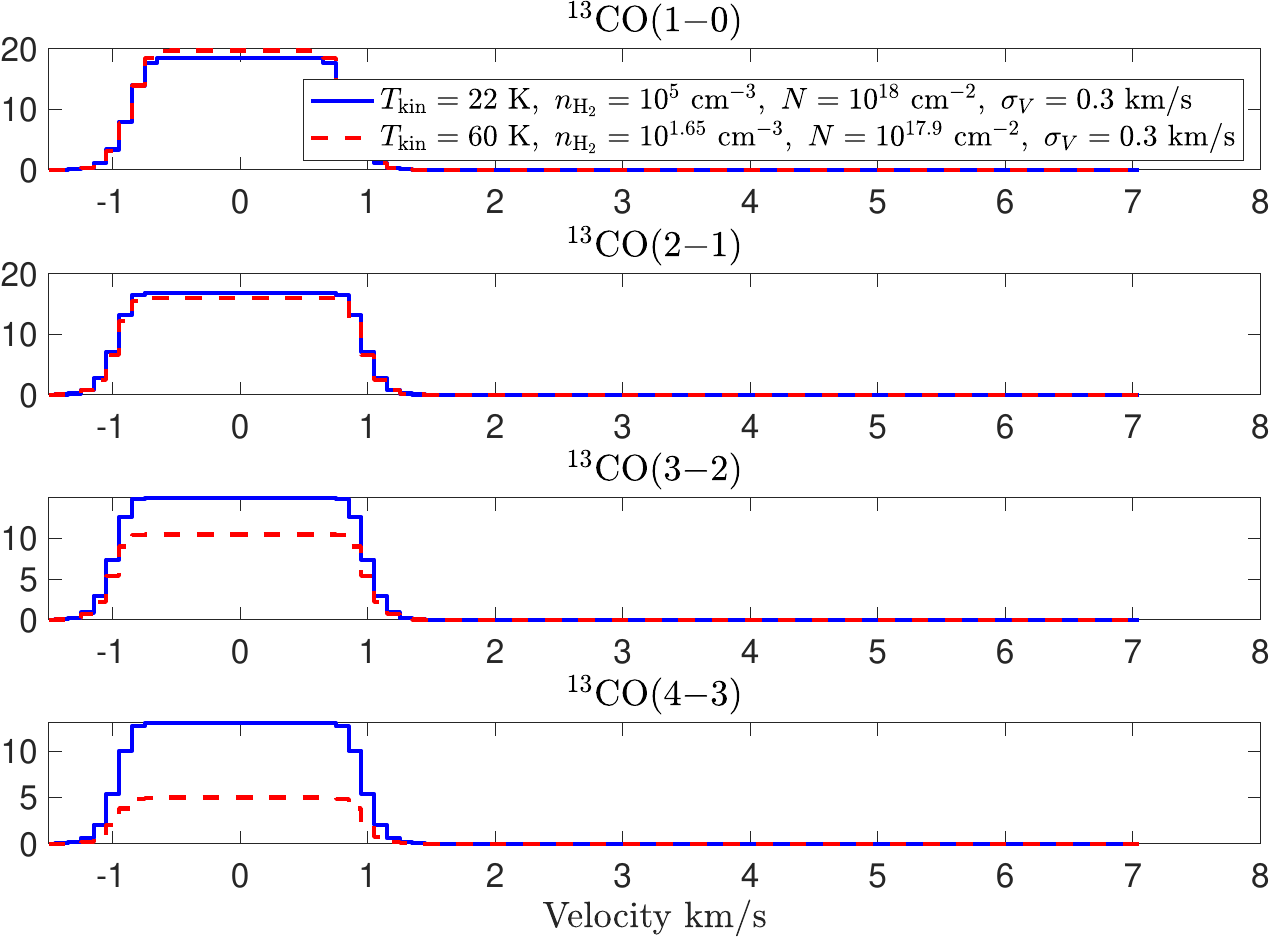}
    \caption{Comparison of the spectral shape of the \Jone{} \textbf{(top)}
      to \Jfour{} \textbf{(bottom)} lines of \thCO{} for two different
      combinations of kinetic temperature, volume density, and column
      density. The plain blue and dashed red histograms show the spectra
      for $(22\K, 10^5\pccm, 10^{18}\pscm)$ and
      $(60\K, 10^{1.65}\pccm, 10^{17.9}\pscm)$, respectively. The lines
      share the same centroid velocity $C_V = 0\kms$ and velocity
      dispersion \mbox{$\sigma_V = 0.32\kms$.}  The two sets of physical
      conditions deliver degenerate spectral profiles within the noise
      level for the \Jone{} and \Jtwo{} lines. Additional information,
      e.g., the \Jthree{} or \Jfour{} lines, is needed to lift this
      degeneracy.}
    \label{fig:crb:spectra}
  \end{figure}
}

\newcommand{\FigCRBoneORtwoSpecies}{%
  \begin{figure}
    \centering %
    \includegraphics[width=\linewidth]{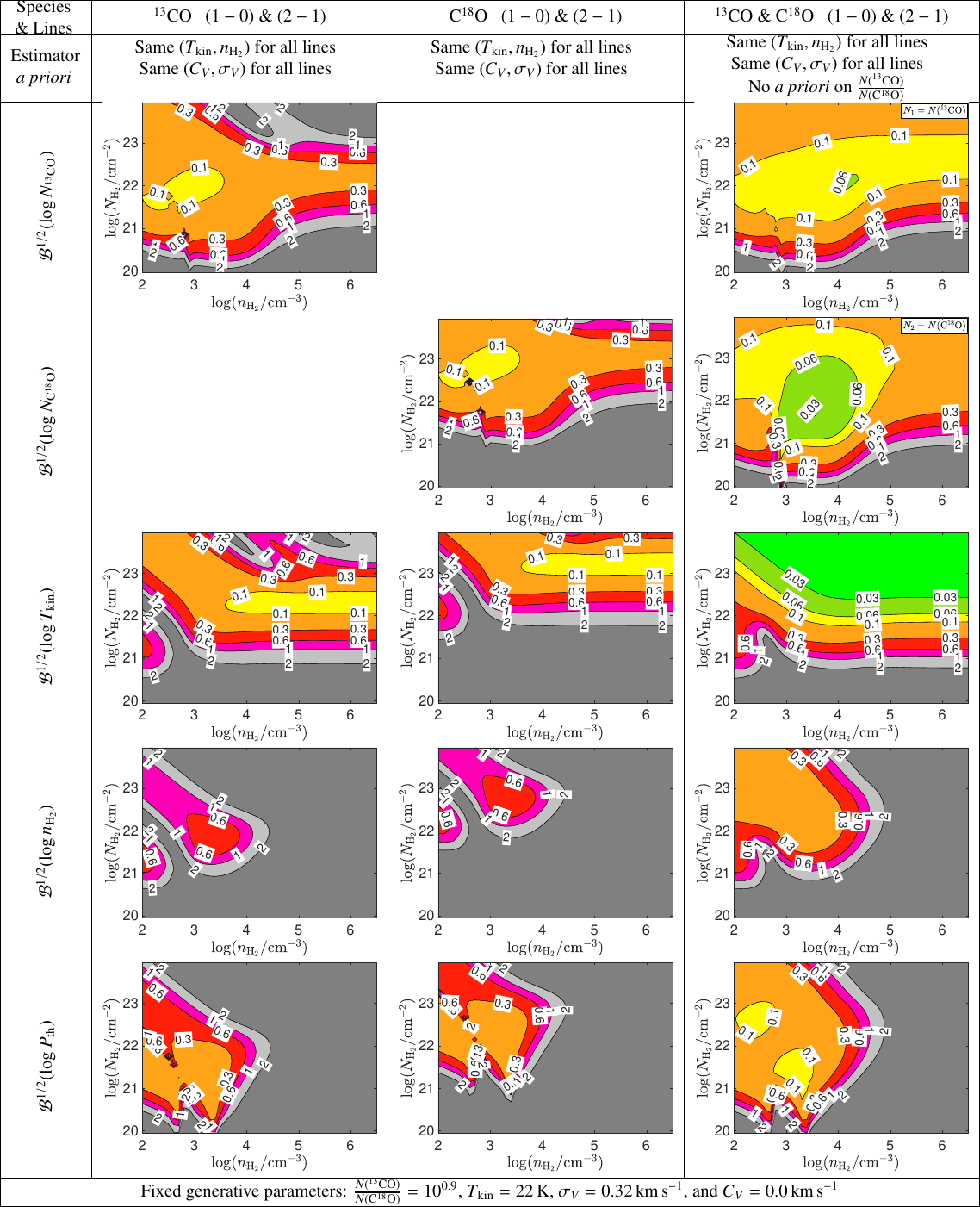}
    \caption{Comparison of the CRB precisions when combining the first two
      $J$ transitions of one or two CO isotopologues.  These reference
      precisions are plotted for different values of column densities
      ($y$-axes) and volume densities ($x$-axes).  The values of the
      parameters that are fixed when generating the RADEX spectra are
      listed in the bottom row.  The precision depends on the a priori
      knowledges that are taken into account by the estimator.  These are
      listed in the second row.  The first two columns show the precisions
      for a single CO isotopologue (\thCO{} or \CeiO{}, while the last
      column shows the precisions when studying the \thCO{} and \CeiO{}
      isotopologues simultaneously.  From top to bottom, the panels show
      the precisions on the \thCO{} and \CeiO{} column densities, kinetic
      temperature, volume density, and thermal pressure.  The CRB
      precisions are color coded as listed in Table~\ref{tab:conversion}.}
    \label{fig:crb:one:or:two:species}
  \end{figure}
}

\newcommand{\FigCRBtwoSpecies}{%
  \begin{figure}
    \centering %
    \includegraphics[width=\linewidth]{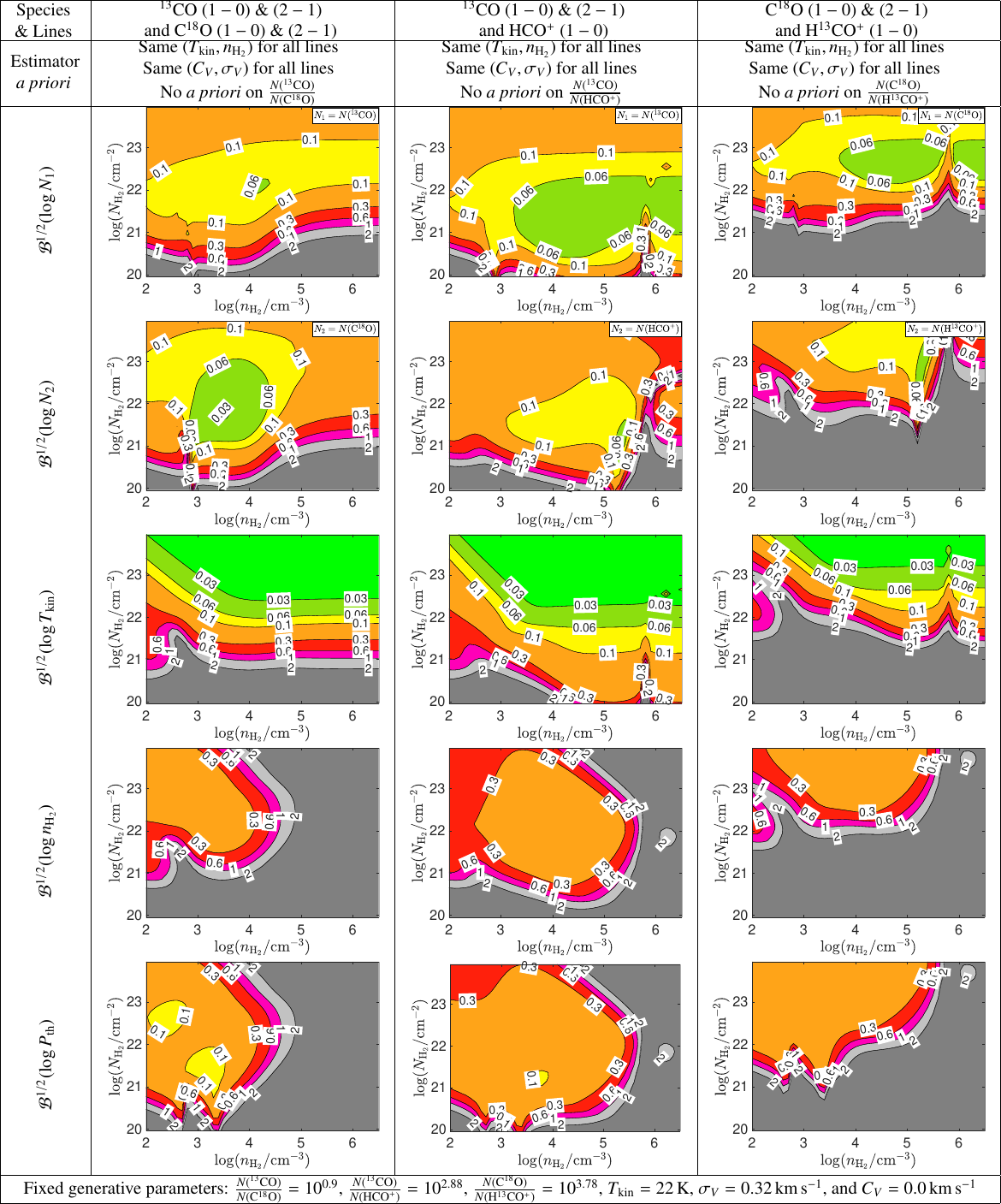}
    \caption{Comparison of the CRB reference precisions for different
      combinations of two species and all their available lines. The layout
      of this figure is identical to Fig.~\ref{fig:crb:one:or:two:species},
      except for the combination of studied lines: The first column shows
      the precision when the \Jone{} and \Jtwo{} lines of the \thCO{} and
      \CeiO{} isotopologues are used. The last two columns study a
      combination of the one CO isotopologue with one \HCOp{} isotopologue:
      \thCO{} and \chem{H^{12}CO^+} on the second column and \CeiO{} and
      \HthCOp{} on the third column.}
    \label{fig:crb:two:species}
  \end{figure}
}

\newcommand{\FigCRBtwoSpeciesPlusCOpeak}{%
  \begin{figure*}
    \centering %
    \includegraphics[width=0.8\linewidth]{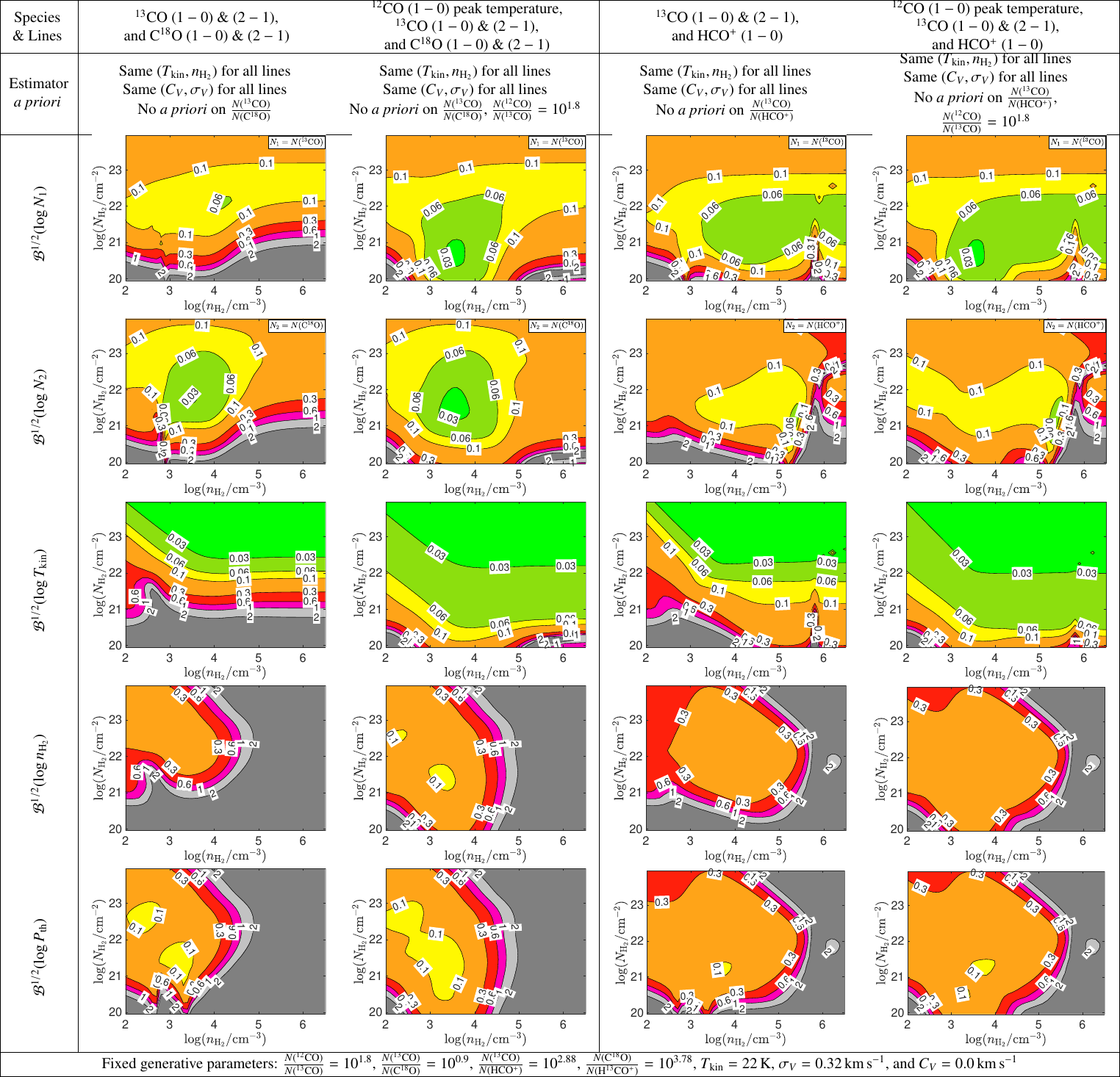}
    \caption{Comparison of the CRB reference precisions for all the
      available lines of different combinations of two species, plus the
      constraint from \twCO{} \Jone{} peak temperature.  The reminder of
      the layout is identical to Fig.~\ref{fig:crb:two:species}.}
    \label{fig:crb:two:species:plus:co:peak}
  \end{figure*}
}

\newcommand{\FigCRBVariations}{%
  \begin{figure*}
    \centering %
    \includegraphics[width=\linewidth]{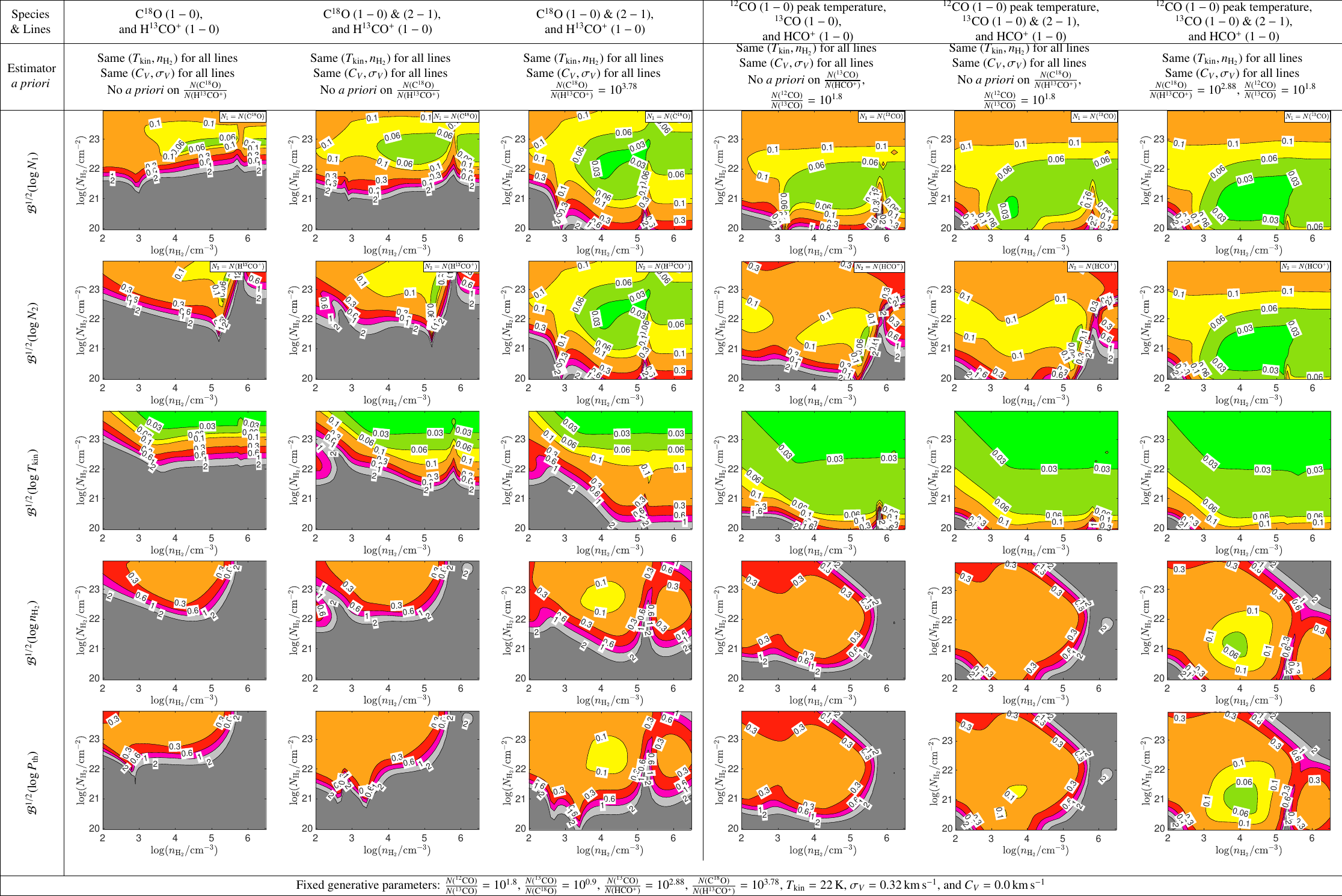}
    \caption{Comparison of the CRB reference precisions when using either
      only \Jone{} lines or adding the \Jtwo{} lines for \thCO{} and
      \CeiO{} and influence of the a priori on the abundance ratio.}
    \label{fig:crb:variations}
  \end{figure*}
}

\newcommand{\TabConversion}{%
  \begin{table*}
    \caption{Considered precisions on $\log x$, and $x$, plus the
      associated color codes systematically used in Sect.~\ref{sec:CRB}.} %
    \centering %
    \begin{tabular}{lccccccc}
      \hline
      \hline
      $\CRB^{1/2}(\log x) = \sigma$ & $0.03$        & $0.06$        & $0.1$         & $0.3$       &  $0.6$       & $1$        &  $2$ \\
      \hline
      $[10^{-\sigma},10^{+\sigma}]$ & $[0.93,1.07]$ & $[0.87,1.15]$ & $[0.79,1.25]$ & $[0.5,2.0]$ & $[0.25,4.0]$ & $[0.1,10]$ & $[0.01,100]$ \\
      Color code                    & Light green   & Green         & Yellow        & Orange      & Red          & Pink       & Grey         \\
      \hline
    \end{tabular}
    \label{tab:conversion}
  \end{table*}
}

\newcommand{\TabthCO}{%
  \begin{table}
    \centering{}
    \caption{Characteristics of \thCO{} lines shown in
      Fig.~\ref{fig:crb:spectra} where $\ncritthin$ and $\ncriteff$ are the
      critical density of $\pHH{}$. In both case
      $\sigma_V=0.3\unit{km/s}$.}
    \begin{tabular}{cccccc}
      \hline
      Line & $\tau$ & $\Tex$  & $\Tpeak$   & \ncritthin & \ncriteff \\ 
           &     --     & \unit{K} & \unit{K}   & \pccm                    & \pccm \\
      \hline
      \hline
      \Jone{} & 50    & 22.0  & 18.5 & $10^{3.29}$ & $10^{1.77}$ \\
      \Jtwo{} &  140  & 22.0 & 16.9  & $10^{3.81}$ & $10^{1.84}$ \\
      \Jthree{} &  174  & 22.0  & 15 & $10^{4.24}$ & $10^{2.17}$ \\
      \Jfour{} &  135 & 22.0 & 13 & $10^{4.56}$ & $10^{2.6}$ \\
      \multicolumn{6}{c}{For $\Tkin=22~\unit{K}$, $\nHH=10^5\unit{cm^{-3}}$, $N=10^{18}\unit{cm^{-2}}$.}
      \\
      \hline
      \hline
      \Jone{} & 43    & 23.2 & 19.7 & $10^{3.27}$ & $10^{1.81}$ \\
      \Jtwo{} &  134  & 21.1  & 16 & $10^{3.82}$ & $10^{1.87}$ \\
      \Jthree{} &  185  & 17.2 & 10.5 & $10^{4.25}$ & $10^{2.15}$ \\
      \Jfour{} &  132 & 12.8 & 5 & $10^{4.56}$ & $10^{2.62}$ \\
      \multicolumn{6}{c}{For
      $\Tkin=60~\unit{K}$, $\nHH=10^{1.65}\unit{cm^{-3}}$, $N=10^{17.9}\unit{cm^{-2}}$.}
      \\
      \hline
    \end{tabular}
    \label{tab:13CO}
  \end{table}
}

\newcommand{\TabThreeSets}{%
  \begin{table}
    \caption{Physical regime for which $\CRB^{1/2}(\log \nHH)<0.1$ and
      $\CRB^{1/2}(\log \Tkin)<0.1$, for $\Tkin=22\K$ and
      $\sigma_V=0.32\kms$. The a priori knowledge and number of unknowns
      associated to \set{\twCO{},\thCO{},\CeiO}, \set{\thCO{},\HCOp} and
      \set{\CeiO,\HthCOp} are defined in Table~\ref{tab_set_of_species}.}
    \begin{tabular}{lcc}
      \hline
      \hline
      Set of species & Column density & \nHH{}  \\
                     & \pscm{}        & \pccm{} \\
      \hline
      $\set{\twCO{},\thCO{},\CeiO}$ & $N(\thCO{})=10^{15.9\pm 0.7}$ & $10^{3.4\pm 0.4}$ \\
      $\set{\thCO{},\HCOp}$         & $N(\thCO{})=10^{16\pm 0.5}$   & $10^{4.0\pm 0.5}$ \\
      $\set{\CeiO,\HthCOp}$         & $N(\CeiO)=10^{16.25\pm 0.5}$  & $10^{4.0\pm 0.5}$ \\
      \hline
    \end{tabular}
    \label{tab:3:sets}
  \end{table}
}

\section{Parametric study of the reference precisions on the column
  densities, kinetic temperature, volume density and thermal pressure}
\label{sec:CRB}

\TabConversion{} %

The Cramér-Rao bound gives the reference precision that can be achieved by
an unbiased and efficient estimator for a given physical model (here RADEX)
and observational setup (combination of lines and noise levels). In the
previous section, we studied two specific combinations of physical
conditions for a molecular cloud. We now extend this quantitative study to
a large domain of physical conditions that can be constrained with a
suitable precision for different combinations of species, associated lines,
and a priori knowledge linking them.

We will restrict the combination of species to the CO and \HCOp{} main
isotopologues and the combination of lines to the \Jone{} transition for
all species, plus the \Jtwo{} transition for the \thCO{} and \CeiO{}
species. In more details, we will study the information that can be
extracted in the following sets of observations: 1) We analyze in depth the
simplest case, i.e., when only the \Jone{} and \Jtwo{} lines of \thCO{} are
available.  2) We then quantify the gain in precision when adding the same
lines of the rarer \CeiO{} isotopologue.  3) We also compare the benefits
of two different combinations of one CO isotopologue with one isotopologue
of \HCOp{}. We will combine \CeiO{} and \HthCOp{} that are easy to observe
in dense cores, but also \thCO{} and \chem{H^{12}CO^+} that are better
suited to probe the lower volume density envelope.  4) We quantify the gain
in precision when adding the constraint from the peak temperature of the
\twCO{} \Jone{} lines to the previous combinations of species and lines.
5) We then quantify the loss of information when only the lowest $J$ line
of \thCO{} and \CeiO{} are available. This is of particular interest for
the ORION-B project that mapped a large fraction of the Orion B cloud
between 70 and 116\GHz.  6) We finally check the potential advantage of
fixing the relative abundances.
 
For these studies, we will explore the variations of the precision in the
two-dimensional space: column density of molecular hydrogen
($10^{20} \le N(\HH) \le 10^{24}\pscm$) vs its volume density
($10^2 \le \nHH \le 10^6\pccm$). The other parameters to generate the model
and compute the CRB are as follows.
\begin{description}
\item[\textbf{Physical parameters}] A kinetic temperature of 22\K, a
  velocity dispersion of 0.32\kms, and a centroid velocity of 0\kms.
\item[\textbf{Chemical abundances}] We use those defined in
  Eq.~\ref{eq_abundance} in Sect.~\ref{sec:data}, plus the following
  \thCO{} abundance relative to molecular hydrogen
  \begin{equation}
    \frac{N(\thCO)}{N(\HH)} = 2.5\,10^{-6}=10^{-5.6}.
  \end{equation}
  The range of studied column densities for the different species has been
  adapted so that it fits the range of molecular hydrogen column densities,
  given the chosen relative abundances.
\end{description}
  
\subsection{Typical figure layout}
\label{sec:CRB:color:codes}

The first column of Fig.~\ref{fig:crb:one:species}, as well as
Fig.~\ref{fig:crb:one:or:two:species}, \ref{fig:crb:two:species},
and~\ref{fig:crb:two:species:plus:co:peak} compare the CRB reference
precisions when combining different low-$J$ lines of different
species. This precision is plotted for different values of column densities
($y$-axes) and volume densities ($x$-axes).  The values of the parameters
that are fixed when generating the RADEX spectra are listed in the bottom
row.  The reference precision depends on the a priori assumptions
that are taken into account by the estimator. These are listed in the
second row from the top.  From top to bottom, the panels show the
precisions on the species column densities, and gas kinetic temperature,
volume density, and thermal pressure.

For any parameter $x$, $\CRB^{1/2}(\log x)$ is equal to the standard
deviation of any unbiased and efficient estimators of $\log
x$. Table~\ref{tab:conversion} lists the relationship between
$\CRB^{1/2}(\log x)$ and the associated relative interval of confidence on
$x$ and the adopted color code. We are going from a $\pm 1\sigma$ interval
to a multiplicative factor interval of $[10^{-\sigma},10^{+\sigma}]$.  A
$\pm 1\sigma$ value of 0.03 gives a relative confidence interval of the
same order of magnitude as the calibration error (5-10\%). A $\pm 1\sigma$
value of 0.1 corresponds to a multiplicative uncertainty of the order of
20\%. This is considered a good precision for the column density and
kinetic temperature.  Values of 0.3 and 0.6 for $\pm 1\sigma$ give a
precision within a factor of 2 and 4, respectively. Getting a reliable
estimate of the volume density at this precision level would already be a
great achievement. Values of 1 and 2 for $\pm 1\sigma$ deliver an
uncertainty that covers between two and four orders of magnitude. The
associated parameters are too uncertain to be useful.

In order to emphasize these different ranges of relative precisions, the
maps of $\CRB^{1/2}(\log x)$ are systematically color-coded throughout the
paper as follows: Green means good to excellent relative precisions, yellow
and orange standard relative precisions, and red to pink poor relative
precisions. The grey color is used to indicate the parameter space where
the precision is so low that we consider that the lines deliver no
information on the considered parameter.

\subsection{A simple case: \thCO{} \Jone{} and \Jtwo{}}
\label{sec:crb:simple-case}

Figure~\ref{fig:crb:one:species} compares the CRB reference precisions with
the main characteristics of the \Jone{} and \Jtwo{} lines for \thCO{}. The
precisions depend much on the considered physical parameter.

\subsubsection{Precision of the column density}

\FigCRBoneSpecies{} %
\FigCRBspectra{} %

The column density is the easiest parameter to estimate.  It can be
estimated within a factor of two for a large variety of column and volume
densities. However the range of column densities with a good precision
decreases from three orders of magnitude at low volume density to 0.5 order
of magnitude at high volume density.

A precision better than 25\% can only be reached for volume densities lower
than $5\times 10^3\pccm$ and a column density between $10^{21.5}$ and
$10^{22.5}\pscm$.  The latter conditions imply $\tau_\Jone{}<10$,
$\tau_\Jtwo{}>1$, and $\nHH < 10\,\ncriteff$ for the two lines.  The
influence of the opacity can be interpreted as a trade-off between having a
sufficiently high S/N while avoiding a loss of information on the column
density because the line intensities saturate. The influence of
$\nhd/\ncriteff$ is more complex but can be related to the radiative
pumping effect at high column densities and low volume densities.

\subsubsection{Precision on the kinetic temperature}

The kinetic temperature can be estimated within a factor of two in a
narrower range of conditions that typically spans one order of magnitude
for the column density around $N(\HH) = 10^{22.3}\pscm$ when the volume
density is larger than the optically thin critical density
$\sim 10^3\pccm$.

A precision better than 25\% can only be reached in a different parameter
space from the one that maximimizes the precision on the column density. It
corresponds to a column density $10^{22.1}< N(\HH) < 10^{22.6}\pscm$ for
$\nHH > 10^{3.6}\pccm$. This implies similar conditions for the line
opacities ($\tau_\Jone<10$, $\tau_\Jtwo>1$), but a different condition on
the line critical density ($\nHH > 5\,\ncriteff$).

For high opacity, Eq.~\ref{eq_s} simplifies at its peak value to
$s\simeq J(\Tex)-J(\TCMB)$. Moreover, when $\nHH$ is sufficiently above
$\ncriteff$, the local thermodynamic equilibrium is reached, and
$\Tex = \Tkin$ for both lines. It thus makes sense that the estimation of
$\Tkin$ can be accurate when these conditions are met. It is
counter-intuitive that the precision decreases when the $\thCO{}$ column
density increases above $10^{17}\pscm$ and $\nHH \ga 10^{4}\pccm$. Indeed,
the approximation $s\simeq J(\Tkin)-J(\TCMB)$ is expected to be valid under
these conditions, and the gas is expected to be at local thermodynamic
equilibrium.

To understand this phenomenon, we compare in Fig.~\ref{fig:crb:spectra} the
spectra of the lines for the \Jone{} to \Jfour{} transitions of \thCO{} for
two different pairs of (\Tkin,\nHH) values, $(22\K,10^5\pccm)$ and
$(60\K,10^{1.65}\pccm)$, and logarithms of the column densities that differ
by the precision we aim for, i.e., 0.1: $N(\thCO) = 10^{18}\pscm$ or
$10^{17.9}\pscm$. The \Jone{} and \Jtwo{} line have similar profiles within
the calibration uncertainty for these two sets of conditions, implying
uncertain kinetic temperatures and column densities. An ambiguity therefore
remains between a high density, low kinetic temperature case, and a low
density, high kinetic temperature case, even for a large molecular column
density\footnote{The fact that at least one high-$J$ CO line is required to
  break the degeneracy between these two regimes is true even though
  flat-top CO line profiles are absent in actual observations.}.
Table~\ref{tab:13CO} lists the main parameters of the lines shown in
Fig.~\ref{fig:crb:spectra}. It shows that only $\Tex$ are significantly
different between the blue and red profiles, and LTE is not reached for the
red profiles.

\TabthCO{} %

This example shows the power of the Cramér-Rao bound. It quantifies the
precision that any unbiased efficient estimator will reach for a given
model, independent on the physical insight we can have.  Indeed, assuming
that the peak temperature of the spectra is a good approximation of the
kinetic temperature for optically thick lines is an a priori
knowledge that would change the model of Eq.~\ref{eq_s}, and thus the CRB
computation. When the RADEX model is used for fitting noisy data without
this a priori knowledge, the precision on the kinetic temperature
and column density decreases significantly. Two other possibilities to
increase the precision would be to average many observations of the \Jone{}
or \Jtwo{} lines to decrease the noise level and the uncertainty of the
calibration or to add the observations of the \Jthree{} or \Jfour{} lines
with enough S/N because their profiles are much more sensitive to the
variations of the physical conditions. A gain of a factor 10 on the
precision of the estimation of $\sigma_b$ is approximately required to gain
the same factor on the estimation of the kinetic temperature.  Adding lines
is thus a better strategy than increasing the integration time because the
required integration time to actually detect a new line is often
significantly lower than the additional time needed to improve the S/N and
calibration uncertainty on an already detected line.

\subsubsection{Precision of the volume density or thermal pressure}

The precision of the volume density is at best a factor of 4 for a small
region around $\nHH \sim 10^{3.5}\pccm$ and $N(\HH) \sim 10^{22}\pscm$.  As
expected, when $\Tex\simeq \Tkin$, the local thermal equilibrium is
reached, and the line loses any sensitivity to the volume density. When
$\nHH \la \ncriteff$, the line becomes sensitive either to the volume
density or to the thermal pressure. It is well known that when the lines
become optically thin, the excitation is controlled by the product of the
volume density and kinetic temperature, namely, the thermal pressure. The
independent determination of the volume density and the kinetic temperature
is difficult (the shape of their CRB are similar at low densities), while
the thermal pressure can be determined to within a factor of two. It can
therefore be interesting to derive thermal pressure maps of molecular
clouds, which can be compared with maps of the magnetic or turbulent
pressure.

\subsubsection{Effective critical density versus excitation temperature}

While the ratios of the volume density to the effective critical density
show similar behaviors for the \Jone{} and \Jtwo{} lines, the ratios of the
line excitation temperatures to the gas kinetic temperature behave quite
differently for $\nHH \ga 10^{3.3}\pccm$ and $N(\HH) \le 10^{22}\pscm$.  In
this region $\Tex\Jone \ga \Tkin$. This is the sign that the $J=1$ energy
level becomes super-thermally excited. While this inversion of population
effect is moderate, it happens in a region of $(\Tkin,\nHH)$ that is well
represented in the Horsehead pillar.

\subsection{One or two CO isotopologues}

\FigCRBoneORtwoSpecies{} %
\FigCRBtwoSpecies{} %
\FigCRBtwoSpeciesPlusCOpeak{} %
\FigCRBVariations{} %

Figure~\ref{fig:crb:one:or:two:species} compares the precision that can be
reached when studying either a single CO isotopologue (either \thCO{} or
\CeiO{}) or their combinations. In all three cases we use their \Jone{}
and/or \Jtwo{} lines, but we make no assumption on their abundances.

The two first columns compare the cases when using the two lines of either
\thCO{} or \CeiO{}.  As \CeiO{} and \thCO{} have similar values of $\Eup$,
$\nu_l$ and $A$ (see Table~\ref{tab_lines}), the shapes of the CRB
variations in the $N(\HH),\nHH$ space are similar, but the results for the
reference precisions with \CeiO{} are vertically shifted by the fixed
abundance ratio $N(\thCO{})/N(\CeiO) = 10^{0.9}$ with respect to the
reference precisions with \thCO{}. For a given $N(\HH)$ column density and
noise level, this makes two major differences. First, the \CeiO{} lines
will have lower S/N. This difference will only have an impact when the S/N
becomes lower than 10 as this is our noise saturation level. Second, the
\CeiO{} lines will often be optically thin when the \thCO{} lines will have
a higher opacity, approaching the optically thick regime.  These two
effects imply that the \CeiO{} lines will be more suited to constrain gas
with larger column densities than the \thCO{} lines for the same velocity
dispersion.

When we assume that the emission of \thCO{} and \CeiO{} is produced in gas
with the same $\nHH$, $\Tkin$, and velocity dispersion, the combinations of
different opacities and S/Ns for the same rotational transition of two
isotopologues will considerably modify the shape of the precision in the
$N(\HH)$ vs $\nHH$ space. This corresponds to the physical situation where
\thCO{} and \CeiO{} are co-located spatially. The last column of
Fig.~\ref{fig:crb:one:or:two:species} shows the CRB reference precisions
when using both isotopologues together.

The most spectacular change on precision happens for the kinetic
temperature. The low precision for $\Tkin$ at
$N(\HH{})\simeq 10^{23.6}\pscm$ and $\nHH \simeq 10^5\pccm$ is now replaced
by an excellent precision.  Adding an isotopologue therefore helps to
resolve the ambiguity between low density-high temperature and high
density-low temperature gas as discussed above. As the \CeiO{} lines have
smaller opacities than the \thCO{} ones, the associated effective critical
densities $\ncriteff$ are closer to the optically thin critical densities,
and we can expect that $\nHH$ is more precisely estimated. Indeed, the best
precision achieved on $\nHH$ increased by a factor 2.

\subsection{Two CO isotopologues versus one CO plus one \HCOp{}
  isotopologues}

Figure~\ref{fig:crb:two:species} compares the CRB reference precisions when
studying two molecular species among four. The first column shows the
precision for the two low-$J$ lines of \thCO{} and \CeiO{}, while the last
two columns combines the two low-$J$ lines of one of the CO isotopologues
with the \Jone{} line of one of the \HCOp{} isotopologues.

As the effective critical density, $\ncriteff$, of \HCOp{} (resp. \HthCOp)
is significantly larger than the one of \thCO{} (resp. \CeiO, see
Table~\ref{tab:exp:1}), we expect a significant gain in precision for
$\nHH \ga 10^4\pccm$ compared to the estimations using only the \thCO{} and
\CeiO{} lines. This is indeed the case for $N(\HH)$ column densities above
$10^{21.1}\pscm$ and below $10^{22.6}\pscm$ for the \set{\thCO,\HCOp} set
of species, and for $N(\HH)$ column densities above $10^{22.2}\pscm$ for
the \set{\CeiO,\HthCOp} set. This is due to the lower abundance of \CeiO{}
and \HthCOp{} that turns into lower opacities, and thus a greater
sensitivity to physical conditions at large column densities. For the
\set{\thCO,\HCOp} set of species, the precision for the kinetic
temperature, volume density and pressure is now reasonable for almost all
the space of covered $N(\HH),\nHH$ values. A precision better than 25\%
seems reachable for the conditions encountered in the Horsehead nebula,
which corresponds to $\log(\nHH)=4\pm 0.5\pccm$ and $\Tkin\in[15,\,40]\K$
as shown in Fig.\ref{fig:data:histo:WM:horsehead:nT}.

\subsection{Adding the constraint from the \twCO{} \Jone{} peak
  temperature}

Figure~\ref{fig:crb:two:species:plus:co:peak} compares the same
combinations of species (\set{\thCO,\CeiO} or \set{\thCO,\HCOp}) and their
associated available transitions with the additional information coming
from the peak temperature (line temperature at the velocity channel that
maximizes the intensity) of the \twCO{} \Jone{} line.

As expected by the fact that the \twCO{} \Jone{} line is highly optically
thick, this additional constraint enables one to reach an estimation of
\Tkin{} within 10\% for almost all the covered space of physical
conditions. Adding this constraint also has a significant impact on the
$\nHH$ accuracy for the \set{\thCO,\CeiO} set.  The impact remains modest
for the other set, \set{\thCO,\HCOp}.

\subsection{Only using the \Jone{} lines}

\citet{roueff21} showed in the LTE framework that using the two lowest $J$
transition for \thCO{} or \CeiO{} considerably increases the space of
$N(\HH),\nHH$ values for which the CRB reference precision is
reasonable. We find a similar effect when only studying \thCO{} or \CeiO{}
alone or even their combination.  However this result is linked to the
association of species (\thCO{} and \CeiO) having similar collisional and
radiative rate coefficients. We here consider the benefit of using only the
\Jone{} lines or both the \Jone{} and \Jtwo{} lines of \thCO{} and \CeiO{}
when studying the associations with one of the isotopologues of \HCOp{}
that have a much higher dipole moment, and thus different radiative rate
coefficients.

Columns~1 and~2 (resp.~4 and~5) of Fig.~\ref{fig:crb:variations} allow us
to quantify the benefit of using the \Jone{} and \Jtwo{} lines over only
using the \Jone{} line for the \set{\CeiO,\HthCOp}
(resp. \set{\twCO,\thCO,\HCOp}) set. In contrast with the cases using only
the CO isotopologues, the loss of precision remains weak when combining a
ground state transition from a CO isotopologue with the same line from an
\HCOp{} isotopologue. The region of $N(\HH),\nHH$ values for which the CRB
stays relatively constant only slightly decreases. This suggests that
getting several low-$J$ lines of selected species with 1) different
excitation requirements and 2) a sufficiently well understood chemistry,
would be a good surrogate to observations of higher $J$ transitions of the
same species, which lie in atmospheric windows that require more stringent
weather conditions to get good observations.

\subsection{Fixing the relative abundances of the CO and \HCOp{}
  isotopologues}

Up to now, we only studied cases where the relative abundance of the CO and
\HCOp{} isotopologues have been unconstrained. We finally study the impact
of fixing the abundance of either \thCO/\HCOp{} or \CeiO/\HthCOp{} on the
CRB reference precision.  Column \#2 and \#3 (resp. \#5 and \#6) of
Fig.~\ref{fig:crb:variations} allow us to quantify the benefit of fixing
the relative abundances at constant number of transitions for the
\set{\CeiO,\HthCOp} (resp. \set{\twCO,\thCO,\HCOp}) set. The increase of
precision is important for the column densities as expected because
information on the abundances has been added and we thus have two
measurements at different opacities of the same quantity (the column
density). However, the increase of the precision is also obvious for
kinetic temperature, volume density, and thermal pressure, in the case of
the \set{\CeiO,\HthCOp} set. For the \set{\twCO,\thCO,\HCOp} case, only the
volume density and thermal pressure show an increased precision as the
kinetic temperature is mostly constrained by the \twCO{} \Jone{} peak
intensity. However, these gains in precision require an accurate a priori
on the abundance ratio. Otherwise it induces biases as shown in
Sect.~\ref{sec:MC}. Leaving the relative abundance as free as possible is
thus a better solution as long as their knowledges is imprecise.

%%%%%%%%%%%%%
% DISCUSSSION
%%%%%%%%%%%%%

\newcommand{\FigDataHistoWMHorseheadN}{%
  \begin{figure*}
    \centering %
    \includegraphics[width=0.95\linewidth]{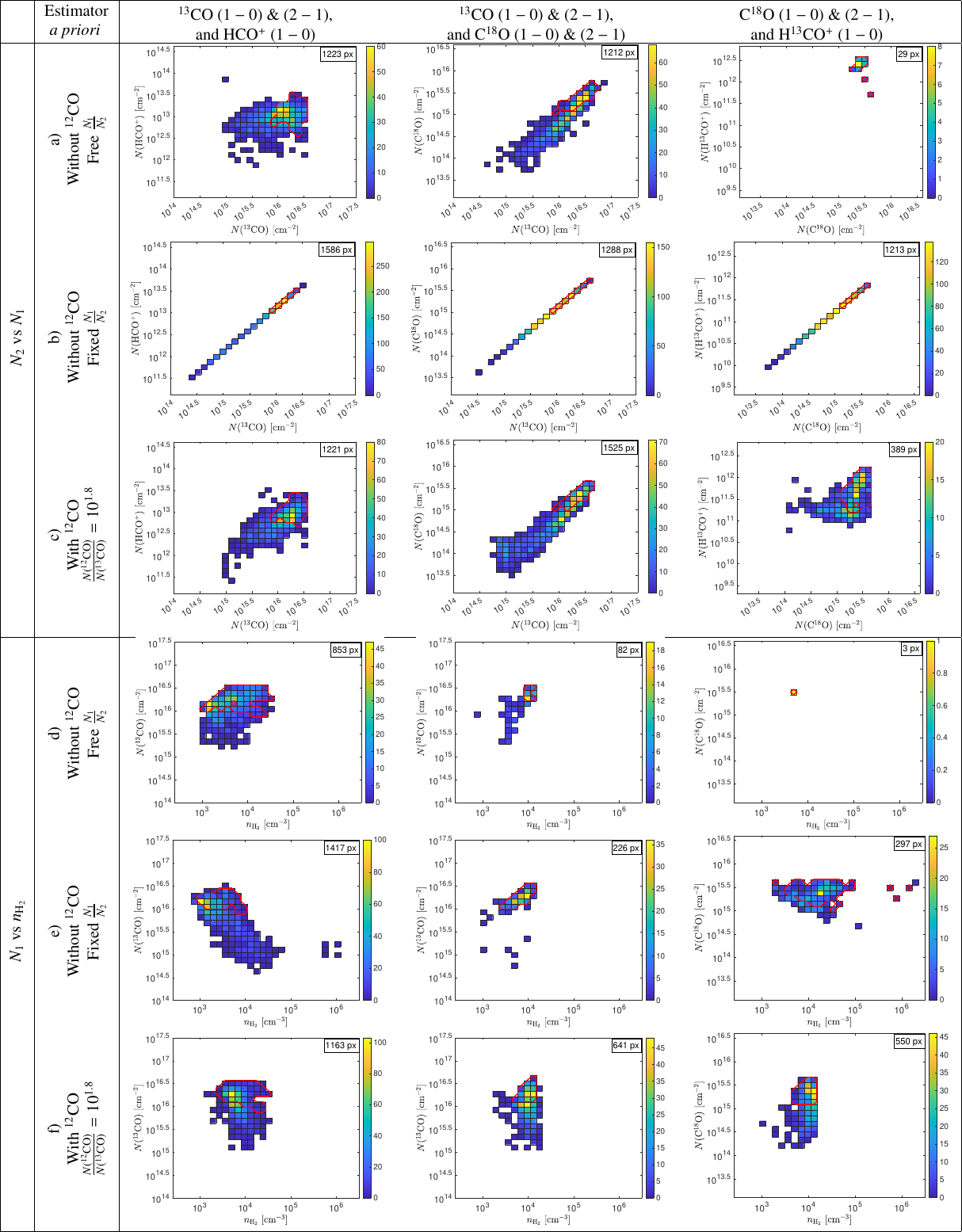}
    \caption{Comparison of the column density and physical conditions
      obtained towards the Horsehead nebula using different sets of
      molecular lines. The red contour delineates the pixels associated
      with dense cores and where \HthCOp{} \Jone{} is detected at the
      3$\sigma$ level.  When abundances or abundance ratios are fixed, the
      values are given in Eq.~\ref{eq_abundance}. Only estimations with a
      good reference precision (CRB) have been kept (see
      Eq.~\ref{eq:bad:CRB}).}
    \label{fig:data:histo:WM:horsehead:N}
  \end{figure*}
}

\newcommand{\FigDataHistoWMHorseheadnT}{%
  \begin{figure*}
    \centering %
    \includegraphics[width=0.95\linewidth]{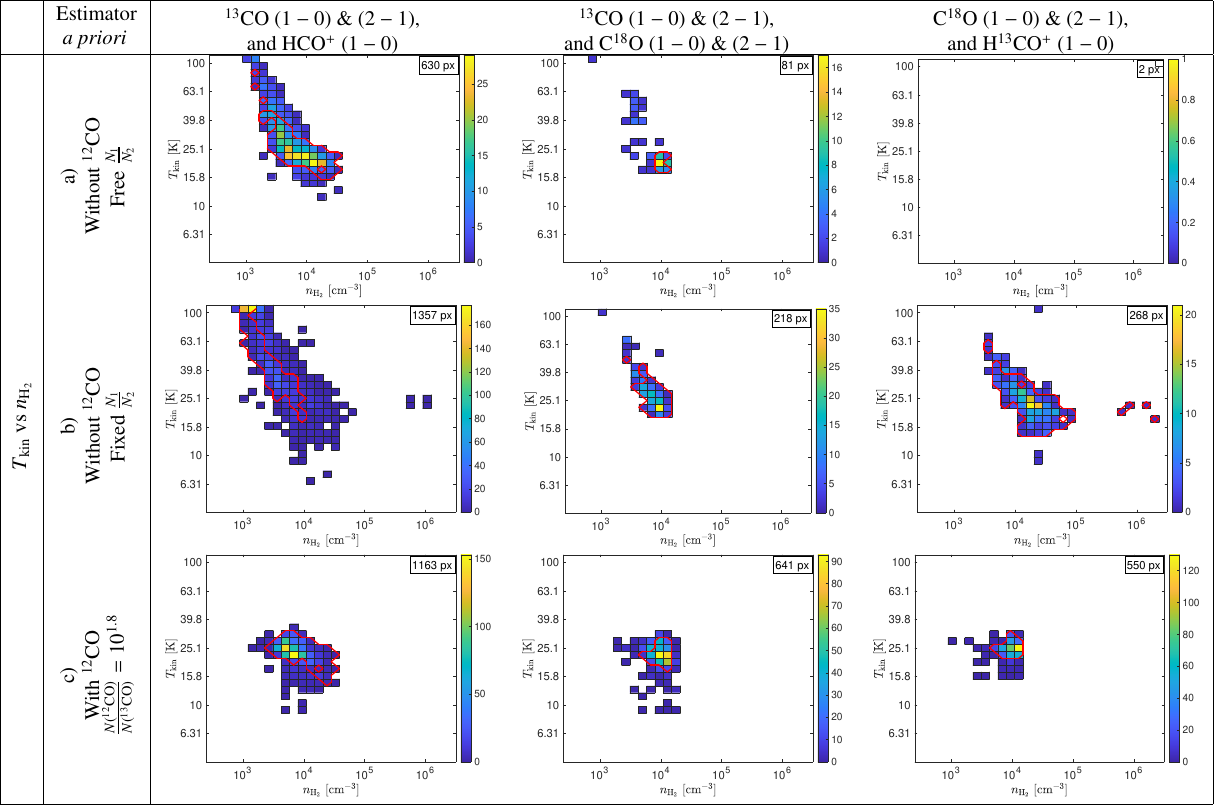}
    \caption{Continuation of Fig.~\ref{fig:data:histo:WM:horsehead:N}
      showing the (\nHH,\Tkin) estimations.}
    \label{fig:data:histo:WM:horsehead:nT}
  \end{figure*}
}

\newcommand{\FigSummary}{%
  \begin{figure*}
    \centering %
    \includegraphics[width=0.5\linewidth,angle=-90]{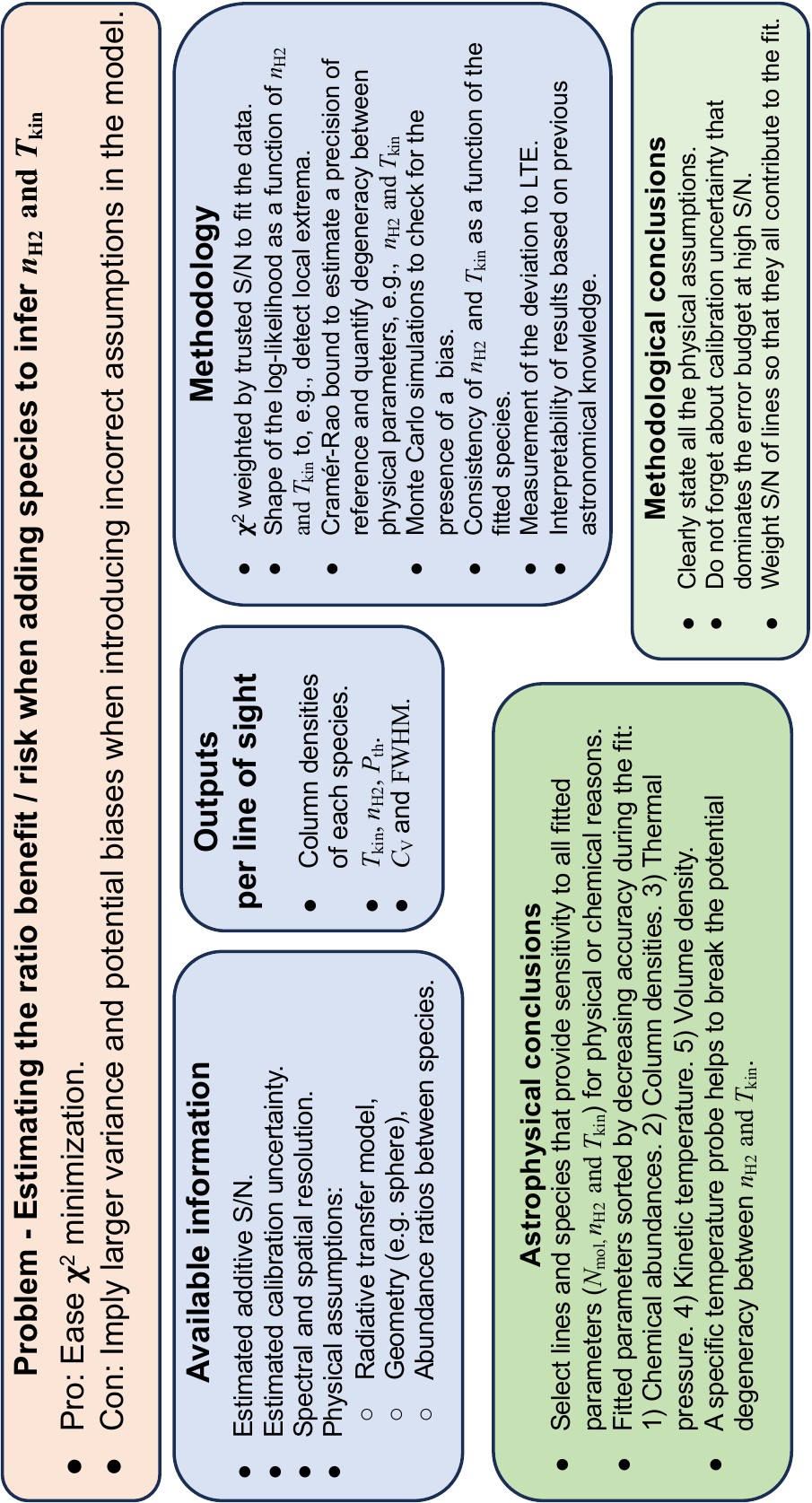}
    \caption{Methodological and astrophysical summary.}
    \label{fig:summary}
  \end{figure*}
}

\section{Astrophysical consequences and perspectives}
\label{sec:discussion}

\FigDataHistoWMHorseheadN{} %
\FigDataHistoWMHorseheadnT{} %

In the previous sections we have analyzed the accuracy that can be achieved
on the column density and physical conditions, depending on the choice of
considered spectral lines, the S/N, and domain of parameters.
Figures~\ref{fig:data:histo:WM:horsehead:N}
and~\ref{fig:data:histo:WM:horsehead:nT} summarize the fits obtained on
molecular maps of the Horsehead nebula, in the framework of a single
emitting layer with uniform physical conditions and fixed or free relative
abundances. Three sets of molecular lines are used: \thCO{} and \HCOp{},
\thCO{} and \CeiO{}, and \CeiO{} and \HthCOp{}. The first one is adapted to
more diffuse lines of sight, while the third one is more suited to denser
ones. Each set of estimations has been performed with or without the
additional constraint from the peak \twCO{} intensity. The used abundance
ratio values, when fixed, are the ones listed in Eq.~\ref{eq_abundance}.

\subsection{Column density}

The temperature information brought by the \twCO{} \Jone{} peak intensity
is essential to keep the kinetic temperature in an acceptable range, below
$100\K$. However this constraint may bias the results on column densities
in the presence of a temperature gradient along the LoS. As the temperature
probed by the \twCO{} \Jone{} line corresponds to the outer layers where
the line is emitted, it may be invalid for the deeper and more shielded
layers.

A comparison of the middle column of rows (a) and (c) in
Fig.~\ref{fig:data:histo:WM:horsehead:N} shows that the \thCO{} and \CeiO{}
column densities derived with or without using the \twCO{} peak temperature
agree well with each other. This suggests that this bias is acceptable for
the CO isotopologues. The flattening of the N(\CeiO{}) vs N(\thCO{})
histogram for low values may reflect the effect of noise in the relatively
faint \CeiO{} lines since a positive fluctuation of the noise may lead to
an apparently stronger line given the relatively low line intensity in
these regions, $\le 0.5\Kkms$.

The situation is more complex for the high dipole moment species \HCOp{}
and \HthCOp{}. For both species, the scatter is quite large reaching about
0.5 dex or a factor of three. The flattening of the relation at low column
densities in the case of N(\HCOp{}) vs N(\CeiO{}), and the presence of high
\HCOp{} column density pixels for moderate values of N(\thCO{}) in the case
of N(\HCOp{}) vs N(\thCO{}), probably indicate a bias induced by our
hypotheses in addition to the effect of noise.  For instance, in regions
with relatively faint \thCO{} emission, the hypothesis of optically thick
\twCO{} \Jone{} emission may not be fully valid leading to a bias in the
assumed kinetic temperature. When the \thCO{} \Jone \, line is not fully
optically thick, its peak temperature is expected to be lower than the
kinetic temperature, which implies that the kinetic temperature could be
biased towards somewhat lower values.

An independent result is that the \HthCOp{} column densities derived
without fixing the abundance ratio are somewhat higher (by a factor of
about two) than when fixing it. The scatter thus probably indicates real
variation of the relative abundances of these two species.

\subsection{Kinetic Temperature}

The effect of the assumptions used to derive the kinetic temperature is
clearly seen in rows b and c of
Fig.~\ref{fig:data:histo:WM:horsehead:nT}. Without the information from
\twCO{}($1-0)$, the fitted physical conditions are tightly correlated. This
reflects the shape of the log-likelihood function shown in
Fig.~\ref{fig:mc:co:iso} to~\ref{fig:mc:c18o:h13cop}. Indeed the strong
degeneracy between a low density/high temperature and high density/low
temperature models implies that the derived parameters explore the valley
of low negative log-likelihood in the (\nHH{},\Tkin{}) space, depending on
the line intensities, line profiles, and noise level of each pixel.

The situation is slightly better when combining the \Jone{} and \Jtwo{}
lines of both \thCO{} and \CeiO{} because the lower opacity of the \CeiO{}
lines as compared to \thCO{} implies a higher sensitivity of the level
population to the \HH{} density up to the \CeiO{} \Jtwo{} critical density
of $\sim 10^4\pccm$.

The combination of \CeiO{} and \HthCOp{} data also leads to an exploration
of the shape of the log-likelihood function by the different pixels. The
results towards dense core pixels reach $10^6\pccm$, a somewhat
unrealistically high value at the spatial scale probed by the studied
observations, $0.05 - 0.1\pc$. Since these pixels have high \CeiO{} column
densities and a good CRB, this incorrect result suggests a model mismatch,
or an impossibility to separate the different minima of the negative
log-likelihood function.

When the kinetic temperature is constrained by the \twCO{} peak
temperature, the range of derived temperatures and densities becomes
smaller and more consistent between the different ensembles of molecular
lines. However, the small scatter of the kinetic temperature map may be
directly related to the hypotheses (single layer, use of \twCO{}) and may
not reflect the real physical conditions.

\subsection{Volume density}

The bias and variance analyses presented in Sect.~\ref{sec:MC}
and~\ref{sec:CRB} have shown that it is difficult to obtain a good
precision on molecular gas density because 1) of degeneracies between the
volume density, the kinetic temperature, and the column densities, or 2) of
the potential presence of more than one local minimum in the log-likelihood
function. These degeneracies are clearly seen in the plots displaying the
variation of the derived column densities as a function of the density when
\twCO{} \Jone{} is not used (see Fig.~\ref{fig:data:histo:WM:horsehead:N}.d
and e). The anti-correlation between the density and column density when
using the \thCO{} and \HCOp{} lines seems nonphysical, as high column
density regions are usually associated with high density regions.  The
lower density pixels appear to have a high kinetic temperature, a
consequence of the degeneracy between the density and kinetic
temperature. The same behavior is seen for the two other sets of lines,
with densities shifted towards somewhat higher values as these sets of
lines include transitions with higher effective critical densities than the
first set of lines.

Adding a constraint on the kinetic temperature from \twCO{} breaks the
degeneracy. The derived densities occupy a rather narrow range of values,
between $\sim 3 \times 10^3\pccm$ and $\sim 3 \times 10^4\pccm$ for all
sets of lines. There is no difference in the volume density between the
pixels associated with the dense cores (defined as regions with strong
\HthCOp{} emission) and pixels elsewhere in the map, while it would have
been expected that the density increases in the dense cores.

When looking at the volume density maps displayed in
Figure~\ref{fig:data:n} more differences appear. Using the \twCO{} peak
temperature is essential for the first set of lines as the kinetic
temperatures (and densities) become too high (resp. low) without this
constraint. The combination of \thCO{} and \CeiO{} lines performs quite
well without \twCO{} in the high column density region of the horsehead
neck, but many pixels remain with nonphysical estimations. For these two
sets of lines, using \twCO{} without adding a constraint on the relative
abundances appears to be the best choice. Because the \HthCOp{}\Jone{}
emission is only detected in restricted areas, the different hypotheses
have a strong impact on the derived parameters as some hypotheses allow for
the use of pixels with marginal detections while others do not. Fixing the
relative abundances between \CeiO{} and \HthCOp{} allows us to obtain
reasonable values for the density and kinetic temperature in the \HthCOp{}
cores with a good CRB. This avoids the risk of biasing the temperature by
using a temperature probe associated with the cloud envelope. However, the
analysis is limited to the small set of pixels with a good S/N detection of
\HthCOp{}\Jone{}.

\subsection{Pressure}

Although the estimations of density and temperature may be biased, the
estimation of the thermal pressure is expected to be more accurate as the
minimum of the negative log-likelihood function is nearly coincident with
the locus of constant thermal pressure.  Nevertheless, a detailed analysis
of the accuracy of the pressure determination remains necessary, as the
different degeneracies described above still play a role.  In addition, the
determination of the thermal pressure results from a fit of the molecular
line emission under some hypotheses (e.g., uniform medium, co-localized
molecular species, etc).  Therefore, it is recommended to clearly describe
the assumptions leading to the estimation of the parameters of interest as
different sets of hypotheses may lead to different values. Overall, the CO
isotopologue lines can be used to obtain rather good estimations of the
thermal pressure, an interesting quantity for the energy budget of
molecular gas when all hypotheses are clearly stated as recommended above.

\subsection{Generalization}

\FigSummary{} %

From the analysis of the Horsehead nebula, we can draw some recommendations
for the analysis of the selected molecular emission lines: low $J$ lines of
the CO and \HCOp{} isotopologues using radiative transfer models of a
uniform cloud. Figure~\ref{fig:summary} summarizes these
recommendations. These lines are among the strongest seen in Galactic
molecular clouds and nearby galaxies. It is therefore interesting to
optimize their analysis and extract most of information from such
observations.

In many conditions, even the excitation of the CO isotopologues departs
from the local thermodynamic equilibrium (see
Fig.~\ref{fig:crb:line:characteristics}), and a non-LTE model like RADEX
should be preferred for the analysis of the molecular line
emission. However, it is important that the hypotheses involved in the
non-LTE calculations are clearly spelled out.

The determination of column densities is accurate for \thCO{} and \CeiO{}.
It does not need additional hypotheses than the presence of all species in
the same volume.  Column densities for the higher dipole moment species
\HCOp{} and \HthCOp{} are more sensitive to the set of hypotheses as these
species are sub-thermally excited in most of the explored parameter space,
hence their emission is sensitive to the physical conditions.  The lower
S/N of the faint emission lines also implies a lower accuracy of the
determined parameters.

The determination of the molecular hydrogen volume density and kinetic
temperature is more difficult to achieve with a good precision.  As
discussed in Sect.~\ref{sec:mc:experiment:2}, because of the strong
degeneracy between the density and kinetic temperature in determining the
CO isotopologue excitation, an independent constraint on the kinetic
temperature is needed. Using the peak value of the \twCO{} \Jone{} line
works well for that purposes and allows us to at least partially break the
degeneracy. This constraint is useful for the determination of the kinetic
temperature but ambiguities may remain for the molecular hydrogen
density. Given the achievable S/N and calibration accuracy, different
physical conditions may lead to very similar emerging spectra as
illustrated in Fig.~\ref{fig:crb:spectra} and discussed in
Sect.~\ref{sec:crb:simple-case}. Adding a line from a high dipole moment
species is useful for constraining the volume density, provided the line is
observed with a good S/N and this species abundance relative to that of one CO
isotopologue can be used to constrain the fit. Otherwise, adding another
line is less beneficial because of the degeneracy between density and
column density. An error in fixing the relative abundance of the high
dipole moment species relative to the CO isotopologue leads to a bias in
the derivation of the density but this bias remains tolerable when compared
to the uncertainties when only one CO isotopologue is used.

Using different molecular lines from different species brings more
constraints for deriving the physical conditions and column densities, at
the expense of the need for a strong hypothesis: the presence of all
species within the same volume of gas with uniform conditions. Such a
strong hypothesis may be far from the reality of interstellar clouds which
exhibit gradients of density, temperature and molecular abundances across
their volumes. In particular, towards lines of sight where dense cores are
present, the range of physical conditions is broad and the uniform cloud
hypothesis is expected to lead to poor results. One example of such poor
results is the determination of the kinetic temperature using the \twCO{}
\Jone{} line which does not differentiate pixels within dense cores from
pixels outside cores. In fact, it is known from more detailed analyses that
the kinetic temperature decreases from the extended molecular cloud
envelopes to cores~\citep[e.g.,][]{Hocuk:2017,Rodriguez:2021}. Instead of
uniform models, more sophisticated geometrical layouts are needed for
specific environments. For instance, dense cores that are embedded in
filaments, and in the molecular gas envelope clearly need a multi-layer
model, in which a cold and dense region is surrounded by a warmer and more
diffuse envelope. This will be the subject of another paper. Another case
which would deserve a multi-layer model could be a dense, non edge-on, UV
illuminated photodissociation region (PDR) where a strong temperature
gradient is present due to the attenuation of the UV radiation when
entering the molecular gas. Such PDRs often have a nearly isobaric equation
of state \citep[e.g., ][for the Horsehead nebula]{Hernandez:2023}, hence an
isobaric model would be better suited than a uniform model with a single
density and temperature value.

For all cases, having a temperature probe independent of the lines used for
the column density and volume density estimations has been shown to be
useful for breaking the degeneracies in the excitation conditions. The peak
temperature of the \twCO{} \Jone{} line is adequate for the bulk of the
molecular emission, but overestimates the temperature of the dense and
shielded regions. In these regions, the ratio of the $(1,1)$ and $(2,2)$
inversion lines of NH$_3$ near 23\GHz{} is known to perform well as a
temperature probe \citep{Maret2009}, but getting such data requires
combining observations from different telescopes. Efforts should be made to
find another, easily accessible, temperature probe based on molecular
lines, that could be used in the analysis of well shielded
regions. \citet{Hacar2020} proposed to use the ratio of the HCN and HNC
intensities to probe the kinetic temperature between about 15 and
60\K. However in their extensive analysis of the Orion B cloud,
\citet{santa-maria2023} showed that the HCN/HNC ratio is more sensitive to
the radiation field than to the kinetic temperature.  Another possibility
to probe the kinetic temperature of shielded regions could be using the
dust temperature, but this also requires observations over a wide range of
wavelengths to break the degeneracy between the dust temperature and column
density.  \citet{Rodriguez:2021} have used radiative transfer calculations
and showed that the difference between the kinetic temperature and the dust
temperature remains lower than 1\K{} along lines of sight with an
extinction larger than 8 magnitudes (4\,mag on each side from the cloud
edge) for the conditions corresponding to the sources probed in the GEMS
IRAM large program, dense cores with an external FUV radiation field
between 5 and 60 times the Inter-Stellar Radiation Field and a thermal
pressure of $5\times 10^5\Kpccm$).  Further studies over a wider parameter
space and including the comparison of the predicted dust temperature and
its estimation from the fit of the spectral energy distribution are needed
to establish the conditions where the dust temperature is a good proxy of
the kinetic temperature, and thus the bias which could be introduced by
this hypothesis.

%%%%%%%%%%%%%
% CONCLUSION
%%%%%%%%%%%%

\section{Conclusion}

In this article, we study the fit of multi-species molecular lines with a
non-LTE radiative transfer model and we propose a method to test the
derived results, including the possibility of identifying potential model
misspecifications.  Figure~\ref{fig:summary} summarizes our main
recommendations and their implications for further studies.  We list  a
few general points below.
\begin{itemize}
\item While LTE models are simple to run and interpret, the excitation of
  CO isotopologues does not always reach this limit and non-LTE models
  should be preferred, especially when the analysis of CO isotopologue
  lines is combined with lines from higher dipole moment species.
\item It is important to take into account the S/N of the transition
  studied, including the thermal noise and calibration uncertainties, when
  fitting radiative transfer models to observed molecular emission. This
  can be done for instance by limiting the S/N ratio of the strongest lines
  to mimic the effect of calibration uncertainties.
\item When fitting radiative transfer models to observations, it is
  important to consider the information from the full line profile and not
  just from the integrated intensity, because this allows us to consider
  the broadening of the line profile when the opacity increases.
\item Because different physical conditions may produce very similar line
  profiles for some transitions, it is important to account for such
  potential degeneracies when exploring the parameter space.
\item Combining molecular species with different excitation requirements,
  for instance, high dipole (high critical density) and low dipole moment species
  (low critical density), allows for the degeneracy in the physical
  condition parameter space to be broken, provided the relative abundances of the
  considered species are known from independent observations or from
  models.
\item The escape probability model is adequate for studying large scale
  molecular line maps. For detailed studies of specific objects of known
  geometry, more sophisticated models may be required to better take into
  account the radiative coupling along the line of sight.
\end{itemize}

Based on the precision analysis, we make the following additional
recommendations.
\begin{itemize}
\item To increase the precision of the estimations, it is more interesting
  to combine lines with different excitation requirements than to increase
  the S/N ratio of a targeted spectral line. This is because the needed integration
  time to detect a new line if often significantly lower than the
  additional time needed to improve the S/N ratio on an already detected
  line.
\item With the availability of broad band receivers, combining lines from
  different species accessible with the same receiver usually allows for a more
  efficient use of the observing time than accessing lines from different
  energy levels, which often require stringent weather conditions to get
  good observations.
\item The thermal pressure of the emitting medium is often determined with
  a better precision than the kinetic temperature of volume density taken
  separately. It can therefore be interesting to derive thermal pressure
  maps of molecular clouds, which can be compared with maps of the magnetic
  or turbulent pressure.
\end{itemize}

This work is focused on the combination of CO and \HCOp{} isotopologue
lines. Other high-dipole moment species could be considered in the future. For instance, CS, CN, HNC or N$_2$H$^+$ have bright lines that are easily
detectable in large scale maps of molecular clouds. While this study is also
focused on a single homogeneous layer, it is clear that this hypothesis is
not fully suited to lines of sight towards dense cores.  In a following
paper, we will study a multi-layer model and compare its performances with
the homogeneous model.

\begin{acknowledgements}
  % Observations
  This work is based on observations carried out under project numbers
  019-13, 022-14, 145-14, 122-15, 018-16, and finally the large program
  number 124-16 with the IRAM 30m telescope. IRAM is supported by INSU/CNRS
  (France), MPG (Germany) and IGN (Spain).
  % ANR DAOISM and PCMI
  This work was supported by the French Agence Nationale de la Recherche
  through the DAOISM grant ANR-21-CE31-0010, and by the Programme National
  ``Physique et Chimie du Milieu Interstellaire'' (PCMI) of CNRS/INSU with
  INC/INP, co-funded by CEA and CNES.
  % PEPS
  This project has received financial support from the CNRS through the
  MITI interdisciplinary programs.
  % Javier
  JRG and MGSM thank the Spanish MCINN for funding support under grant
  PID2019-106110G-100.
  % Darek
  Part of the research was carried out at the Jet Propulsion Laboratory,
  California Institute of Technology, under a contract with the National
  Aeronautics and Space Administration (80NM0018D0004). D.C.L. was
  supported by USRA through a grant for SOFIA Program 09-0015.
  % EMAA and LAMDA
  This research has made use of spectroscopic and collisional data from the
  EMAA (\url{https://emaa.osug.fr} and
  \url{https://dx.doi.org/10.17178/EMAA}) and LAMDA
  (\url{https://home.strw.leidenuniv.nl/~moldata/}) databases.  EMAA is
  supported by the Observatoire des Sciences de l’Univers de Grenoble
  (OSUG).  The LAMDA database is supported by the Netherlands Organization
  for Scientific Research (NWO), the Netherlands Research School for
  Astronomy (NOVA), and the Swedish Research Council.
\end{acknowledgements}

\bibliographystyle{aa} %
\bibliography{ms} %

\begin{appendix}
  
%%%%%%%%%% 
% FISHER
%%%%%%%%%%

\section{Fisher information matrix} 
\label{sec_Fisher_calc}

To compute Cramér-Rao bound associated to the unknown parameters $\btheta$,
we need to calculate the Fisher matrix of the $L$ observed lines
$\bx_{1:L}=\{\bx_1...,\bx_L\}$.  Nevertheless, since the noise on the
different lines are statistically independent, the Fisher matrix of
$\bx_{1:L}$ is simply the sum of the Fisher matrix of each line
\eq{
  \bI_F(\btheta;
  \bx_{1:L})
  =\sum_{l=1}^L \bI_F(\btheta,
  \bx_l).
}
Below we detail the calculation of the Fisher matrix for a single line
and the vector of unknown parameter is simply
$\btheta=[\log \Tkin,\log n_{H_2},\log N,\sigma_V,C_V]$. The
generalization of these computations to the case where we also fit the
logarithm of some of the column densities; for instance,
$\btheta=[\log \Tkin,\log n_{H_2},\log N(\thCO) ,\log
N(\CeiO),\sigma_V,C_V]$, is straightforward. For didactic reasons, the
presentation is organized in two parts.  In
Sect.~\ref{sec_Fisher_nocal}, the multiplicative noise is omitted. It
is taken into account in Sect.~\ref{sec_Fisher_cal}.

\subsection{Case without multiplicative noise}
\label{sec_Fisher_nocal}

In this section, we assume only additive noise:
\eq{
  x_n=s_n+b_n.
  \label{eq_x_n} 
}
where $b_n$ is a white Gaussian noise with variance $\sigma_{b,l}^2$
and $s_n$ is a sampled version of Eq.~\ref{eq2_s}:
\eq{
s_n=
\left\{ J(\Texl,\nu_l)-J(\TCMB,\nu_l)\right\}
\left[1-\exp(-\Psi_n)\right].
\label{eq_s_n}
}
Thus, $\Psi_n$ is a sampled version of Eq.~\ref{eq_Psi}:
\eq{
\Psi_n = 
  \tau_l\,\exp\left(-\frac{(V_n-C_V)^2}{2\sigma_V^2}\right),
  \label{eq_Psi_n}
}
where $V_n$ is the velocity at channel $n$.
Assuming that the noise is white, the log-likelihood of the sampled
observation at line $l$ is
\eq{
  \calL(\btheta;\bx_l)
  =cte-\sum_{n}
\frac{\left(x_n-s_n\right)^2}{2\sigma_{b,l}^2},
\label{eq_calL}
}
and the Fisher matrix $\bI_F$ \citep[see][]{sto05} is
\eq{
  \forall (i,j) \quad
  \bI_F(\btheta, \bx_l)=
\frac{1}{\sigma_{b,l}^2}\sum_{n}
 \frac{\partial s_{n}}{\partial \theta_i}\frac{\partial
   s_{n}}{\partial \theta_j}.
 \label{eq_Fisher}
}
Thus, to get $\bI_F(\btheta, \bx_l)$, we simply need to evaluate the
gradient $\frac{\partial s_n}{\partial \theta_i}$, where $\theta_i$ is the
$i^{th}$ component of $\btheta$. This is done in the following.

\subsubsection{Calculation of $\frac{\partial s_n}{\partial \log\Tkin}$}

We note that
\eq{
  \frac{\partial s_n}{\partial \log\Tkin}
  =\frac{\partial s_n}{\partial \Tkin}\ln(10)\Tkin.
}
Moreover, based on Eq.~\ref{eq_s_n}, we have\ 
\eq{
\begin{array}{ll}
\frac{\partial s_n}{\partial \Tkin}=
&
\frac{\partial J(\Texl,\nu_l)}{\partial \Texl}
\frac{\partial \Texl}{\partial \Tkin}
\left[1-\exp(-\Psi_n)\right]
\\&
+
\left\{ J(\Texl,\nu_l)-J(\TCMB,\nu_l)\right\}
\frac{\partial \Psi_n}{\partial \Tkin}
\exp(-\Psi_n).
\end{array}
\label{eq_gradient_s}
}
In practice, we numerically estimated $\frac{\partial \Texl}{\partial
  \Tkin}$ with a finite difference technique using RADEX with a step
of $\Tkin/1024$.

\subsubsection{Calculation of $\frac{\partial J(T,\nu)}{\partial T}$}

\eq{
J(T,\nu) = \frac{h \nu}{k}
\frac{1}{\exp{\frac{h \nu}{k T}} - 1}.
}
Thus
\eq{
\frac{\partial J(T,\nu)}{\partial T}=
\frac{h \nu}{k}
\frac{
\frac{h \nu}{k T^2}
\exp{\frac{h \nu}{k T}}
}{\left(\exp{\frac{h \nu}{k T}} - 1\right)^2}
=
\frac{h^2 \nu^2}{k^2 T^2}
\frac{\exp{\frac{h \nu}{k T}}}
{\left(\exp{\frac{h \nu}{k T}} - 1\right)^2},
}
and finally
\eq{
\frac{\partial J(T,\nu)}{\partial T}=
\frac{h^2 \nu^2}{k^2 T^2}
\frac{1}
{\exp{\frac{h \nu}{k T}} -2 +\exp{-\frac{h \nu}{k T}}}.
}

\subsubsection{Calculation of $\frac{\partial \Psi_n}{\partial \Tkin}$}

Based on Eq.~\ref{eq_Psi_n},
\eq{
\frac{\partial \Psi_n}{\partial \Tkin}=
\frac{\partial \tau_l}{\partial \Tkin}
\exp\left(-\frac{(V_n-C_V)^2}{2\sigma_V^2}\right),
}
where the term $\frac{\partial \tau_l}{\partial \Tkin}$ is also numerically 
estimated with finite difference as in Eq.~\ref{eq_gradient_s}.

\subsubsection{Calculation of $\frac{x_n}{\partial
    \log n_{H2}}$ and $\frac{x_n}{\partial
    \log N}$}

The calculation of $\frac{\partial s_n}{\partial \log n_{H_2}}$ and
$\frac{\partial s_n}{\partial \log N}$ are straightforward adaptations
from $\frac{x_n}{\partial\log N}$. We only need to specify that the
finite difference technique is applied for both $\log n_{H_2}$ and $\log N$
with a step of 0.001.

\subsubsection{Calculation of $\frac{\partial s_n}{\partial \sigma_V}$}

To get
\eq{
\begin{array}{ll}
\frac{\partial s_n}{\partial \sigma_V}=
&
\frac{\partial J(\Texl,\nu_l)}{\partial \Texl}
\frac{\partial \Texl}{\partial \sigma_V}
\left[1-\exp(-\Psi_n)\right]
\\&
+
\left\{ J(\Texl,\nu_l)-J(\TCMB,\nu_l)\right\}
\frac{\partial \Psi_n}{\partial \sigma_V}
\exp(-\Psi_n),
\end{array}
}
we need to detail the calculation of $\frac{\partial
  \Psi_n}{\partial \sigma_V}$.

\subsubsection{$\frac{\partial \Psi_n}{\partial \sigma_V}$}

Based on Eq.~\ref{eq_Psi_n}, one has
\eq{
  \begin{array}{ll}
    \frac{\partial \Psi_n}{\partial \sigma_V}=&
\frac{\partial \tau_l}{\partial \sigma_V}
\exp\left(-\frac{(V_n-C_V)^2}{2\sigma_V^2}\right)
    \\&
    +\tau_l
\frac{(V_n-C_V)^2}{\sigma_V^3} 
\exp\left(-\frac{(V_n-C_V)^2}{2\sigma_V^2}\right),
\end{array}
}
where the term $\frac{\partial \tau_l}{\partial \sigma_V}$ is numerically 
estimated with a finite difference technique with a step of
$\FWHM/128$.

\subsubsection{Gradient calculation along $\Delta_{V}$}

\eq{
\frac{\partial s_n}{\partial \Delta_{V}}
=
\left(J(\Texl, \nu_l)-J(\TCMB,\nu_l)\right)
\frac{\partial \Psi_n}{\partial \Delta_{V}}
\exp(-\Psi_n).
}

\subsubsection{$\partial \Psi_n /\partial \Delta_{V} $}

Based on Eq.~\ref{eq_Psi_n}, one has
\eq{
\frac{\partial \Psi_n}{\partial \Delta_{V}}=
\tau_l
\frac{V_n-C_V}{\sigma_V^2}
\exp\left(-\frac{(V_n-C_V)^2}{2\sigma_V^2}\right).
}

\subsection{Case with multiplicative noise}
\label{sec_Fisher_cal}

Let us now take into account the multiplicative noise
\eq{
  \forall n \quad x_n=c.s_n+b_n,
  \label{eq_x_n_cal} 
}
where in addition of $b_n$, one has $c\sim\calN(1,\sigma_{c,l}^2)$,
which is assumed independent $b_n$. The column vector
$\bx=[x_1,\,x_2,\,...]^T$ is the sum of two Gaussian random vectors,
thus it remains a Gaussian vector. Its mean is
$\bs=[s_1,\,s_2,\,...]^T$ and it is straightforward to show that is
covariance matrix is 
\eq{
  \bSigma=\sigma_{b,l}^2\,\bI_d+\sigma_{c,l}^2\,\bs\bs^T,
}
where $\bI_d$ is the identity matrix. The log-likelihood of $\bx$ is
\eq{
  \calL(\btheta;\bx_l)=cte-\frac{1}{2}\log \left|\bSigma\right|
  -\frac{1}{2}(\bx-\bs)^T \bSigma^{-1}(\bx-\bs),
  \label{eq_calL_C}
}
which leads, after tedious but straightforward calculations, to
\eq{
  \calL(\btheta;\bx_l)=cte-\frac{1}{2}\left(\log \sigma_l^2
    +\frac{\bx^T\bx}{\sigma_{b,l}^2}
    +\frac{\bs^T\bs}{\sigma_l^2}
    -\frac{(\bx^T\bs)^2\sigma_{c,l}^2}{\sigma_b^2\sigma_l^2}
    -\frac{2\bx^T\bs}{\sigma_l^2}
  \right),
}
where $\sigma_l^2=\sigma_{b,l}^2+\sigma_{c,l}^2(\bs^T\bs)$.
The Slepian-Bang formula \cite{sto05} gives us the following relations,\eq{
  \left[\bI_F\right]_{ij}=
  \frac{\partial \bs^T}{\partial\theta_i}\bSigma^{-1}
  \frac{\partial \bs}{\partial\theta_j}
  +\frac{1}{2}\tr\left(
    \bSigma^{-1}
    \frac{\partial \bSigma}{\partial\theta_i}
    \bSigma^{-1}
    \frac{\partial \bSigma}{\partial\theta_j}
    \right),
}
which leads, after tedious but straightforward calculations to
\eq{
    \begin{array}{lll}
      \left[\bI_F\right]_{ij}&=&
                              \left(\frac{\sigma_l^2+\sigma_{c,l}^4(\bs^T\bs)}{\sigma_{b,l}^2\sigma_l^2}\right)
                              \left(\frac{\partial \bs^T}{\partial\theta_i}\frac{\partial
                              \bs}{\partial\theta_j}\right)
      \\
                           &+&
                               \left(
                               \frac{\sigma_{b,l}^2\sigma_{c,l}^4-\sigma_{c,l}^6(\bs^T\bs)}{\sigma_{b,l}^2\sigma_l^4}
                               -\frac{\sigma_{c,l}^2}{\sigma_{b,l}^2\sigma_l^2}
                               \right)
                               \left(\bs^T\frac{\partial\bs}{\partial\theta_i}\right)
                               \left(\bs^T\frac{\partial\bs}{\partial\theta_j}\right).
    \end{array}
}

%%%%%%
% MLE
%%%%%%

\newcommand{\TabSaturation}{%
  \begin{table}
    \caption{Minimum and maximum parameter bounds, where estimations are
      saturated to prevent RADEX from crashing}
    \begin{tabular}{cccc}
      \hline
      $\Tkin~[\unit{K}]$ &
      $\log(\nHH/\unit{cm}^3)$ &
      $\log (N(\thCO)/\unit{cm}^2)$ &
      $\sigma_V~[\unit{km/s}]$ \\
      \hline
      $[4,\, 113]$ & 
      $[2,\, 6.5]$ & 
      $[13,\,18.6]$ &
      $[0.25,\,2]$ \\ 
      \hline
    \end{tabular}
    \label{tab:min_max}
  \end{table}
}

\section{Details on the used maximum likelihood estimator}
\label{sec:mle}

The following subsections provide details on the implementation of the
maximum likelihood estimator we use in this article.  We first explain
how we localize and detect the Gaussian opacity profile
(Eq.~\ref{eq_Psi}) on high S/N and optically thin lines. This allows
us to provide an initial estimation of the centroid velocity. We then
discuss 1) the initialization of the other parameters and 2) the
iterative algorithm to refine their estimation.

\subsection{Signal detection and initial estimation of the centroid
  velocity}
\label{sec:mle:detection}

We start by computing the noise standard deviation on baseline channels by
using the range of channels that are devoid of signal on the studied field
of view. These channels can be selected manually from raw data (see the
right part of Figure~\ref{fig:data:peak:area:spectra}).  We then estimate
the centroid velocity $C_V$. To do this, we first increase the S/N by
adding all the lines from \CeiO{} and \thCO{} after resampling them on the
same velocity grid. Although the \twCO{} \Jone{} line is bright, its
high opacity makes it mostly sensitive to the kinematics of surface of the
cloud instead of the bulk of the flow. On the opposite, the S/N on $\HCOp$
and $\HthCOp$ is significantly lower, which make them useless to estimate
centroid velocity.  We then apply a matched filter by convolving the summed
spectra with a Gaussian shape of $\emr{FWHM} = 0.35\unit{km/s}$. This shape
corresponds to an optically thin approximation of a line. Selecting the
velocity where the intensity of the filtered signal is maximum thus
delivers an initial estimation of the centroid velocity $C_V$ for all the
lines.

Afterwards, we additionally check the presence of the velocity component
in a majority of the studied lines. More precisely, the pixel is further
processed only when more than half of the lines are detected. Else, the
estimated parameters are nullified.  To do this, we integrate the profile
within a velocity interval of $1\kms$ centered on the estimated centroid
velocity, $C_V$.  The line is considered detected when the signal is larger
than $3\sqrt{N}\,\sigma_b$, where $N$ is the number of velocity channels in
$1\kms$, and $\sigma_b$ the associated noise standard deviation.

\subsection{Initializing the fit estimation}
\label{sec:mle:initialization}

\TabSaturation{} %

The signal detection step provides an initial estimate of the centroid
velocity for all the lines. In order to initialize the other
parameters, we compute the negative log-likelihood (NLL) on a fixed
grid of models computed with RADEX. This grid sample the parameter
space in logarithmic steps for all parameters whose range of plausible
values cover at least one order of magnitude. In practice, only the
velocity dispersion is sampled linearly. In details,
\begin{itemize}
\item The velocity dispersion, $\FWHM$, is sampled linearly between $0.25$
  and $2\kms$ in steps of $0.05\kms$.
\item The kinetic temperature, $\Tkin$, is sampled between $4$ and $113\K$
  in logarithmic steps of $0.05$.
\item The \HH{} volume density, $n(\HH)$ is sampled between $10^{2}$ and
  $10^{6.5}\pccm$ in logarithmic steps of $0.1$.
\item $N(\thCO)$ is sampled between $10^{13}$ and $10^{18}\pscm$ in
  logarithmic steps of $0.05$.
\item The column densities of \twCO, \CeiO, \HCOp, and \HthCOp{} are
  sampled with the same grid as \thCO, shifted to take into account the
  relative abundances defined in Eq.~\ref{eq_abundance}.
\end{itemize}
Using this grid, for each species, we thus get a 4D tensor of negative
log-likelihood (NLL) values. The way we explore this tensor, depends on the
a priori information that are available. For instance, let us
consider the $\set{\thCO{},\HCOp{}}$ estimation problem (see
Table~\ref{tab_set_of_species}). We need to estimate the column density of
these two species independently with the constraint that the triplet
$(\Tkin,\,\nhd,\,\sigma_V)$ is identical for all species. On the other
hand, with $\set{\thCO{},\HCOp{};\calA}$, the situation is slightly
different since we need to impose the abundance ratio between $\thCO{}$ and
$\HCOp{}$. Adapting the implemented estimator to every possible a priori
knowledge is thus required. At the end of this process, we get an initial
estimation noted $\btheta^{(0)}$.

\subsection{Iterative gradient descent}
\label{sec:mle:gradient}

To refine our estimation of the negative log-likelihood, we use the
following iterative estimation in which the Hessian is the Fisher
matrix starting from $\btheta^{(0)}$
\begin{equation}
  \widehat \btheta^{(i+1)}=\widehat \btheta^{(i)} - \bI_F^{-1}(\widehat
  \btheta^{(i)}) \nabla_{\btheta} (\widehat \btheta^{(i)}),
  \label{eq:scoring}  
\end{equation}
where $i$ is the ith iteration, $\bI_F(\btheta)$ is the Fisher matrix as a
function of $\btheta$, and $\nabla_{\btheta} (\btheta)$ is the gradient of
the negative log-likelihood
\begin{equation}
  \forall j
  \quad
  \left[\nabla_{\btheta} (\btheta)\right]_{j} =
  \sum_{l=1}^L \frac{1}{\sigma_{b,l}^2}\sum_{n=1}^K \frac{\partial
    s_{n,l}}{\partial \theta_j} (s_{n,l}-x_{n,l}),
\end{equation}
where $j$ is the index of the estimated parameter.  The inversion of Fisher
matrix in Eq.~\ref{eq:scoring} may be numerically unstable. When the
ratio between the highest and the smallest eigenvalue is higher than
$10^9$, we use the Moore Penrose pseudo-inverse instead of the inverse in
Eq.~\ref{eq:scoring}.

Furthermore, when Eq.~\ref{eq:scoring} would lead to parameters out
of the initial grid, we saturate these parameters to prevent RADEX from
crashing (see Table~\ref{tab:min_max}). For example, we impose $\nHH$
to be smaller than $10^{6.5}\pccm$. While a volume density at
$10^{6.5}\pccm$ cannot be trusted in this case, the column density
may still be reliable. These conditions add some a priori
knowledge that is not available in the computation of the Cramér-Rao
bound.

%%%%%%%%%%%%%%%%
% DATA SUPPLEMENT
%%%%%%%%%%%%%%%%%
  
%Column densities

\newcommand{\FigDataNCeiO}{%
  \begin{figure*}
    \centering %
    \includegraphics[width=0.8\linewidth]{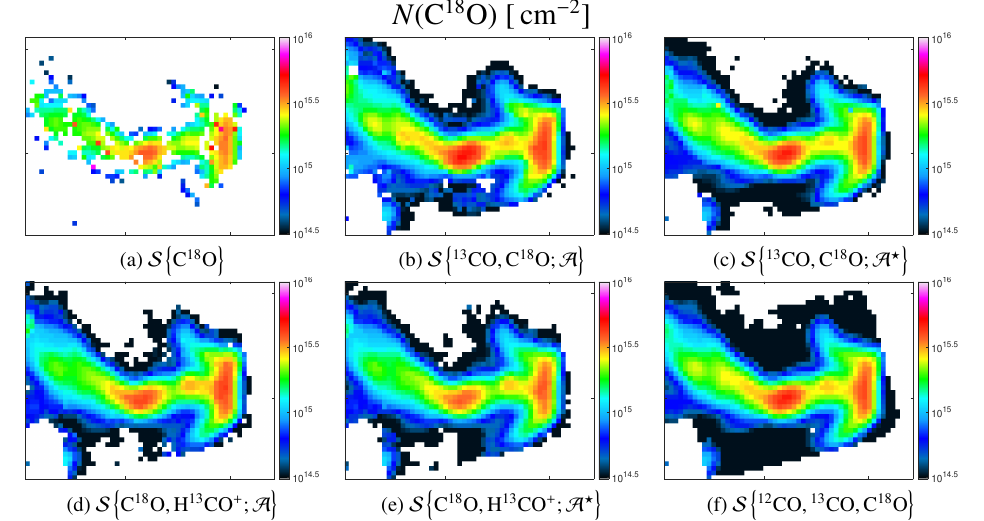}
    \\[\bigskipamount]
    \includegraphics[width=0.9\linewidth]{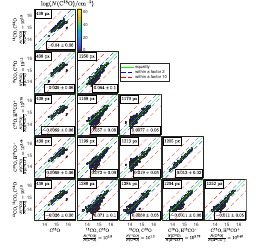}
    \caption{Same as Fig.~\ref{fig:data:N13CO} but for $N(\CeiO)$.}
    \label{fig:data:NC18O}
  \end{figure*}
}

\newcommand{\FigDataNHCOpNHthCOp}{%
  \begin{figure*}
    \centering %
    \includegraphics[width=\linewidth]{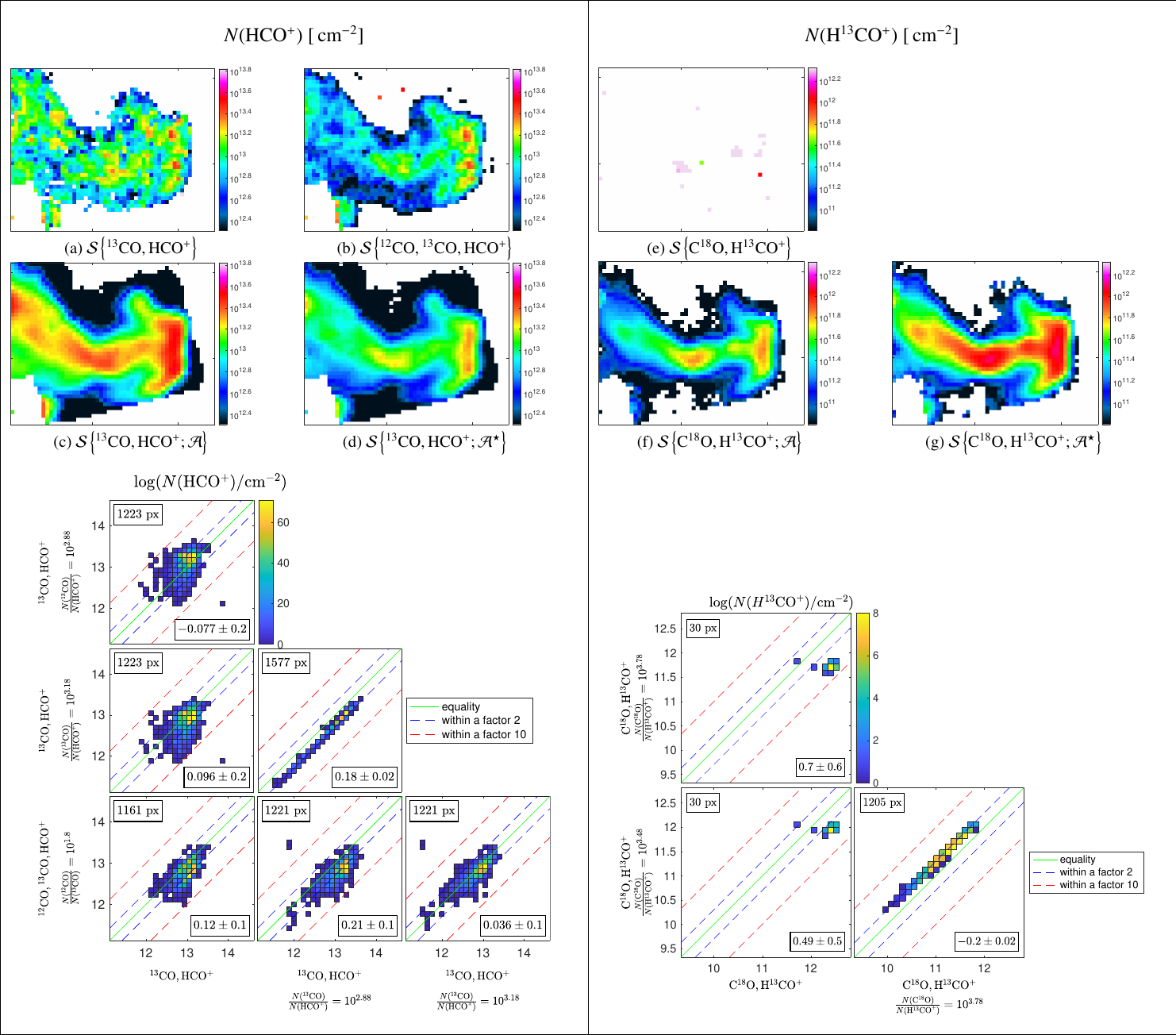}
    \caption{Same as Fig.~\ref{fig:data:N13CO} but for $N(\HCOp)$
      \textbf{(left)} and $N(\HthCOp)$ \textbf{(right)}.}
    \label{fig:data:NHCOp:N13HCOp}
  \end{figure*}
}

% Kinetic temperature

\newcommand{\FigDataTkin}{%
  \begin{figure*}
    \centering %
    \includegraphics[width=0.9\linewidth]{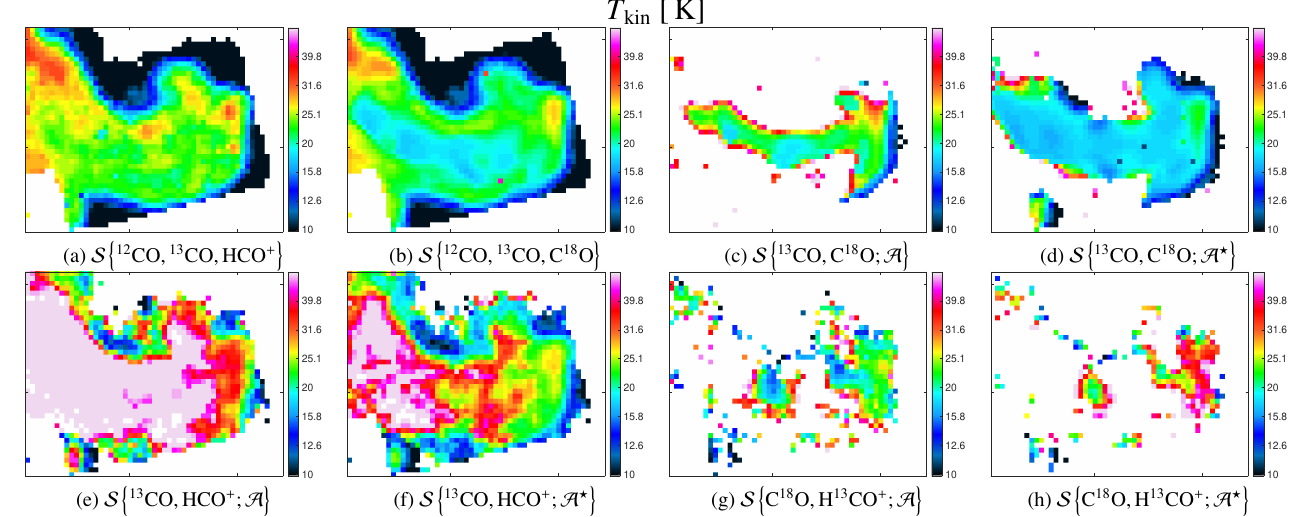}
    \\[\bigskipamount]
    \includegraphics[width=0.95\linewidth]{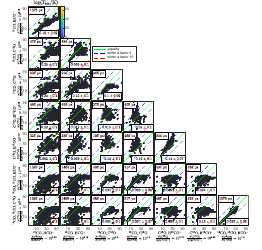}
    \caption{Same as Fig.~\ref{fig:data:N13CO} but for $\Tkin$.}
    \label{fig:data:Tkin}
  \end{figure*}
}

% Volume density

\newcommand{\FigDatan}{%
  \begin{figure*}
    \centering %
    \includegraphics[width=0.9\linewidth]{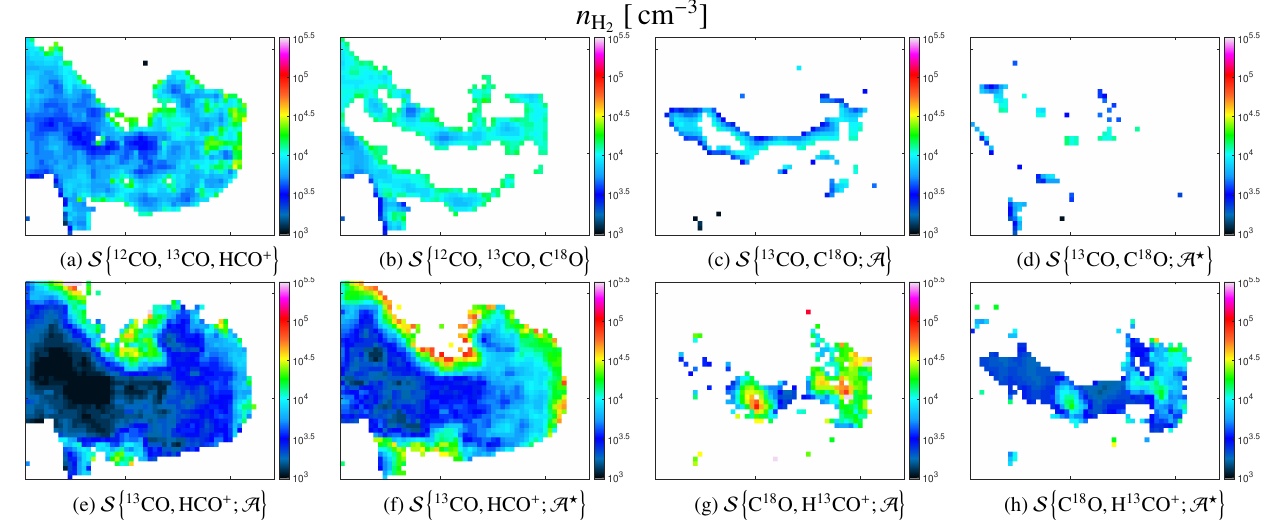}
    \\[\bigskipamount]
    \includegraphics[width=0.95\linewidth]{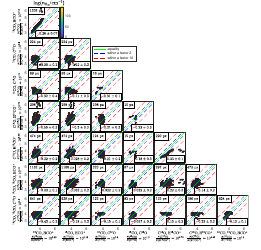}
    \caption{Same as Fig.~\ref{fig:data:N13CO} but for $\nHH$.}
    \label{fig:data:n}
  \end{figure*}
}

% Pressure

\newcommand{\FigDataPth}{%
  \begin{figure*}
    \centering %
    \includegraphics[width=0.9\linewidth]{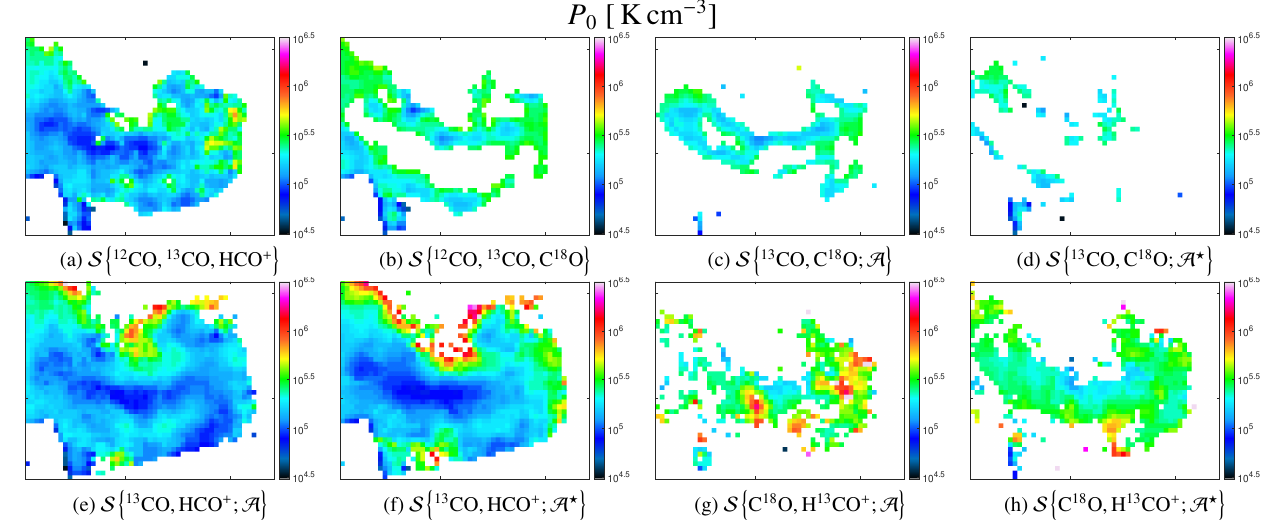}
    \\[\bigskipamount]
    \includegraphics[width=0.95\linewidth]{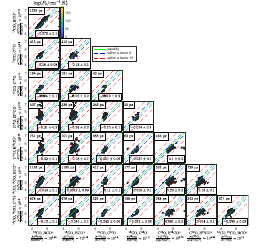}
    \caption{Same as Fig.~\ref{fig:data:N13CO} but for $P_\emr{th}$.}
    \label{fig:data:Pth}
  \end{figure*}
}

\section{Supplemental figures on data fit}
\label{app:data:supplement}

% Column density
\FigDataNCeiO{}%
\FigDataNHCOpNHthCOp

% Tkin, n, Pth
\FigDataTkin{} %
\FigDatan{} %
\FigDataPth{} %

\clearpage{}

%%%%%%%%%%%%%%%
% CRB SUPPLEMENT
%%%%%%%%%%%%%%%

\newcommand{\FigCRBLineCharacteristics}{%
  \begin{figure*}
    \centering %
    \includegraphics[width=1\linewidth]{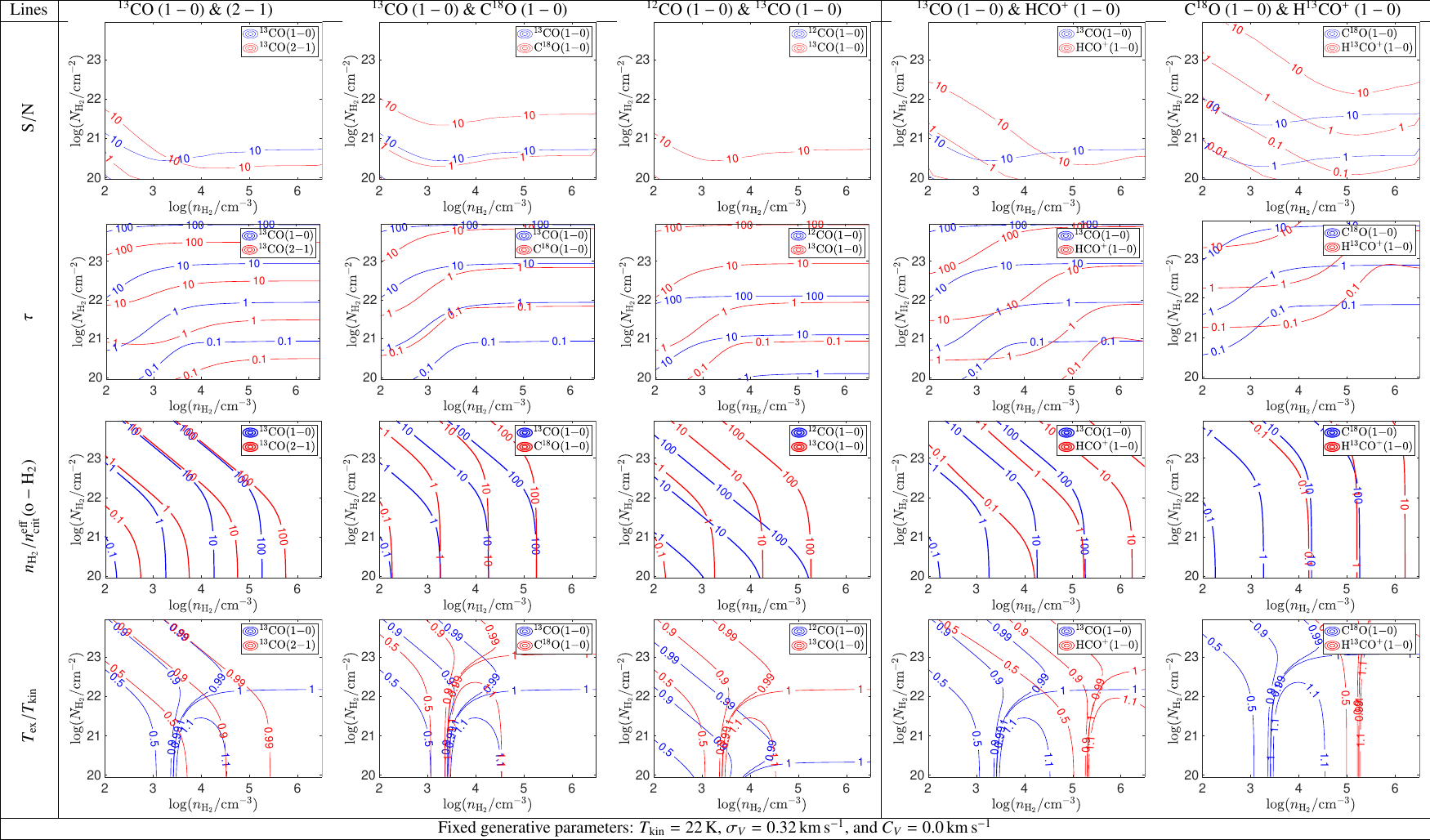}
    \caption{ Line main characteristics: S/Ns, opacities, ratios of the
      volume density to the effective critical densities, and ratios of the
      excitation and kinetic temperatures. The blue and red curves show the
      results for the \Jone{} and \Jtwo{} lines, respectively. The values
      of the parameters that are fixed when generating the RADEX spectra
      are listed in the bottom row, except for the abundances that are
      given in Eq.~\ref{eq_abundance}.}
    \label{fig:crb:line:characteristics}
  \end{figure*}
}

\section{Supplemental figure for the parametric study of the reference
  precisions}
\label{app:crb:supplement}

Figure~\ref{fig:crb:line:characteristics} shows the main characteristics
(signal-to-noise ratios, opacities, ratios of the volume density to the
effective critical densities, and ratios of the excitation temperature over
the kinetic temperature) of all the lines used in this paper for the
same range of volume and column densities.

\FigCRBLineCharacteristics{} %

\end{appendix}

\end{document}